\documentstyle[12pt]{article}

\catcode`\@=11
\long\def\@makefntext#1{ 
\protect\noindent \hbox to 3.2pt {\hskip-.9pt
$^{{\ninerm\@thefnmark}}$\hfil}#1\hfill} 

\def\thefootnote{\fnsymbol{footnote}}
 \def\@makefnmark{\hbox to 0pt{$^{\@thefnmark}$\hss}}  

\def\ps@myheadings{\let\@mkboth\@gobbletwo
\def\@oddhead{\hbox{} 
\rightmark\hfil\ninerm\thepage}
\def\@oddfoot{}\def\@evenhead{\ninerm\thepage\hfil 
\leftmark\hbox{}}\def\@evenfoot{}
\def\sectionmark##1{}\def\subsectionmark##1{}}

\begin{document}

\newcommand{\symbolfootnote}{\renewcommand{\thefootnote}
	{\fnsymbol{footnote}}}
\renewcommand{\thefootnote}{\fnsymbol{footnote}}
\newcommand{\alphfootnote}
	{\setcounter{footnote}{0}
	 \renewcommand{\thefootnote}{\sevenrm\alph{footnote}}}

\newcounter{sectionc}\newcounter{subsectionc}\newcounter{subsubsectionc}
\renewcommand{\section}[1] {\vspace{0.6cm}\addtocounter{sectionc}{1}
\setcounter{subsectionc}{0}\setcounter{subsubsectionc}{0}\noindent
	{\bf\thesectionc. #1}\par\vspace{0.4cm}}
\renewcommand{\subsection}[1] {\vspace{0.6cm}\addtocounter{subsectionc}{1}
	\setcounter{subsubsectionc}{0}\noindent
	{\it\thesectionc.\thesubsectionc. #1}\par\vspace{0.4cm}}
\renewcommand{\subsubsection}[1] {\vspace{0.6cm}\addtocounter{subsubsectionc}{1}
	\noindent {\rm\thesectionc.\thesubsectionc.\thesubsubsectionc.
	#1}\par\vspace{0.4cm}}
\newcommand{\nonumsection}[1] {\vspace{0.6cm}\noindent{\bf #1}
	\par\vspace{0.4cm}}

\newcounter{appendixc}
\newcounter{subappendixc}[appendixc]
\newcounter{subsubappendixc}[subappendixc]
\renewcommand{\thesubappendixc}{\Alph{appendixc}.\arabic{subappendixc}}
\renewcommand{\thesubsubappendixc}
	{\Alph{appendixc}.\arabic{subappendixc}.\arabic{subsubappendixc}}

\renewcommand{\appendix}[1] {\vspace{0.6cm}
        \refstepcounter{appendixc}
        \setcounter{figure}{0}
        \setcounter{table}{0}
        \setcounter{equation}{0}
        \renewcommand{\thefigure}{\Alph{appendixc}.\arabic{figure}}
        \renewcommand{\thetable}{\Alph{appendixc}.\arabic{table}}
        \renewcommand{\theappendixc}{\Alph{appendixc}}
        \renewcommand{\theequation}{\Alph{appendixc}.\arabic{equation}}
        \noindent{\bf Appendix \theappendixc #1}\par\vspace{0.4cm}}
\newcommand{\subappendix}[1] {\vspace{0.6cm}
        \refstepcounter{subappendixc}
        \noindent{\bf Appendix \thesubappendixc. #1}\par\vspace{0.4cm}}
\newcommand{\subsubappendix}[1] {\vspace{0.6cm}
        \refstepcounter{subsubappendixc}
        \noindent{\it Appendix \thesubsubappendixc. #1}
	\par\vspace{0.4cm}}

\def\abstracts#1{{
	\centering{\begin{minipage}{30pc}\tenrm\baselineskip=12pt\noindent
	\centerline{\tenrm ABSTRACT}\vspace{0.3cm}
	\parindent=0pt #1
	\end{minipage} }\par}}

\newcommand{\bibit}{\it}
\newcommand{\bibbf}{\bf}
\renewenvironment{thebibliography}[1]
	{\begin{list}{\arabic{enumi}.}
	{\usecounter{enumi}\setlength{\parsep}{0pt}
\setlength{\leftmargin 1.25cm}{\rightmargin 0pt}
	 \setlength{\itemsep}{0pt} \settowidth
	{\labelwidth}{#1.}\sloppy}}{\end{list}}

\topsep=0in\parsep=0in\itemsep=0in
\parindent=1.5pc

\newcounter{itemlistc}
\newcounter{romanlistc}
\newcounter{alphlistc}
\newcounter{arabiclistc}
\newenvironment{itemlist}
    	{\setcounter{itemlistc}{0}
	 \begin{list}{$\bullet$}
	{\usecounter{itemlistc}
	 \setlength{\parsep}{0pt}
	 \setlength{\itemsep}{0pt}}}{\end{list}}

\newenvironment{romanlist}
	{\setcounter{romanlistc}{0}
	 \begin{list}{$($\roman{romanlistc}$)$}
	{\usecounter{romanlistc}
	 \setlength{\parsep}{0pt}
	 \setlength{\itemsep}{0pt}}}{\end{list}}

\newenvironment{alphlist}
	{\setcounter{alphlistc}{0}
	 \begin{list}{$($\alph{alphlistc}$)$}
	{\usecounter{alphlistc}
	 \setlength{\parsep}{0pt}
	 \setlength{\itemsep}{0pt}}}{\end{list}}

\newenvironment{arabiclist}
	{\setcounter{arabiclistc}{0}
	 \begin{list}{\arabic{arabiclistc}}
	{\usecounter{arabiclistc}
	 \setlength{\parsep}{0pt}
	 \setlength{\itemsep}{0pt}}}{\end{list}}

\newcommand{\fcaption}[1]{
        \refstepcounter{figure}
        \setbox\@tempboxa = \hbox{\tenrm Fig.~\thefigure. #1}
        \ifdim \wd\@tempboxa > 6in
           {\begin{center}
        \parbox{6in}{\tenrm\baselineskip=12pt Fig.~\thefigure. #1 }
            \end{center}}
        \else
             {\begin{center}
             {\tenrm Fig.~\thefigure. #1}
              \end{center}}
        \fi}

\newcommand{\tcaption}[1]{
        \refstepcounter{table}
        \setbox\@tempboxa = \hbox{\tenrm Table~\thetable. #1}
        \ifdim \wd\@tempboxa > 6in
           {\begin{center}
        \parbox{6in}{\tenrm\baselineskip=12pt Table~\thetable. #1 }
            \end{center}}
        \else
             {\begin{center}
             {\tenrm Table~\thetable. #1}
              \end{center}}
        \fi}

\def\@citex[#1]#2{\if@filesw\immediate\write\@auxout
	{\string\citation{#2}}\fi
\def\@citea{}\@cite{\@for\@citeb:=#2\do
	{\@citea\def\@citea{,}\@ifundefined
	{b@\@citeb}{{\bf ?}\@warning
	{Citation `\@citeb' on page \thepage \space undefined}}
	{\csname b@\@citeb\endcsname}}}{#1}}

\newif\if@cghi
\def\cite{\@cghitrue\@ifnextchar [{\@tempswatrue
	\@citex}{\@tempswafalse\@citex[]}}
\def\citelow{\@cghifalse\@ifnextchar [{\@tempswatrue
	\@citex}{\@tempswafalse\@citex[]}}
\def\@cite#1#2{{$\null^{#1}$\if@tempswa\typeout
	{IJCGA warning: optional citation argument
	ignored: `#2'} \fi}}
\newcommand{\citeup}{\cite}

\def\fnm#1{$^{\mbox{\scriptsize #1}}$}
\def\fnt#1#2{\footnotetext{\kern-.3em
	{$^{\mbox{\sevenrm #1}}$}{#2}}}

\font\twelvebf=cmbx10 scaled\magstep 1
\font\twelverm=cmr10 scaled\magstep 1
\font\twelveit=cmti10 scaled\magstep 1
\font\elevenbfit=cmbxti10 scaled\magstephalf
\font\elevenbf=cmbx10 scaled\magstephalf
\font\elevenrm=cmr10 scaled\magstephalf
\font\elevenit=cmti10 scaled\magstephalf
\font\bfit=cmbxti10
\font\tenbf=cmbx10
\font\tenrm=cmr10
\font\tenit=cmti10
\font\ninebf=cmbx9
\font\ninerm=cmr9
\font\nineit=cmti9
\font\eightbf=cmbx8
\font\eightrm=cmr8
\font\eightit=cmti8

\input psfig
\def \mpla#1#2#3{Mod. Phys. Lett. A {\bf#1}, #2 (#3)}
\def \nc#1#2#3{Nuovo Cim. {\bf#1}, #2 (#3)}
\def \np#1#2#3{Nucl. Phys. {\bf#1}, #2 (#3)}
\def \pisma#1#2#3#4{Pis'ma Zh. Eksp. Teor. Fiz. {\bf#1}, #2 (#3) [JETP Lett.
{\bf#1}, #4 (#3)]}
\def \pl#1#2#3{Phys. Lett. {\bf#1}, #2 (#3)}
\def \plb#1#2#3{Phys. Lett. B {\bf#1}, #2 (#3)}
\def \nim#1#2#3{Nucl. Instr.\& Meth. {\bf#1}, #2 (#3)}
\def \pr#1#2#3{Phys. Rev. {\bf#1}, #2 (#3)}
\def \prd#1#2#3{Phys. Rev. D {\bf#1}, #2 (#3)}
\def \prl#1#2#3{Phys. Rev. Lett. {\bf#1}, #2 (#3)}
\def \prp#1#2#3{Phys. Rep. {\bf#1}, #2 (#3)}
\def \ptp#1#2#3{Prog. Theor. Phys. {\bf#1}, #2 (#3)}
\def \rmp#1#2#3{Rev. Mod. Phys. {\bf#1}, #2 (#3)}
\def \rp#1{~~~~~\ldots\ldots{\rm rp~}{#1}~~~~~}
\def \yaf#1#2#3#4{Yad. Fiz. {\bf#1}, #2 (#3) [Sov. J. Nucl. Phys. {\bf #1},
#4 (#3)]}
\def \zhetf#1#2#3#4#5#6{Zh. Eksp. Teor. Fiz. {\bf #1}, #2 (#3) [Sov. Phys. -
JETP {\bf #4}, #5 (#6)]}
\def \zpc#1#2#3{Zeit. Phys. C {\bf#1}, #2 (#3)}
\def\etal{et al.}
\def \indentt{~~~~}
\def\bra#1{\left\langle #1\right|}
\def\ket#1{\left| #1\right\rangle}

\hspace*{\fill} HEPSY 96-01\linebreak
\hspace*{\fill} October 1996~\linebreak
\centerline{~}
\centerline{~}
\Large
\centerline{PROSPECTS FOR B-PHYSICS IN THE NEXT DECADE}
\centerline{~}
\large
\centerline{Sheldon Stone}
\centerline{Department of Physics}
\centerline{Syracuse Univeristy}
\centerline{Syracuse, N.Y. 13244-1130}
\centerline{Email: Stone@suhep.phy.syr.edu}
\centerline{~}
\centerline{~}
\centerline{~}
\centerline{~}
\vspace*{2cm}
\vspace{2cm}
\abstracts{\normalsize
In these lectures I review what has been learned from studies of $b$-quark
decays, including semileptonic decays ($V_{ub}$ and $V_{cb}$),
$B^o-\overline{B}^o$ mixing and rare $B$ decays. Then a discussion on CP
violation follows, which leads to a summary of plans for future experiments and
what is expected to be learned from them.} 

\normalsize 

\vspace{2.4cm}
\begin{flushleft}
.\dotfill .
\end{flushleft}
\begin{center}
{\it Presented at NATO Advanced Study Institute on Techniques and Concepts of
High Energy Physics, Virgin Islands, July 1996} 
\end{center}
\newpage

\section {INTRODUCTION}

My assignment is to discuss ``Future B Physics Experiments." But to understand 
what results we desire, it is  necessary to understand past accomplishments and have 
a firm theoretical background. In this paper I will give a brief theoretical
introduction to the ``Standard Model," and historical introduction to the study
of $b$ quark decays. Then I will discuss in some detail the physics already
found including: $B$ lifetimes, semileptonic $B$ decays and the CKM couplings
$V_{cb}$ and $V_{ub}$, $B^o-\bar{B^o}$ mixing, rare $b$ decays, and CP violation
in $K^o_L$ decays. Following this is a pedantic discussion on CP violation in 
$B$ decays, which leads into a discussion of future experiments.

\subsection {Theoretical Background}

The physical states of the ``Standard Model"  are comprised of  left-handed 
doublets containing leptons and quarks and right handed singlets\cite{SM} 
\begin{eqnarray}
&\left(\begin{array}{c}u\\d\end{array}\right)_L
\left(\begin{array}{c}c\\s\end{array}\right)_L
\left(\begin{array}{c}t\\b\end{array}\right)_L,&~u_R,~d_R,~c_R,~s_R,~t_R,~b_R\\
&\left(\begin{array}{c}e^-\\\nu_e\end{array}\right)_L
\left(\begin{array}{c}\mu^-\\\nu_{\mu}\end{array}\right)_L
\left(\begin{array}{c}\tau^-\\\nu_{\tau}\end{array}\right)_L,&~e^-_R,
~\mu^-_R,~\tau^-_R,~{\nu_e}_R,~{\nu_{\mu}}_R,~{\nu_{\tau}}_R.
\end{eqnarray}
 
The gauge bosons, $W^{\pm}$, $\gamma$ and $Z^o$ couple to  
mixtures of the physical $d,~ s$ and $b$ states. This mixing is described
by the Cabibbo-Kobayashi-Maskawa (CKM) matrix (see below).\cite{ckm}

The Lagrangian for charged current weak decays is
\begin{equation}
L_{cc}=-{g\over\sqrt{2}}J^{\mu}_{cc}W^{\dagger}_{\mu}+h.c., \label{eq:lagrange}
\end{equation}
where
\begin{equation}
J^{\mu}_{cc} =\left(\bar{\nu}_e,~\bar{\nu}_{\mu},~\bar{\nu}_{\tau}\right)
\gamma^\mu \left(\begin{array}{c}e_L\\ \mu_L\\ \tau_L\\\end{array}\right) +
\left(\bar{u}_L,~\bar{c}_{L},~\bar{t}_{L}\right)\gamma^\mu V_{CKM} 
\left(\begin{array}{c}d_L\\  s_L\\ b_L\\ \end{array}\right) 
\end{equation}
and
\begin{equation}
V_{CKM} =\left(\begin{array}{ccc} 
V_{ud} &  V_{us} & V_{ub} \\
V_{cd} &  V_{cs} & V_{cb} \\
V_{td} &  V_{ts} & V_{tb}  \end{array}\right).\end{equation}

\begin{figure}[hbtp]
\vspace{-.9cm}
\centerline{\psfig{figure=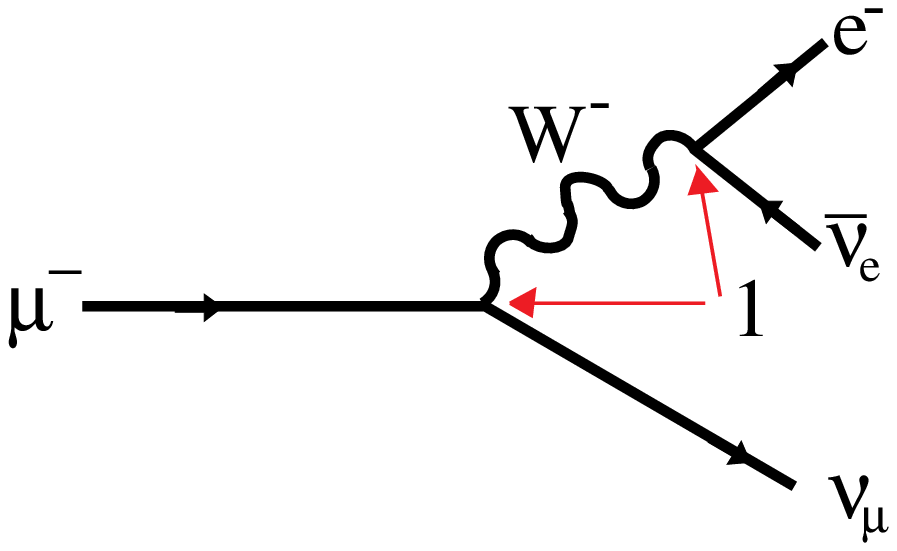,height=1.8in,bbllx=100bp,bblly=500bp,bburx=450bp,bbury=700bp,clip=}}
\vspace{-.8cm}
\fcaption{\label{mu_decay}Diagram for muon decay.}
\end{figure}

Multiplying the mass eigenstates $(d, s, b)$ by the CKM matrix leads to the
weak eigenstates $(d', s', b')$. There are nine complex CKM elements. These 18 
numbers can be reduced to four independent quantities by applying unitarity 
constraints and the fact that the phases of the quark wave functions are arbitrary. 
These four remaining numbers are {\bf fundamental constants} of nature that 
need to be determined from experiment, like any other
fundamental constant such as $\alpha$ or $G$. In the Wolfenstein 
approximation\footnote{In higher order other terms have an imaginary part; in
particular the $V_{cd}$ term becomes $-\lambda-A^2\lambda^5(\rho+i\eta)$, which
is important for CP violation in $K^o_L$ decay.} 
~the matrix is written as\cite{wolf}
\begin{equation}
V_{CKM} = \left(\begin{array}{ccc} 
1-\lambda^2/2 &  \lambda & A\lambda^3(\rho-i\eta) \\
-\lambda &  1-\lambda^2/2 & A\lambda^2 \\
A\lambda^3(1-\rho-i\eta) &  -A\lambda^2& 1  
\end{array}\right)
\end{equation}
The constants $\lambda$ and $A$ are determined from charged-current weak 
decays. To see how this is done, first consider muon decay. The muon decays weakly 
into $\nu_{\mu}e^-\bar{\nu_e}$ as shown in Fig.~\ref{mu_decay}.
The decay width is given by\cite{mudecay}
\begin{equation}
\Gamma_{\mu}={G^2_F\over 192\pi^3}m_{\mu}^5 
\times {\rm (radiative~corrections)} .  \label{eq:Gf}
\end{equation}

The couplings at the vertices are unity for leptons. This process serves to
measure  the weak interaction decay constant (Fermi constant) $G_F$.

\begin{figure}[hbtp]
\vspace{-.4cm}
\centerline{\psfig{figure=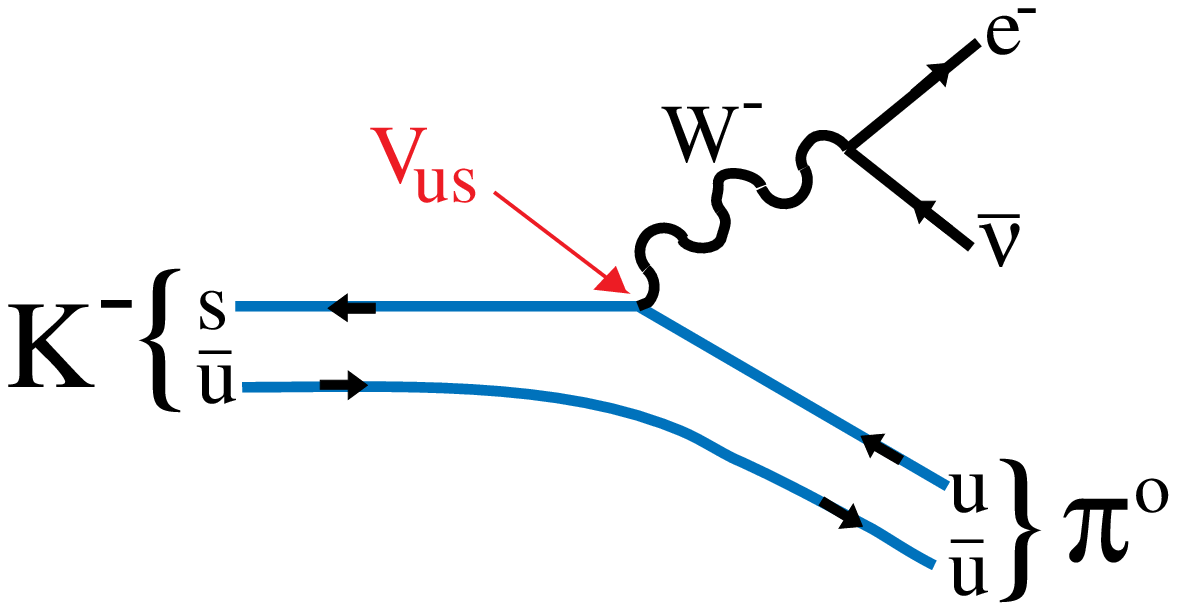,height=2.2in,bbllx=90bp,bblly=450bp,bburx=500bp,bbury=700bp,clip=}}
\vspace{-1.2cm}
\fcaption{\label{k_decay}Semileptonic $K^-$ decay diagram.}
\end{figure}

A charged current decay diagram for strange quark decay is shown in
Fig.~\ref{k_decay}. Here 
the CKM element $V_{us}$ is present. The decay rate is given by a formula similar 
to equation~(\ref{eq:Gf}), with the muon mass replaced by the $s$-quark mass and an 
additional factor of $|V_{us}|^2$. Two  complications arise since we are now 
measuring a decay process involving hadrons, $K^-\to \pi^o e^-\bar{\nu}$ rather 
than elementary constituents.  One is that the $s$-quark mass is not well  defined 
and the other  is that we must make corrections for the probability that the 
$\bar{u}$-spectator-quark indeed forms a $\pi^o$ with the ${u}$-quark  from the 
$s$-quark decay. These considerations will be discussed in greater detail in the 
semileptonic $B$ decays section. For now\cite{PDG} remember that 
$\lambda = V_{us}=0.2205\pm 0.0018$  and, $A \approx 0.8$.
Constraints on $\rho$ and $\eta$ are found from other measurements. These will
also be discussed later.

\subsection{$B$ Decay Mechanisms}

\begin{figure}[hbtp]
\vspace{.1cm}
\centerline{\psfig{figure=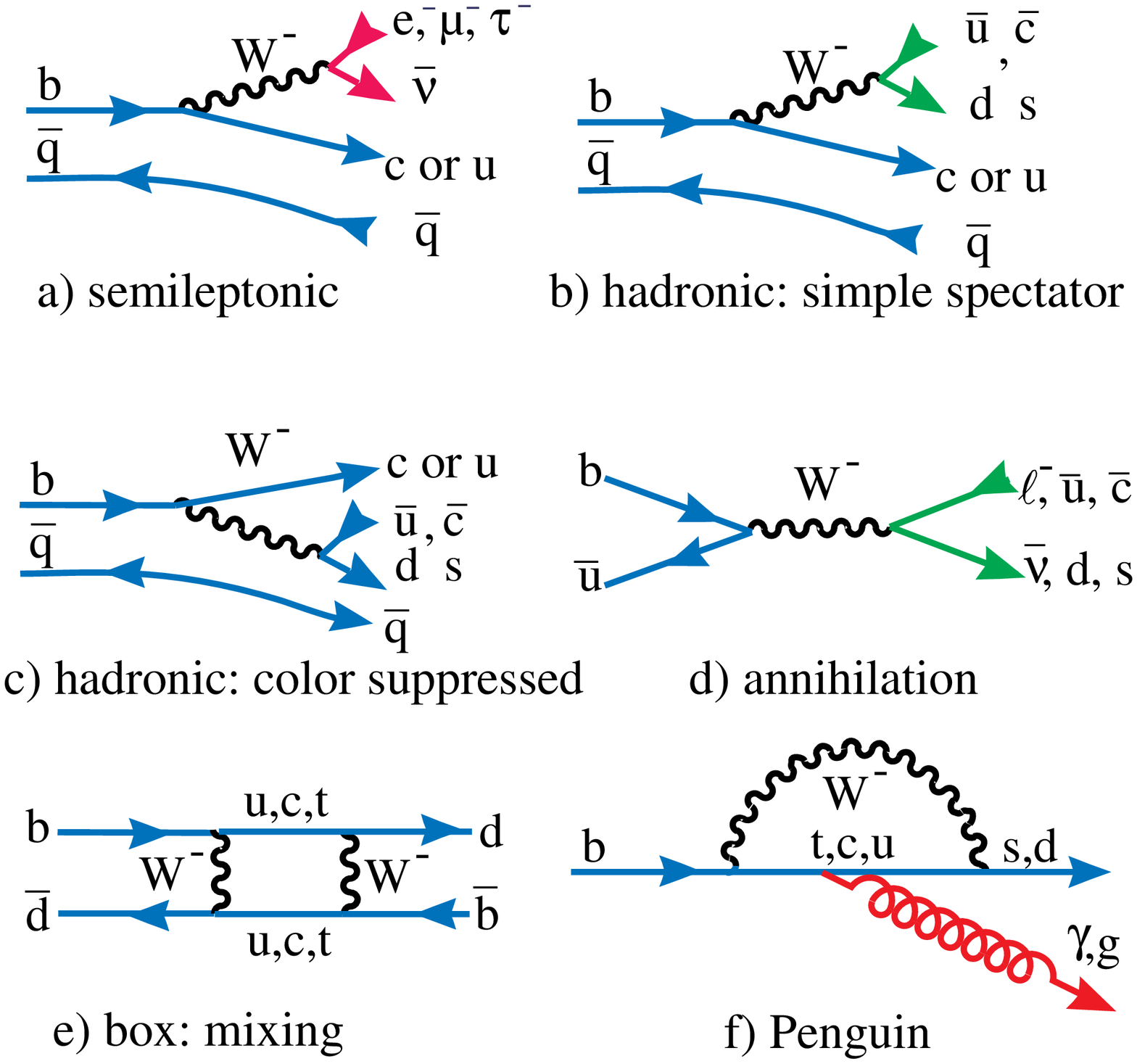,height=8.5in,bbllx=0bp,bblly=0bp,bburx=600bp,bbury=750bp,clip=}}
\vspace{-3.4cm}
\fcaption{\label{Bdecaymech}Various mechanisms for $B$ meson decay.}
\end{figure}

Fig.~\ref{Bdecaymech} shows sample diagrams for $B$ decays. Semileptonic
decays  are shown in Fig.~\ref{Bdecaymech}(a). The name ``semileptonic" is
given, since  there are both hadrons and leptons in the final state.  The
leptons arise from the  virtual $W^-$, while the hadrons come from the coupling
of the spectator anti-quark with either the $c$ or $u$ quark from the $b$ quark
decay.  Note that the $B$ is  massive enough that all three lepton species can
be produced. The simple spectator diagram for hadronic decays
(Fig.~\ref{Bdecaymech}(b)) occurs when the virtual  
$W^-$ materializes as a quark-antiquark pair, 
rather than a lepton pair.  The terminology  {\it simple spectator} comes from
viewing the decay of the $b$ quark, while  ignoring the presence of the {\it
spectator} antiquark. If the colors of the quarks  from the virtual $W^-$ are
the same as the initial $b$ quark, then the color  suppressed diagram, 
Fig.~\ref{Bdecaymech}(c), can occur. While the amount of  color suppression is
not well understood, a good first order guess is that these modes  are
suppressed in amplitude by the color factor 1/3 and thus in rate by 1/9, with 
respect to the non-color suppressed spectator diagram.

The annihilation diagram shown in Fig.~\ref{Bdecaymech}(d) occurs when the $b$ 
quark and spectator anti-quark find themselves in the same space-time region and 
annihilate by coupling to a virtual $W^-$. The probability of such a wave function 
overlap between the $b$ and $\bar{u}$-quarks is proportional to a numerical 
factor called $f_B$.  The decay amplitude is also proportional to the coupling 
$V_{ub}$. The mixing and penguin diagrams will be discussed later.

\section{What is known}
\subsection{Early history}

\begin{figure}[hbtp]
\vspace{-.5cm}
\centerline{\psfig{figure=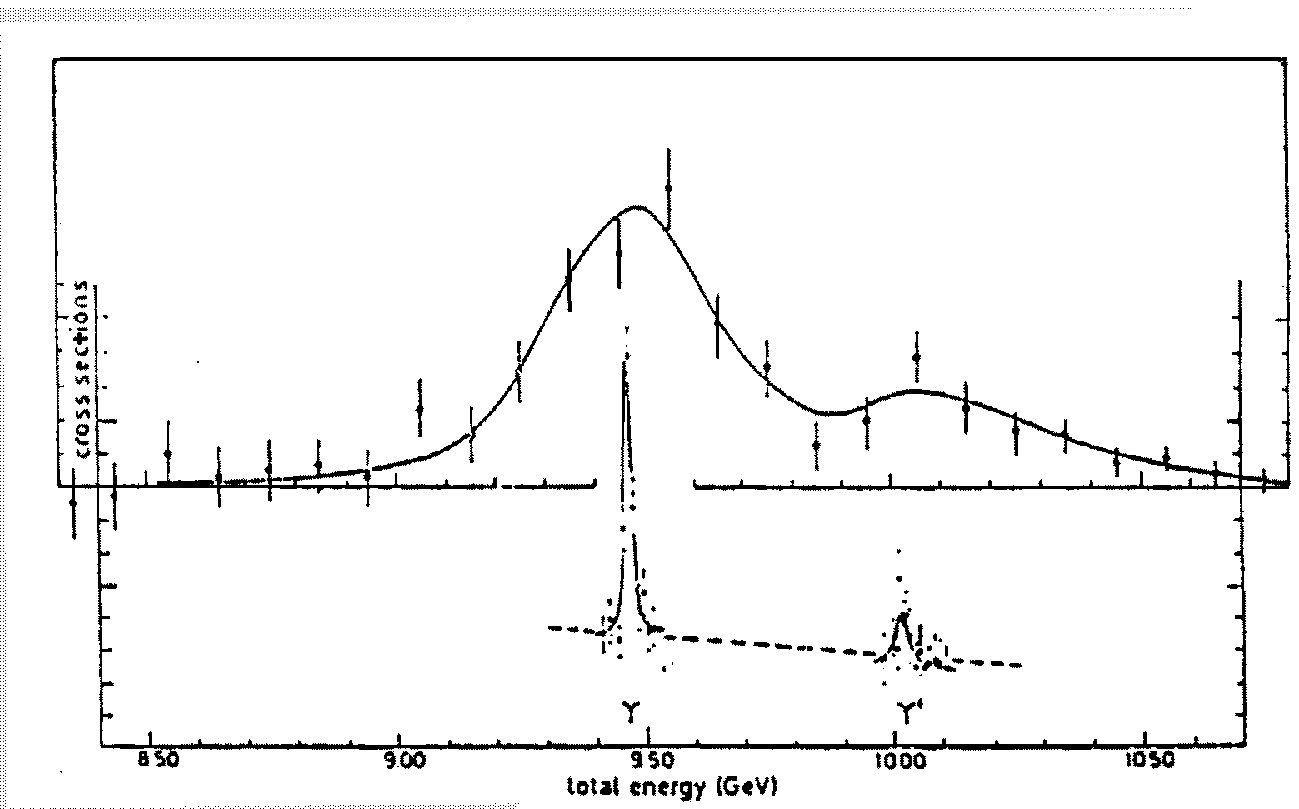,height=2.8in,bbllx=8bp,bblly=3bp,bburx=377bp,bbury=223bp,clip=}}
\vspace{.2cm}
\fcaption{\label{upsilon_old} The data on top is the $\mu^+\mu^-$ invariant mass
from the Columbia-Fermilab-Stony Brook collaboration and the data shown below
is the total $e^+e^-$ cross-section from the DESY-Heidelberg-Hamburg-Munchen
collaboration.}
\end{figure}

The first experimental evidence for $b$ quarks was found at Fermilab by looking
at  high mass dimuon pairs in 800 GeV proton interactions on nuclear 
targets.\cite{leon}  Their results are shown in Fig.~\ref{upsilon_old} along
with  subsequent data from DESY using $e^+e^-$ annihilations which shows
narrow  peaks at the masses of the $\Upsilon$ and $\Upsilon '$ 
resonances.\cite{desy}

The  natural width of the peaks is narrower than the energy resolution of
either  experiment leading to the interpretation that these states are
comprised of  a  bound $b\bar{b}$ quark system. The narrow decay width is
similar to the situation  in charmonium, i.e. the decay width is proportional
to the strong coupling constant $\alpha_s^3$.

\begin{figure}[hbtp]
\vspace{-.3cm}
\centerline{\psfig{figure=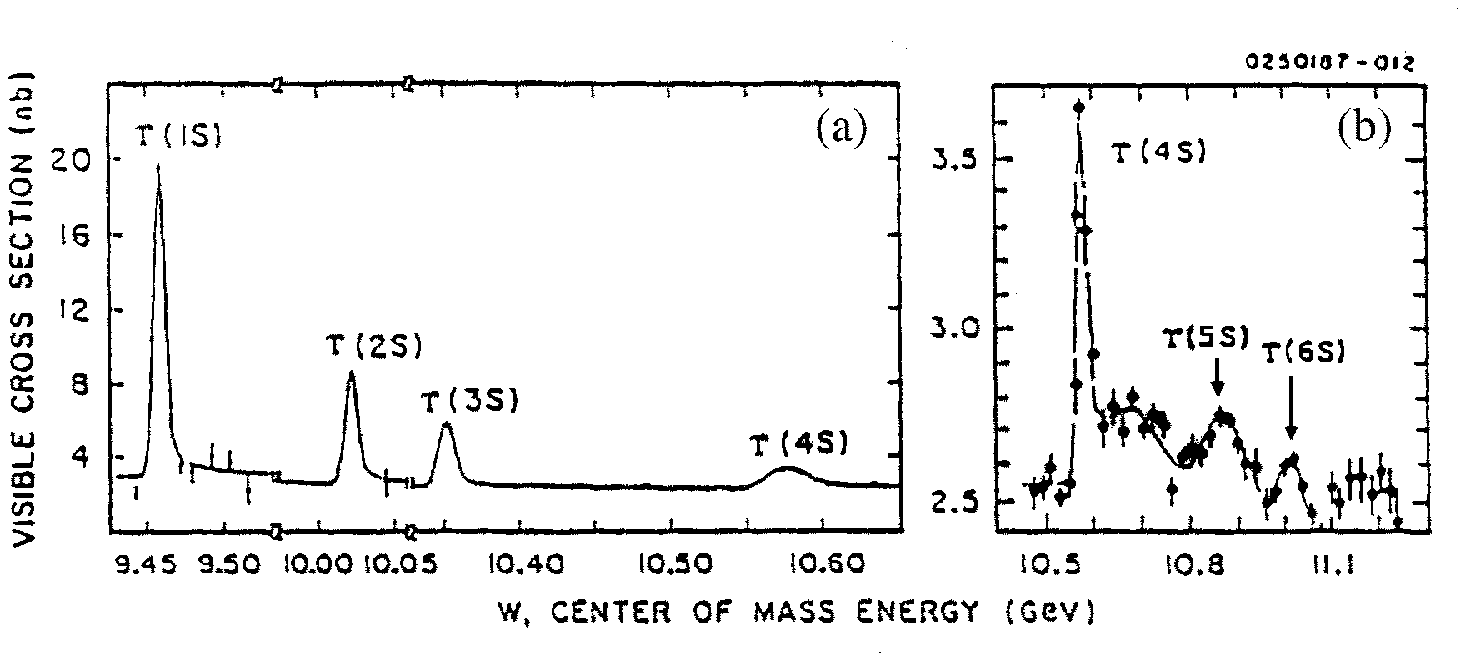,height=2.8in,bbllx=0bp,bblly=0bp,bburx=500bp,bbury=184bp,clip=}}
\vspace{-.5cm}
\fcaption{\label{cleo_scan} Hadronic cross-section scan in the Upsilon
region, (a) shows 1S-4S and (b) region above 4S.}
\end{figure}

As the DESY machine was limited in center-of-mass energy at that time, the
torch  was passed to the CLEO experiment at the CESR  $e^+e^-$ storage ring. An
early  total cross-section scan is shown in Fig.~\ref{cleo_scan}(a). A new narrow
state, the  $\Upsilon ''$ (or $\Upsilon (3S)$), appears along with a state wider
than the  experimental resolution, the $\Upsilon (4S)$. 

\begin{figure}[hbtp]
\vspace{-1.5cm}
\centerline{\psfig{figure=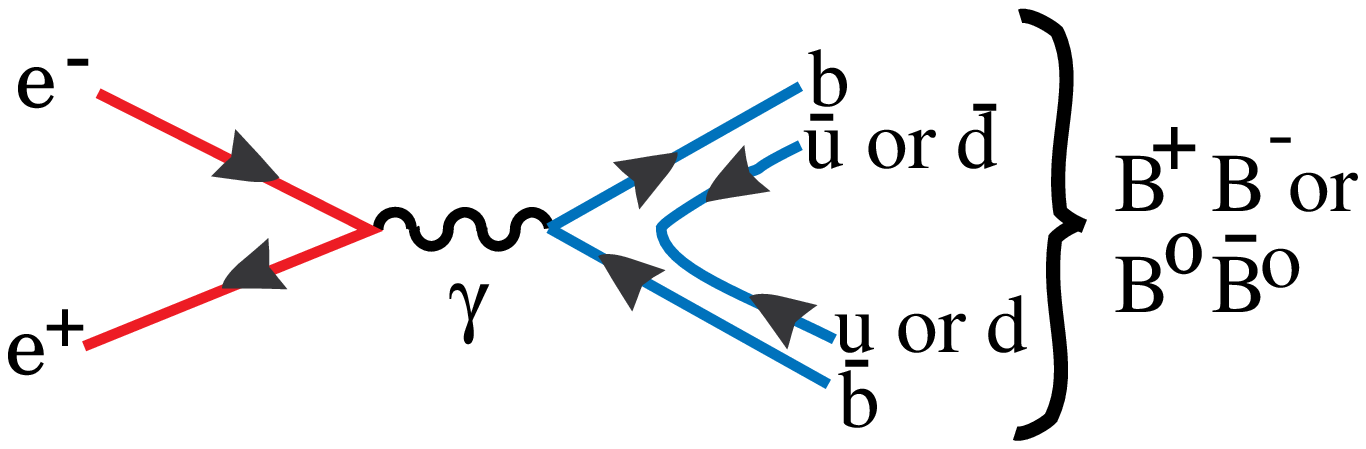,height=2.8in,bbllx=0bp,bblly=300bp,bburx=500bp,bbury=550bp,clip=}}
\vspace{-2.4cm}
\fcaption{\label{epemtoBB} $B$ production mechanism at the $\Upsilon (4S)$.}
\end{figure}

The mechanism of $b$ quark production in $e^+e^-$ collisions and the subsequent 
production of the final states $B^+B^-$ and $B^o\bar{B}^o$ from the 
$\Upsilon(4S)$ are shown in Fig.~\ref{epemtoBB}. Subsequent data shown
in  Fig.~\ref{cleo_scan}(b) shows that the cross-section is $\approx$1 nb and
details  structures in the total cross-section at higher energies.\cite{fives}
 Little data
has  been taken  above the $\Upsilon (4S)$, however.

Many properties of $B$ meson decays have been discovered by two $e^+e^-$ 
experiments operating at the $\Upsilon (4S)$ resonance, CLEO at CESR and 
ARGUS at DESY (the DESY machine group upgraded the energy so they could do this 
physics).  Fully reconstructed $B$ meson decays were first seen by CLEO and the 
$B$ masses determined.\cite{Breconfirst} Now there are several thousand fully reconstructed decays 
in many modes allowing for branching ratio determinations.  A different 
technique is used to reconstruct $B$ mesons at the $\Upsilon (4S)$ than at other 
machines. At this resonance we have
\begin{eqnarray}
e^+e^-\to\Upsilon (4S) & \to B^-B^+ \\ 
&~\to B^o\overline{B}^o~.
\end{eqnarray}
From energy conservation, the energy of each $B$ is equal to the beam energy, 
$E_{beam}$ (the center-of-mass energy is twice $E_{beam}$). To reconstruct 
exclusive $B$ meson decays, we first require the energy of the decay products be 
consistent with the beam energy. Suppose the final state we are considering is 
$D^o\pi^-$. We require that 
\begin{equation}
E_{D^o}+E_{\pi^-} = E_{beam}. \label{eq:econ}
\end{equation}
In practice this means that the difference between the left-hand side and the right-
hand side is less than $\approx$3 times the error on the measured energy sum.

\begin{figure}[hbtp]
\vspace{1.1cm}
\centerline{\psfig{figure=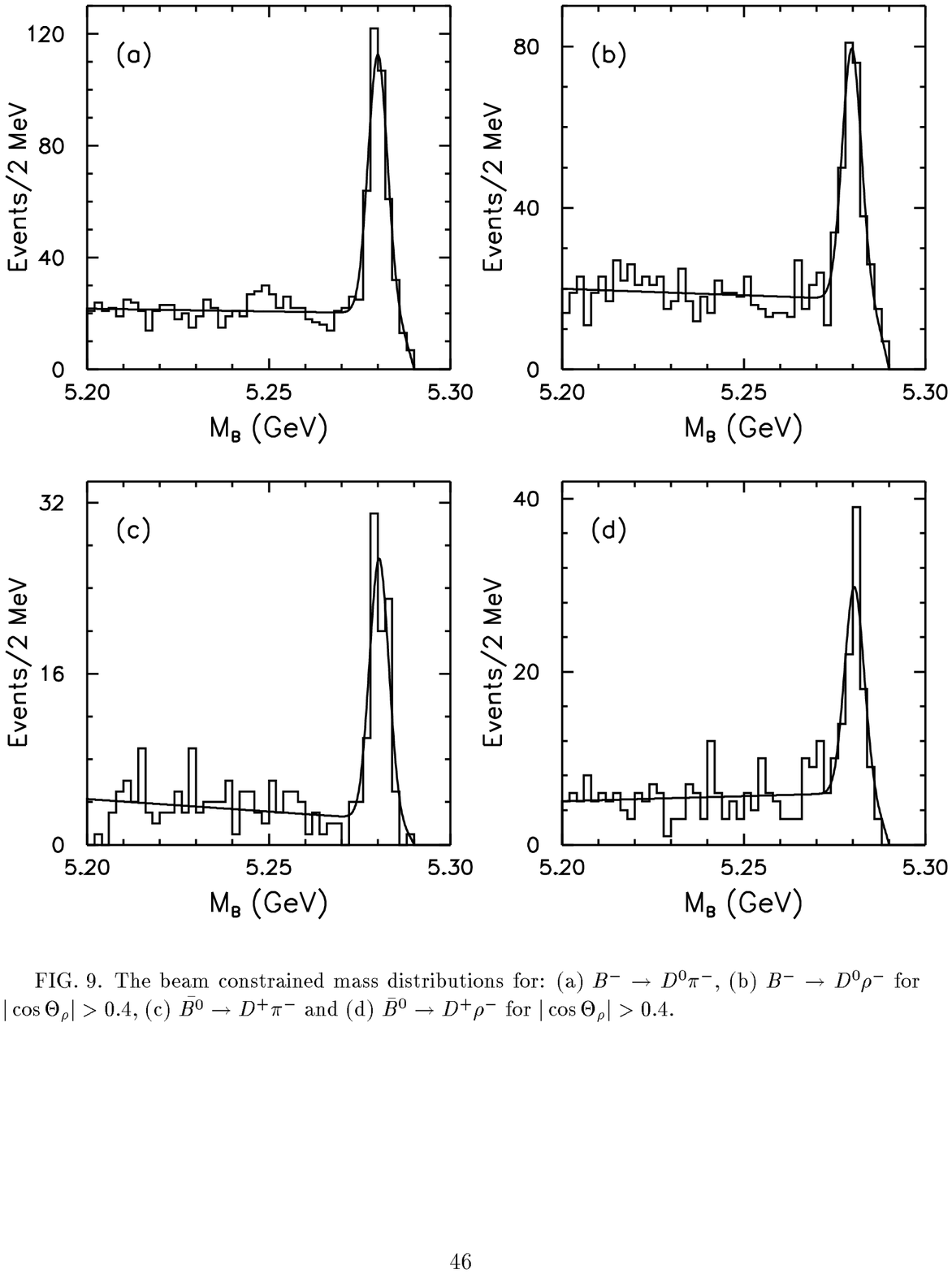,height=6in,bbllx=0bp,bblly=200bp,bburx=600bp,bbury=700bp,clip=}}
\vspace{-.2cm}
\fcaption{\label{m_dpi} Beam constrained mass distribution from CLEO
for (a) $B^-\to D^o\pi^-$, (b) $B^-\to D^o\rho^-$, (c) $\overline{B}^o\to
D^+\pi^-$ and (d) $\overline{B}^o\to D^+\rho^-$.}
\end{figure}

The next step is to compute the invariant mass of the candidate $B^-$ using the 
well known beam energy:
\begin{equation}
m_B = \sqrt{E_{beam}^2 -
(\overrightarrow{p}\!_{D^o}+\overrightarrow{p}\!_{\pi^-})^2}. \label{eq:mcon}
\end{equation}
In practice this technique leads to large background rejections and a $B$ mass 
resolution of $\sigma\approx$2.5 MeV  (at CESR) which is due mostly to the energy 
spread of the beam. A few sample $B$ decay candidate mass plots are shown in
Fig.~\ref{m_dpi} from the CLEO experiment.\cite{Breconnew}

\begin{figure}[hbtp]
\vspace{-3.1cm}
\centerline{\psfig{figure=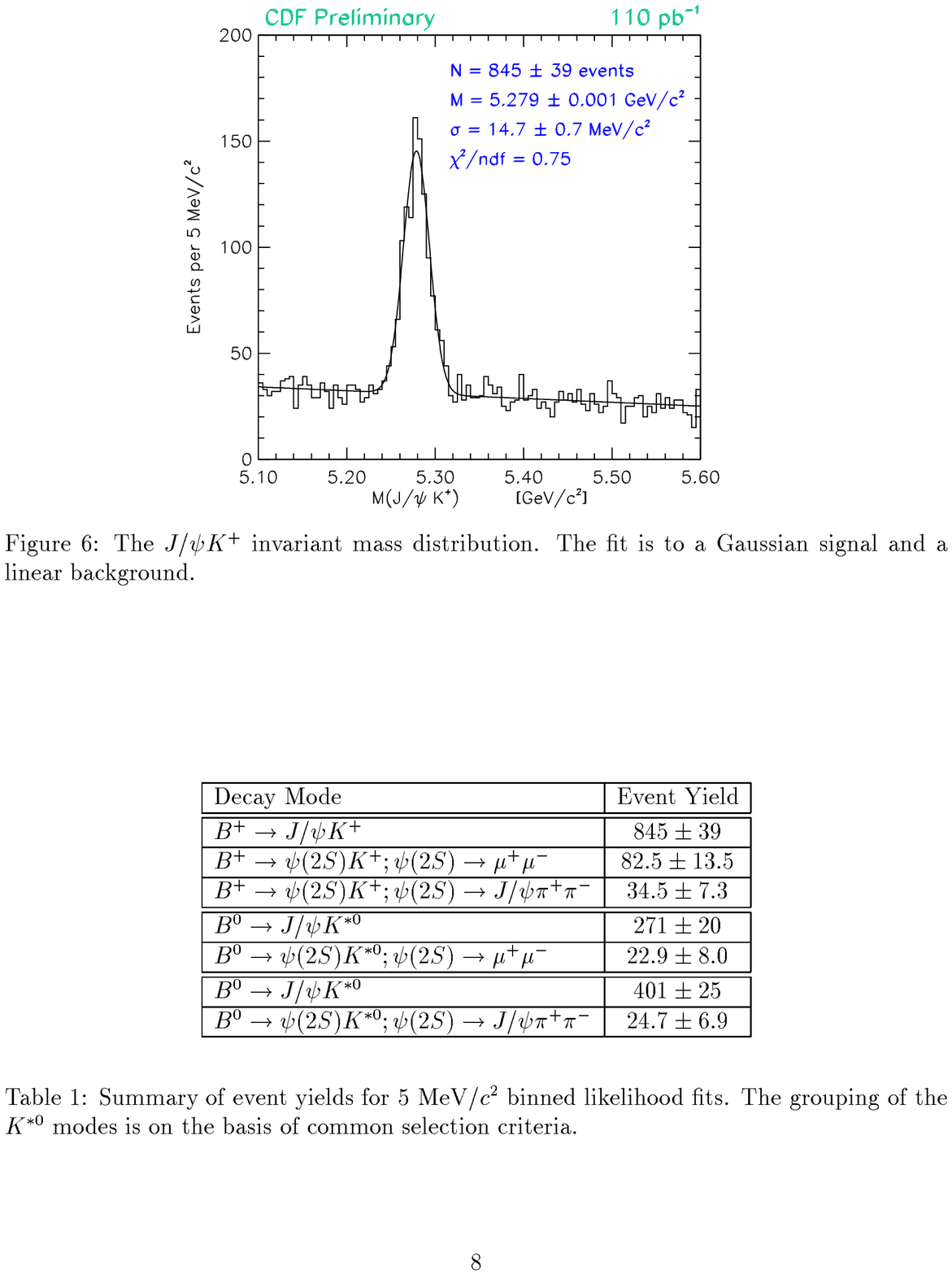,height=3.5in,bbllx=0bp,bblly=436bp,bburx=600bp,bbury=700bp,clip=}}
\centerline{\psfig{figure=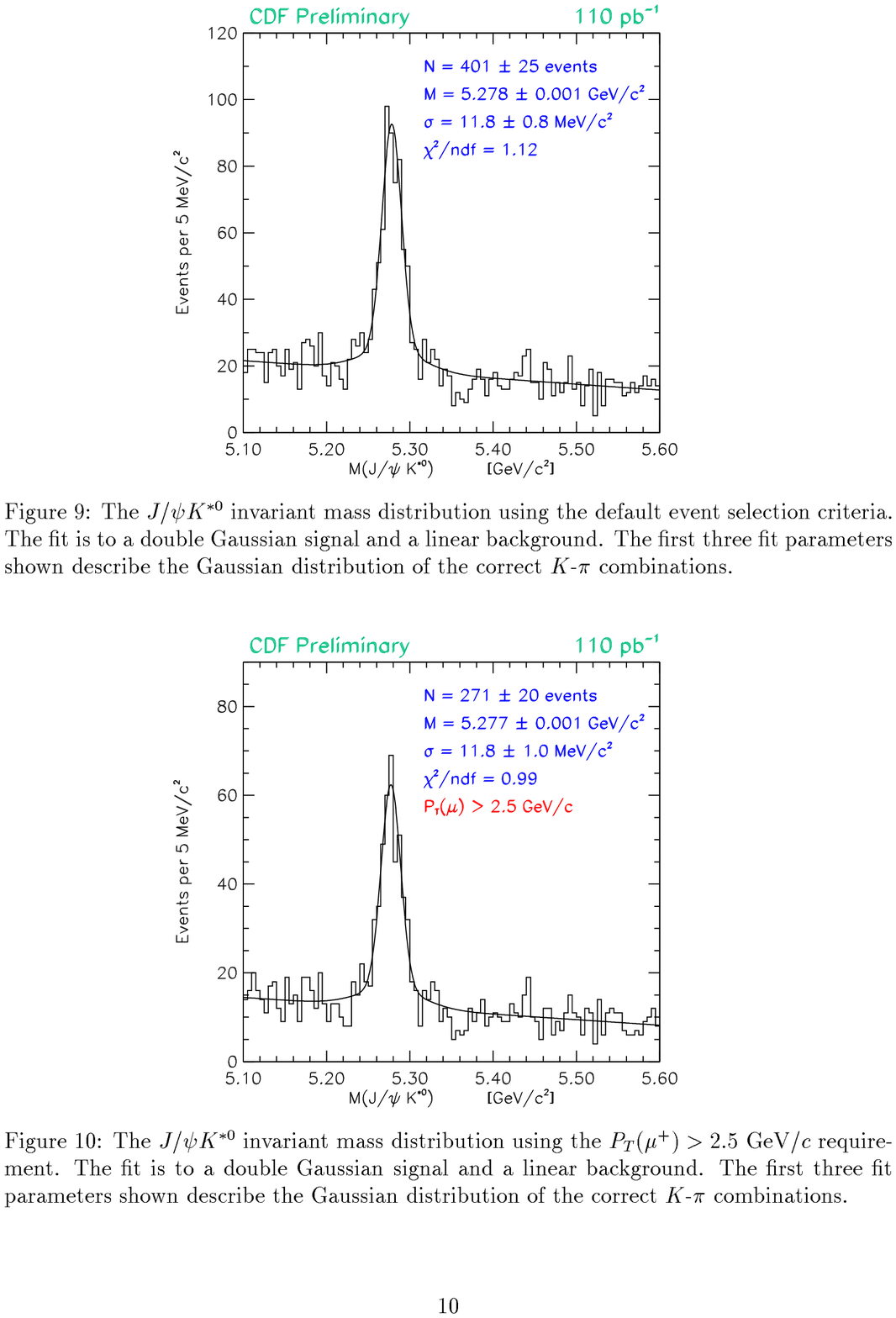,height=3.5in,bbllx=0bp,bblly=500bp,bburx=600bp,bbury=761bp,clip=}}
\fcaption{\label{cdfBmass}Invariant mass spectra from CDF for $\psi K^+$ and
$\psi K^{*o}$ candidates.}
\end{figure}

Hadronic production rates for $b$ quarks have been measured at two $p\bar{p}$ 
colliders, UA1 at the SPS,\cite{UA1cross} and CDF at the
Tevatron.\cite{CDFcross}  E789 has also measured $b$ production using an 800 GeV
proton beam hitting nuclear targets.\cite{E789cross} CDF has reconstructed $B$
meson decays into modes containing a $\psi$ meson. These are
shown\cite{cdf_mass_plts} in Fig.~\ref{cdfBmass}.

\subsection{Lifetimes}

Lifetimes are a fundamental property of elementary particles.
The  $b$ quark lifetime, however, was measured before the individual lifetimes
of $b$  flavored hadrons  at the higher energy $e^+e^-$ machines, PEP
and PETRA.\cite{liferevold}
More recent measurements have come from LEP, SLD and CDF.\cite{liferevnew} The meaning of $b$ quark
lifetime is really the average of the $B$ hadron lifetimes over the kinds
of $B$ hadrons which happen to be produced in the particular environment.
The results are 
summarized in Table ~\ref{table:life}.\cite{blife}
\begin{table}[th]\centering\tcaption{$B$ lifetime measurements (ps)}
\label{table:life}
\begin{tabular}{ccccc}\hline\hline
 & LEP Avg & CDF & SLD& World Avg \\\hline
b quark & 1.54$\pm$0.02 & 1.51$\pm$0.03 & 1.56$\pm$0.05 & 1.53$\pm$0.02 
\\
$B^-$ & 1.63$\pm$0.06 & 1.68$\pm$0.07 &  & 1.65$\pm$0.05 \\
${B}^o$ & 1.52$\pm$0.06 & 1.58$\pm$0.09 &  & 1.55$\pm$0.05 \\
${B}_s^o$ & 1.60$\pm$0.10 & 1.36$\pm$0.12 &  & 1.50$\pm$0.08 \\
$\Lambda_b$ & 1.21$\pm$0.07 & 1.32$\pm$0.17 &  & 1.23$\pm$0.06 \\
$\Xi_b$ & $1.39^{+0.34}_{-0.28}$ & & &   $1.39^{+0.34}_{-0.28}$\\
\hline\hline
\end{tabular}\end{table}

The meson lifetimes are nearly equal implying the dominance of the spectator
diagram. The $\Lambda_b$ lifetime appears to be shorter, which implies the
existence of other diagrams in baryon decay. This is very different from the
situation in charm decay where the $D^o$, $D^+$ and $\Lambda_c$ have different
lifetimes.

\subsection{The CKM element $V_{cb}$}
\vskip -1pc
\subsubsection{Theory of semileptonic decays}

The same type of semileptonic charged current decays used to find $V_{us}$ are
used to find $V_{cb}$ and $V_{ub}$. The basic diagram is shown in 
Fig.~\ref{Bdecaymech}(a). We can use either inclusive decays, where we look
only at the lepton and ignore the hadronic system at the lower vertex, or
exclusive decays where we focus on a particular single hadron. Theory currently
can  predict either the inclusive decay rate, or the exclusive decay rate when
there is only a single  hadron in the final state.  The fraction of
semileptonic decays into exclusive final states containing either a 
pseudoscalar or vector meson is given in Table~\ref{table:frac}.

\begin{table}[th]\centering\tcaption{Fraction of $q\to x\ell\nu$ to $0^-$ or $1^-$ final states}
\label{table:frac}
\begin{tabular}{crl}\hline\hline
$s$ quark & 100\% &$K\to\pi\ell\nu$ \\
$c$ quark & $>$90\% &$D\to (K+K^*)\ell\nu$ \\
        &  ? & $D\to (\pi+\rho)\ell\nu$ \\
$b$ quark & $\approx$66\% &$B\to (D+D^*)\ell\nu$ \\
        &  ? & $B\to (\pi+\rho)\ell\nu$ \\
$t$ quark &  0\% & $t$ does not form hadrons\\                
\hline\hline\normalsize
\end{tabular}\end{table}

Now let us briefly go through the mathematical formalism of semileptonic
decays. Let us start with pseudoscalar to pseudoscalar transitions. The decay
amplitude is given by\cite{semilepform}
\begin{equation}
A(\bar{B}\to me^-\bar{\nu})={G_F\over\sqrt{2}}V_{ij}L^{\mu}H_{\mu}{\rm ,~where} 
\end{equation}
\begin{equation}
L^{\mu}=\bar{u}_e\gamma^{\mu}\left(1-\gamma_5\right)v_{\nu} {\rm ,~and}
\end{equation}
\begin{equation}
H_{\mu}=\langle m|J^{\mu}_{had}(0)|\overline{B}\rangle=
f_+(q^2)(P+p)_{\mu}+f_-(q^2)(P-p)_{\mu},
\end{equation}
where $q^2$ is the four-momentum transfer squared between the $B$ and the $m$,
and $P(p)$ are four-vectors of the $B(m)$.
$H_{\mu}$ is the most general form the hadronic matrix element can have. It
is written in terms of the unknown $f$ functions that are called ``form-factors." 
It turns out that the term multiplying the $f_-(q^2)$ form-factor is the mass
of lepton squared. Thus for electrons and muons (but not $\tau$'s), the decay 
width is given by
\begin{equation}
{d\Gamma_{sl}\over dq^2} = {G_F^2|V_{ij}|^2K^3M_B^2\over 
24\pi^2}|f_+(q^2)|^2, {\rm ~~where}
\end{equation}
\begin{equation}
K={M_B\over 2}\left[\left(1-{{m^2-q^2}\over M^2_B}\right)-4{m^2q^2\over
M^4_B}\right]^{1/2}
\end{equation}
is the momentum of the particle $m$ (with mass $m$) in the $B$ rest frame.
In principle, ${d\Gamma_{sl}/ dq^2}$ can be measured over all $q^2$. Thus
 the shape of $f_+(q^2)$ can be determined experimentally. However, the 
normalization, $f_+(0)$ must be obtained from theory, for $V_{ij}$ to
be measured. In other words,
\begin{equation}
\Gamma_{SL} \propto |V_{ij}|^2|f_+(0)|^2{1\over \tau_B}\int K^3 g(q^2)dq^2,
\end{equation} 
where $g(q^2) =f_+(q^2)/f_+(0)$. Measurements of semileptonic $B$ decays 
give the integral term, while the lifetimes are measured separately, allowing
the product $|V_{ij}|^2|f_+(0)|^2$ to be experimentally determined. 

For pseudoscalar to vector transistions there are three independent
form-factors whose shapes and normalizations must be determined.\cite{threef}

\subsubsection{$\overline{B}^o\to D^+\ell^-\bar{\nu}$}

CLEO has recently measured the branching ratio and form-factor for the
reaction $\bar{B}^o\to D^+\ell^-\bar{\nu}$ using two different 
techniques.\cite{CLEOdplnu}
In the first method the final state is reconstructed finding only lepton
and $D^+$ candidates, where the  $D^+\to K^-\pi^+\pi^+$  decay is used. Then,
using the fact that the $B's$ produced at the $\Upsilon$(4S) are nearly
at rest the missing mass squared ($MM^2$) is calculated as

\begin{eqnarray}
MM^2 &=&E^2_{\nu}-\overrightarrow{P_{\nu}}^2\\ \nonumber
&=&\left(E_B-E_{D^+}-E_{\ell}\right)^2-\left(\overrightarrow{p}\!_B
-\overrightarrow{p}\!_{D^+}-\overrightarrow{p}\!_{\ell}\right)^2 \\ \nonumber
     &\approx &\left(E_B-(E_{D^+}+E_{\ell})\right)^2
-\left(\overrightarrow{p}\!_{D^+}+\overrightarrow{p}\!_{\ell}\right)^2\\
\nonumber
 &\approx& E^2_{beam}+m_B^2+m_{D^+}^2+m^2_{\ell}-
 2\overrightarrow{p}\!_{D^+}\cdot\overrightarrow{p}\!_{\ell},
\end{eqnarray}
where $E$ refers to particle energy, $m$ to mass and $\overrightarrow{p}$ to
three-momentum. The approximation on the third line results from setting $p_B$ to zero. This
approximation causes a widening of the $MM^2$ distribution, giving a r.m.s.
width of 0.2 GeV$^2$. 

\begin{figure}[hbtp]
\vspace{-4.cm}
\centerline{\psfig{figure=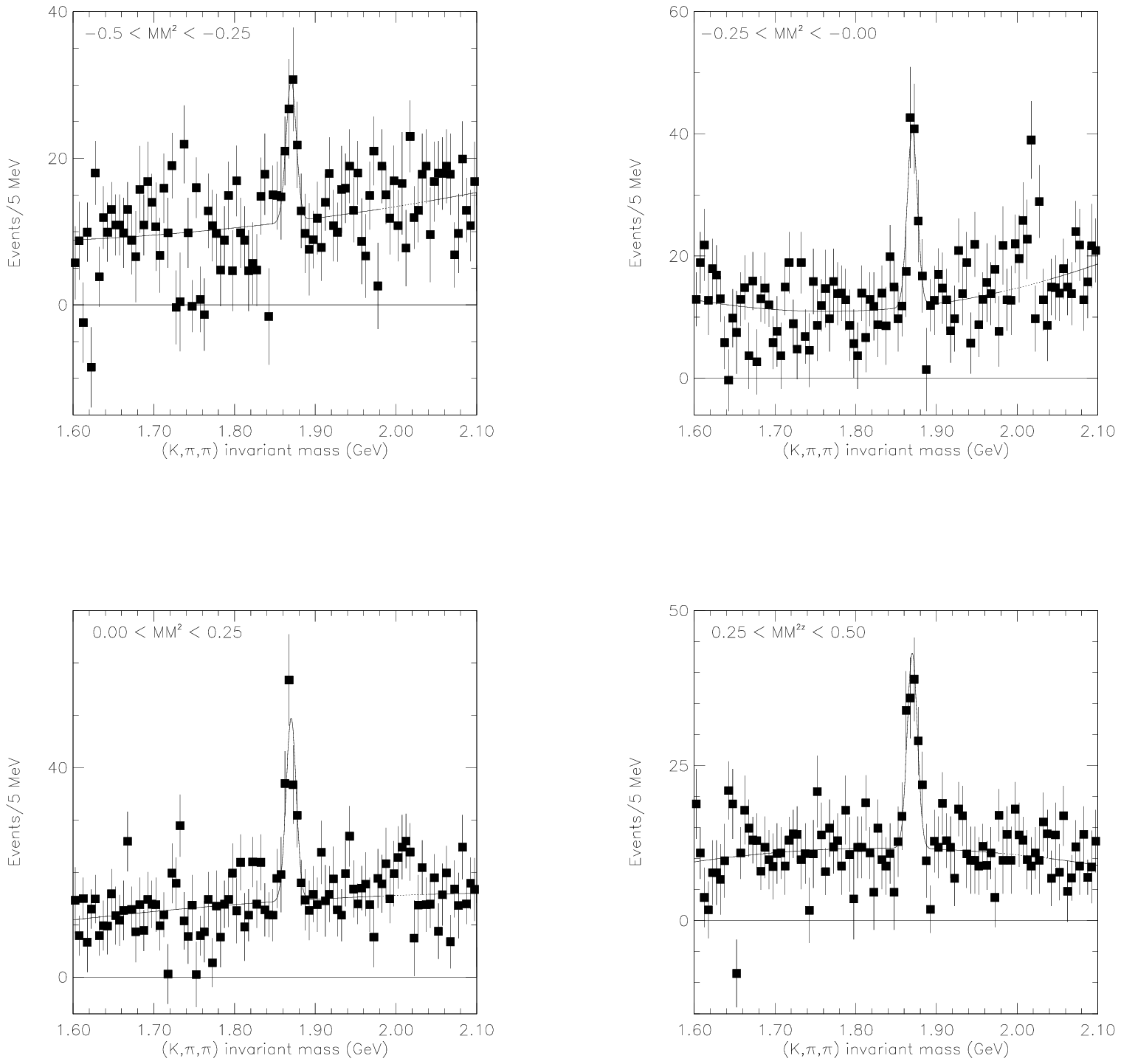,height=7in,bbllx=0bp,bblly=200bp,bburx=700bp,bbury=700bp,clip=}}
\vspace{-.8cm}
\fcaption{\label{mkpipi}Invariant $K^-\pi^+\pi^+$ mass spectra from CLEO for 
events with an opposite sign lepton in the interval $4>q^2>2$ in four
different $MM^2$ slices The curves are  a fit to a Gaussian signal shape summed
with a polynomial background.}
\end{figure}

This analysis is done by finding the number of $D^+$ events with opposite
sign leptons in different $q^2$ and $MM^2$ bins. The $K^-\pi^+\pi^+$ mass
distributions for the interval $4>q^2>2$ GeV$^2$ and several $MM^2$ bins are
shown in Fig.~\ref{mkpipi}. 

\begin{figure}[hbtp]
\vspace{.2cm}
\centerline{\psfig{figure=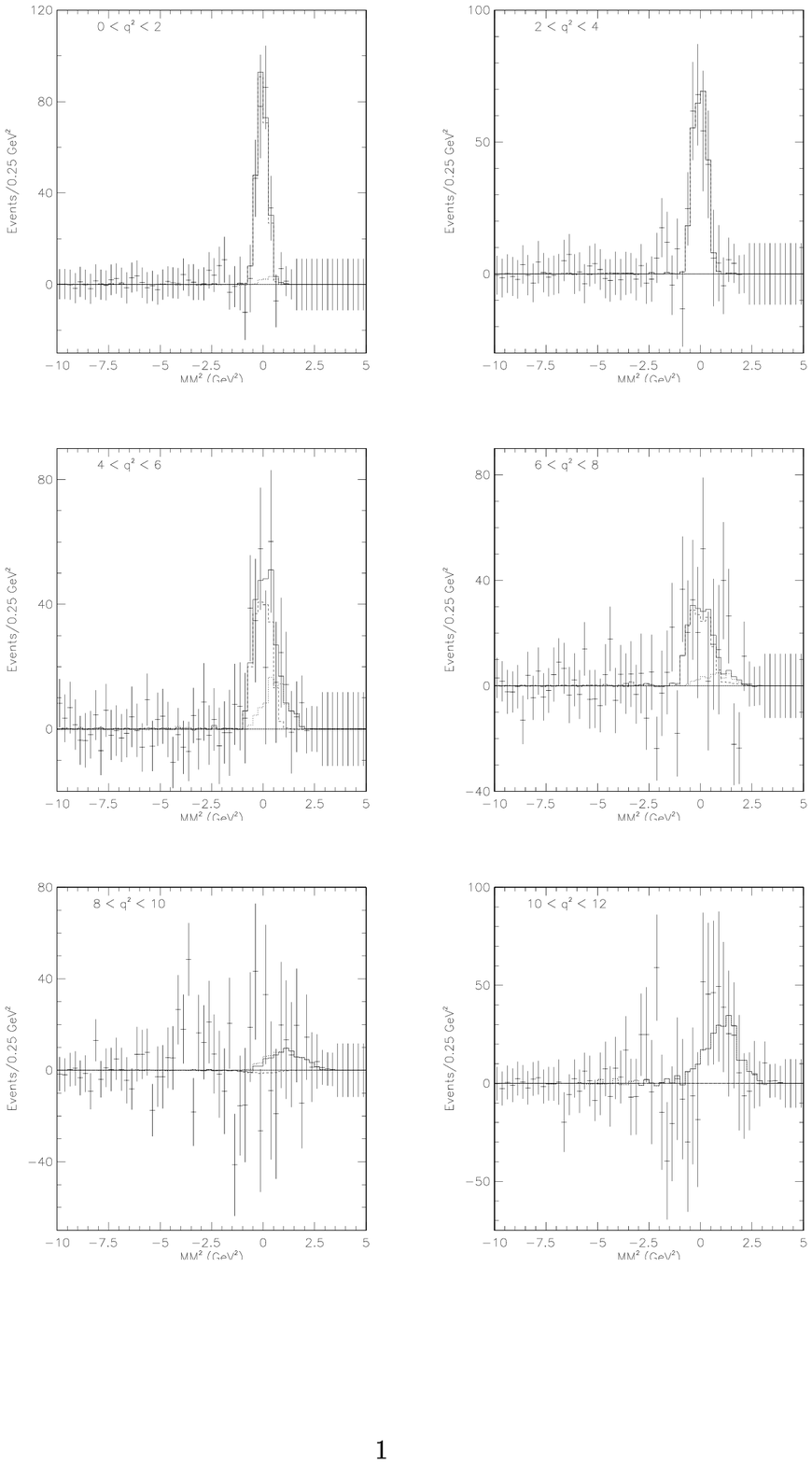,height=8.in,bbllx=0bp,bblly=100bp,bburx=600bp,bbury=700bp,clip=}}
\vspace{.3cm}
\fcaption{\label{allqsq}Fits to the $MM^2$ distribution for the
$D^+\ell^-\bar{\nu}$ (dashed) and $D^+ X\ell^-\bar{\nu}$ (dotted) components
and the sum (solid) in different $q^2$
intervals}
\end{figure}

There is also a large background from $\bar{B}^o\to D^{*+} X\ell^-\bar{\nu}$
decays where the $D^{*+}\to \pi^o D^+$. These events are reconstructed and
their $MM^2$ distribution is directly subtracted (after correcting for
efficiencies) from the candidate signal distribution. We are left with
a sample that contains $D^+\ell^-\bar{\nu}$ decays and also $D^+ X\ell^-\bar{\nu}$,
where $X$ can be a single hadron or hadrons but cannot be the result of final
state with a $D^{*+}$. We ascertain the total number of signal events by fitting
the $MM^2$ distribution in the different $q^2$ bins as shown in
Fig.~\ref{allqsq} to a $D^+\ell^-\bar{\nu}$ signal shape and a background shape
for $D^+ X\ell^-\bar{\nu}$.

\begin{figure}[hbtp]
\vspace{-2.1cm}
\centerline{\psfig{figure=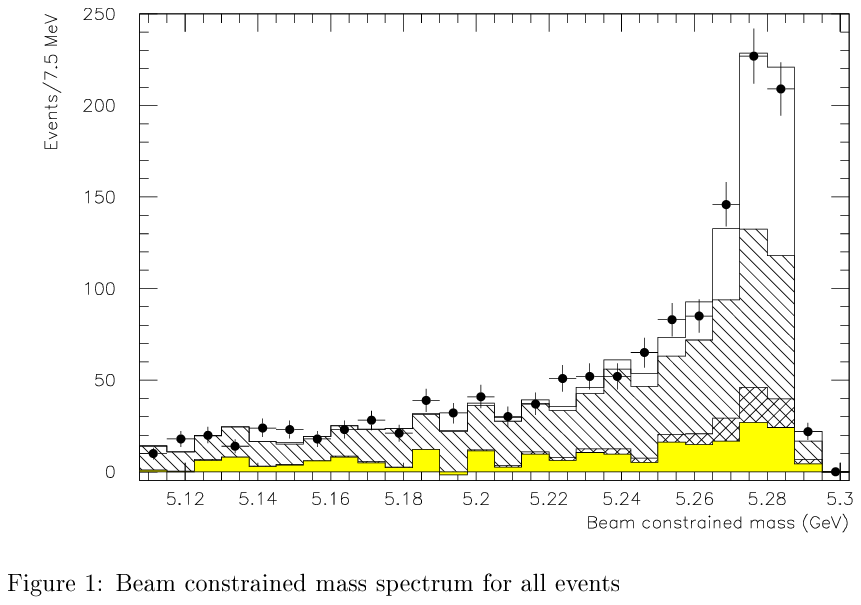,height=3.5in,bbllx=0bp,bblly=464bp,bburx=600bp,bbury=800bp,clip=}}
\vspace{-.05cm}
\fcaption{\label{dplnunu}Beam constrained mass spectrum for all events passing
the cuts. The white area represents the signal events, the hatched area
represents the combinatoric background, the crosshatched area represents the
$D^{*+}\ell^-\bar{\nu}$ and the shaded area represents all the remaining
backgrounds.}
\end{figure}

The second technique reconstructs the neutrino by using missing energy and
momentum measurements. Essentially all charged tracks and photons in the
event are added up and since the total energy must be equal to the center
of mass energy and the total three-momentum must be zero, any difference
is assigned to the neutrino. Events with a second lepton or which do not
conserve charge are eliminated. Furthermore, the momentum and energy
measurements must be consistent. Once the neutrino four-vector is determined,
the $B$ can be reconstructed in the ``usual" way as shown in Fig.~\ref{dplnunu}.

The $MM^2$ technique gives a branching ratio of $(1.75\pm 0.25 \pm 0.20)$\%,
while the neutrino reconstruction gives $(1.89\pm 0.22 \pm 0.35)$\%, giving a 
combined (preliminary) yield of $(1.78\pm 0.20 \pm 0.24)$\%. The statistical
errors in both methods are essentially uncorrelated, while the systematic
error is almost completely correlated.

\begin{figure}[hbtp]
\vspace{-.6cm}
\centerline{\psfig{figure=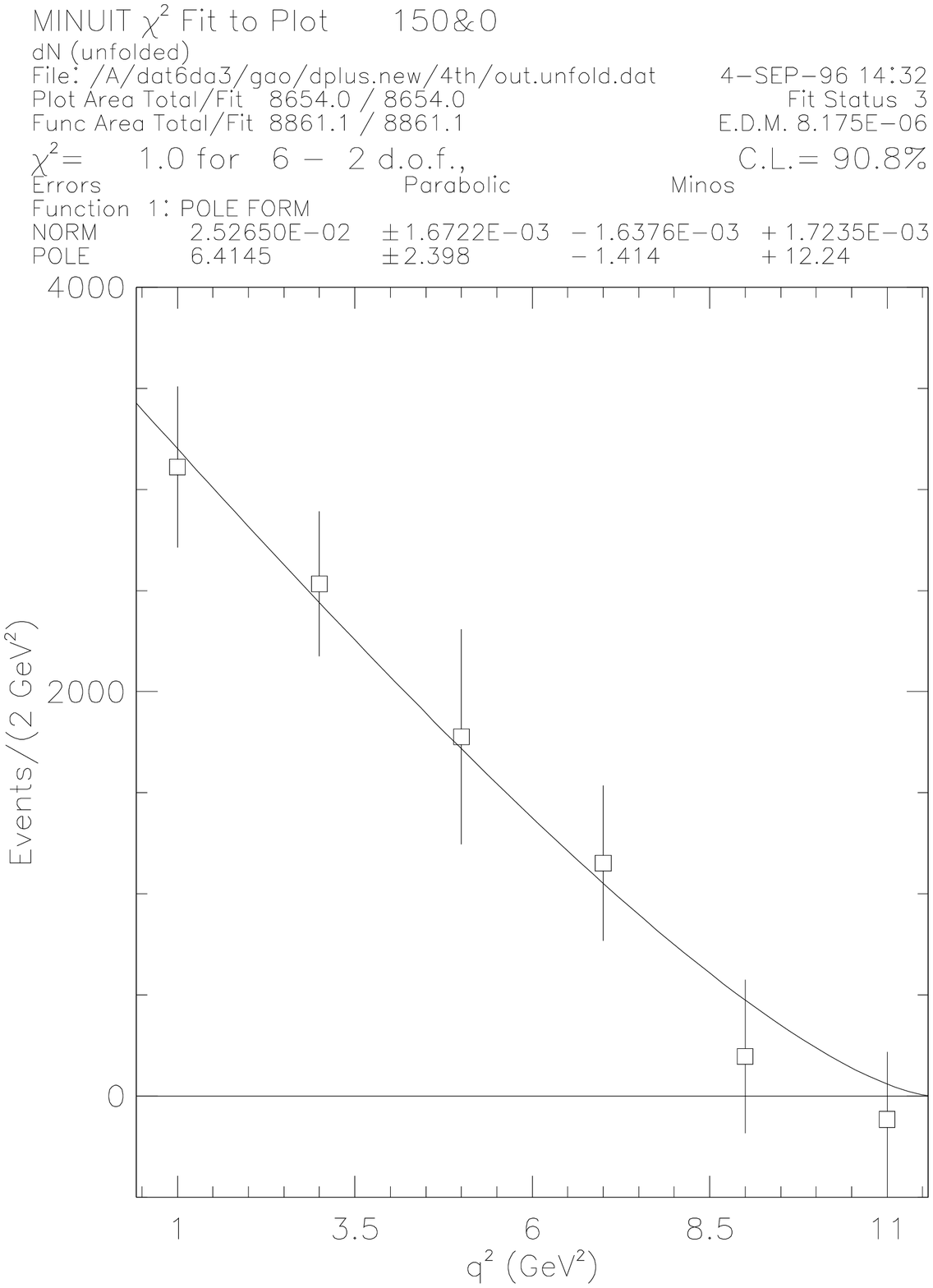,height=3.5in,bbllx=0bp,bblly=0bp,bburx=600bp,bbury=617bp,clip=}}
\vspace{-1cm}
\fcaption{\label{qsq_dp}The $q^2$ distribution for $\bar{B}^o\to D^+\ell^-\bar{\nu}$
from the $MM^2$ analysis.}
\end{figure}

The $q^2$ distribution from the $MM^2$ method is shown in Fig.~\ref{qsq_dp}. The intercept
at $q^2$ of zero is proportional to $|V_{cb}f_+(0)|^2$.  The curve is a fit to a 
functional form 
\begin{equation}
f_+(q^2)= {f_+(0)\over  1-q^2/M^2_V},
\end{equation}
where $M_V$ is left unspecified but is theorized to be the mass of the vector
exchange particle in the $t$ channel, namely the $B^*$. 
The results and comparison
with different models are shown in Table~\ref{table:dpln}.

\begin{table}[th]\centering\tcaption{Results of $\bar{B}^o\to D^+\ell^-\bar{\nu}$ analysis}
\label{table:dpln}
\vspace*{2mm}
\begin{tabular}{lccc}\hline\hline
Model & $f_+(0)$ prediction & $|V_{cb}f_+(0)|\times 10^3$ & $|V_{cb}|\times 
10^3$\\\hline
WSB\cite{WSB} & 0.70 & $25.7\pm 1.4\pm 1.7$ & $37.3\pm 2.0 \pm 2.5$ \\
KS\cite{KS} & 0.69 & $25.7\pm 1.4\pm 1.7$ & $36.7\pm 2.0 \pm 2.5$ \\  
Demchuk$^{\dagger}$\cite{Demchuk} & 0.68 & $24.8\pm 1.1\pm 1.6$ & $36.4\pm 1.6 \pm 2.4$ \\ 
\hline
Average &  &  & $36.9\pm 3.7 \pm 0.5$ \\        
\hline\hline
\multicolumn{4}{l}{${\dagger}$  A smaller statistical error is quoted
 for this model because $M_V$ is specified.}

\end{tabular}\end{table}

For the average value for $V_{cb}$, the first error is the quadrature of the
the  systematic and statistical errors in the data,  and the fact that the
fraction of $B^o$'s  produced in $\Upsilon (4S)$ decay is known only as
0.49$\pm$0.05.\cite{4sfrac} The second error is due only to the  model dependence.

\subsubsection{Branching Ratio of  $\bar{B}^o\to D^{*+}\ell^-\bar{\nu}$}

We next turn to measurements of the branching ratio of the pseudoscalar to
vector transition $\bar{B}^o\to D^{*+}\ell^-\bar{\nu}$, shown in
Table~\ref{table:dsplusbr}.

\begin{table}[th]\centering\tcaption{Measurements of ${\cal{B}}(\bar{B}^o\to D^{*+}\ell^-\bar{\nu})$}
\label{table:dsplusbr}
\vspace*{2mm}
\begin{tabular}{lc}\hline\hline
Experiment & $\cal{B}$(\%)\\\hline
CLEO\cite{oldCLdsbr} & $4.1\pm 0.5\pm 0.7$ \\
ARGUS\cite{ARGdsbr} & $4.7\pm 0.6\pm 0.6$ \\
CLEO II\cite{4sfrac}& $4.50\pm 0.44\pm 0.44 $\\
ALEPH\cite{ALPdsbr} & $5.18\pm 0.30\pm 0.62$ \\
DELPHI\cite{DELPdsbr} & $5.47\pm 0.16\pm 0.67$ \\\hline
Average & $4.90\pm 0.35$ \\
\hline\hline
\end{tabular}\end{table}

The width predictions of a collection of representative models and the resulting
values of $V_{cb}$ are given in Table~\ref{table:dstrvcb}. Here the first error
on the average is the from the error on the measured branching ratio ($\pm$3.6\%) in
quadrature with the error on the lifetime ($\pm$1.6\%) and the second error reflects
the spread in the models ($\pm$5.2\%).

\begin{table}[th]\centering\tcaption{Values of $V_{cb}$ from ${\cal{B}}(\bar{B}^o\to D^+\ell^-\bar{\nu})$}
\label{table:dstrvcb}
\vspace*{2mm}
\begin{tabular}{lcc}\hline\hline
Model &  Predicted $\Gamma(B\to D^*\ell\nu)$ (ps$^{-1}$) & $|V_{cb}|\times 
10^3$\\\hline
ISGW\cite{ISGW} &  25.2$|V_{cb}|^2$  & $35.2\pm 1.4$\\
ISGW II\cite{ISGW2}&  24.8$|V_{cb}|^2$  & $35.5\pm 1.4$\\
KS\cite{KS} &  25.7$|V_{cb}|^2$  & $34.8\pm 1.4$\\
WBS\cite{WSB} &  21.9$|V_{cb}|^2$  & $37.8\pm 1.5$\\
Jaus1\cite{Jaus} &  21.7$|V_{cb}|^2$  & $37.9\pm 1.5$\\
Jaus2\cite{Jaus} &  21.7$|V_{cb}|^2$  & $37.9\pm 1.5$\\\hline
Average & &$36.5\pm  1.5\pm 1.9$ \\
\hline\hline
\end{tabular}\end{table}

\subsubsection{Heavy Quark Effective Theory and  
$\bar{B}\to D^{*}\ell^-\bar{\nu}$}

Our next method for finding $V_{cb}$ uses ``Heavy Quark Effective Theory"  
(HQET).\cite{HQET}
We start with
a quick introduction to this theory. It is difficult to solve QCD at long distances, but 
its
possible at short distances. Asymptotic freedom, the fact that the strong coupling 
constant 
$\alpha_s$ becomes weak in processes with large $q^2$, allows perturbative 
calculations. Large distances are of the order $\sim 1/\Lambda_{QCD}\sim$1 fm, 
since
$\Lambda_{QCD}$ is about 0.2 GeV. Short distances, on the other hand, are of the 
order
of the quark Compton wavelength; $\lambda_Q\sim 1/m_Q$ equals 0.04 fm for the 
$b$
quark and 0.13 fm for the $c$ quark. 

For hadrons, on the order of 1 fm, the light quarks are sensitive only to the heavy 
quark's
color electric field, not the flavor or spin direction. Thus, as $m_Q\to\infty$, 
hadronic
systems which differ only in flavor or heavy quark spin have the same configuration of 
their light degrees of freedom. The following two predictions follow 
immediately
(the actual experimental values are shown below):
\begin{eqnarray}
m_{B_s}-m_{B_d} &= &m_{D_s}-m_{D^+}\\
(90\pm 3) {\rm ~MeV}  &    & (99\pm 1){\rm ~MeV  ~, ~and} \nonumber \\
m^2_{B^*}-m^2_{B} &= &m^2_{D^*}-m^2_{D}  .\\
0.49 {\rm ~GeV^2}  &    & 0.55 {\rm ~GeV^2}. \nonumber
\end{eqnarray}

The agreement is quite good but not exceptional.
Since the charmed quark is not that heavy, there is some heavy quark symmetry 
breaking. This must be accounted for in quantitative predictions, and can
probably explain the discrepancies above. The basic idea is that
if you replace a $b$ quark with a $c$ quark moving at the same {\bf velocity},
there should only be  small and calculable changes.

In lowest order HQET there is only one form-factor function $\xi$ which is a 
function of
the Lorentz invariant four-velocity transfer $y$, where
\begin{equation}
y = {{M^2_B+M^2_{D^*}-q^2}\over {2M_BM_{D^*}}}.
\end{equation}
The point $y$ equals one corresponds to the situation where the $B$ decays to a
$D^*$ which is at rest in the $B$ frame. Here the ``universal" form-factor function
$\xi (y)$ has the value, $\xi (1)=1$, in lowest order. This is the point in phase space 
where the $b$ quark changes to a $c$ quark with zero velocity transfer.  The idea
is to measure the decay rate at this point, since we know the value of the form-factor,
namely unity, and then apply the hopefully small and hopefully well understood 
corrections.  Although this analysis can be applied to $\bar{B}\to D\ell^-\nu$,
the vanishing of the decay rate at $y$ equals 1, ( maximum $q^2$, see
Fig.~\ref{qsq_dp}), makes this inaccurate.\cite{CLEOdplnu}

The corrections are of two types: quark mass, characterized as some
coefficient  times $\Lambda_{QCD}/m_Q$, and hard gluon, characterized as
$\eta_A$. The value of the form-factor can then be expressed as\cite{neub}
\begin{equation}
\xi(1)=\eta_A\left(1+0\cdot \Lambda_{QCD}/m_Q + 
c_2\cdot \left(\Lambda_{QCD}/m_Q\right)^2+....\right)= \eta_A(1+\delta).
\end{equation}
The zero coefficient in front of the $1/m_Q$ term reflects the fact that the
first order correction in quark mass vanishes at $y$ equals one. This is called
Luke's Theorem.\cite{luke} Recent estimates are 0.96$\pm$0.007 and $-0.55\pm$0.025
for $\eta_A$ and $\delta$, respectively. The value predicted for $\xi(1)$ then
is 0.91$\pm$0.03. This is the conclusion of Neubert.\cite{neub} There has been much
controversy surrounding the theoretical prediction of this number.\cite{xione}

\begin{figure}[thb]
\centerline{\psfig{figure=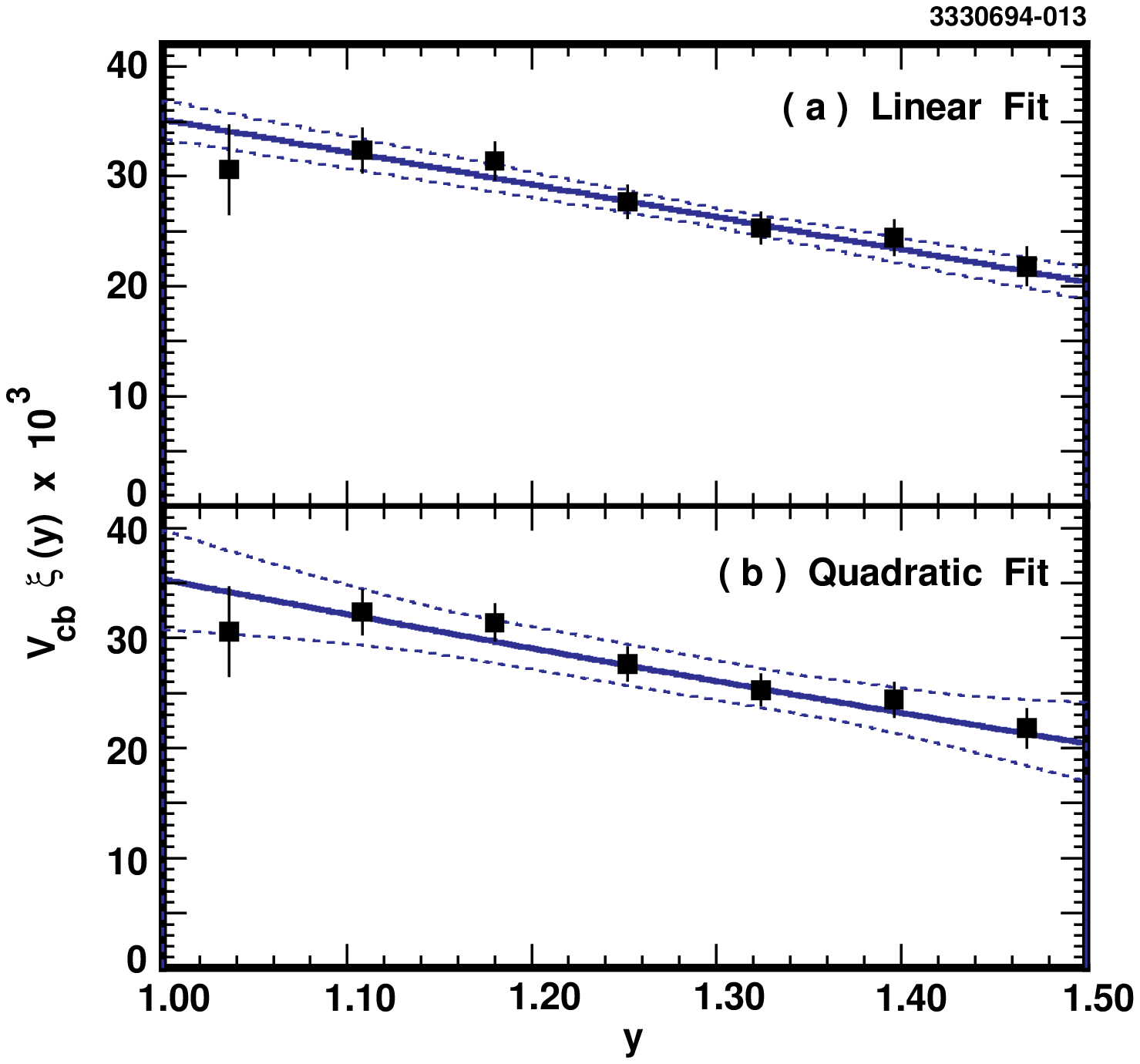,height=4.0in,bbllx=90bp,bblly=300bp,bburx=550bp,bbury=700bp,clip=}}
\vspace{.1cm}
\fcaption{\label{vcbf1f} Linear and quadratic fits to the CLEO data for the
 $D^{*+}\ell^-\bar{{\nu}}$ and $D^{*o}\ell^-\bar{{\nu}}$.}  \end{figure}

To find the value of the decay width at $y$ equals one, it is necessary to
fit data over a finite range in $y$ and extrapolate to $y$ of one. HQET
does not predict the shape of the form-factor; hence the shape of the 
$d\Gamma/dy$ distribution is not specified. Most experimental groups have 
done the simplest thing and used a linear fit. The CLEO results with both
linear and quadratic fits are shown in Fig.~\ref{vcbf1f}. The results from the 
different groups are summarized in Table~\ref{yone}. Also fits of the
slope parameter, $\rho^2$, coming from the linear fit are included.

\begin{table}[th]\centering\tcaption{Values of $|V_{cb}|\xi(1)\times 10^3$ }
\label{yone}
\vspace*{2mm}
\begin{tabular}{lcc}\hline\hline
Experiment & $|V_{cb}|\xi(1)\times 10^3$  & $\rho^2$\\\hline
ARGUS\cite{argus_y1}&  $38.8\pm 4.3\pm 3.5$ & $1.17\pm 0.22\pm 0.06$ \\
CLEO II\cite{cleo_y1} & $35.1\pm 1.9\pm 2.0$ & $0.84\pm 0.12\pm 0.08$ \\
ALEPH\cite{aleph_y1} & $31.4\pm 2.3\pm 2.5$ & $0.39\pm 0.21\pm 0.12$\\
DELPHI\cite{delphi_y1} & $35.0\pm 1.9\pm 2.3$ & $0.81\pm 0.16\pm 0.10$ \\\hline
Average & $34.6\pm 1.6$ & $0.82\pm 0.09$\\
\hline\hline
\end{tabular}\end{table}

\begin{figure}[htb]
\vspace{-1.4cm}
\centerline{\psfig{figure=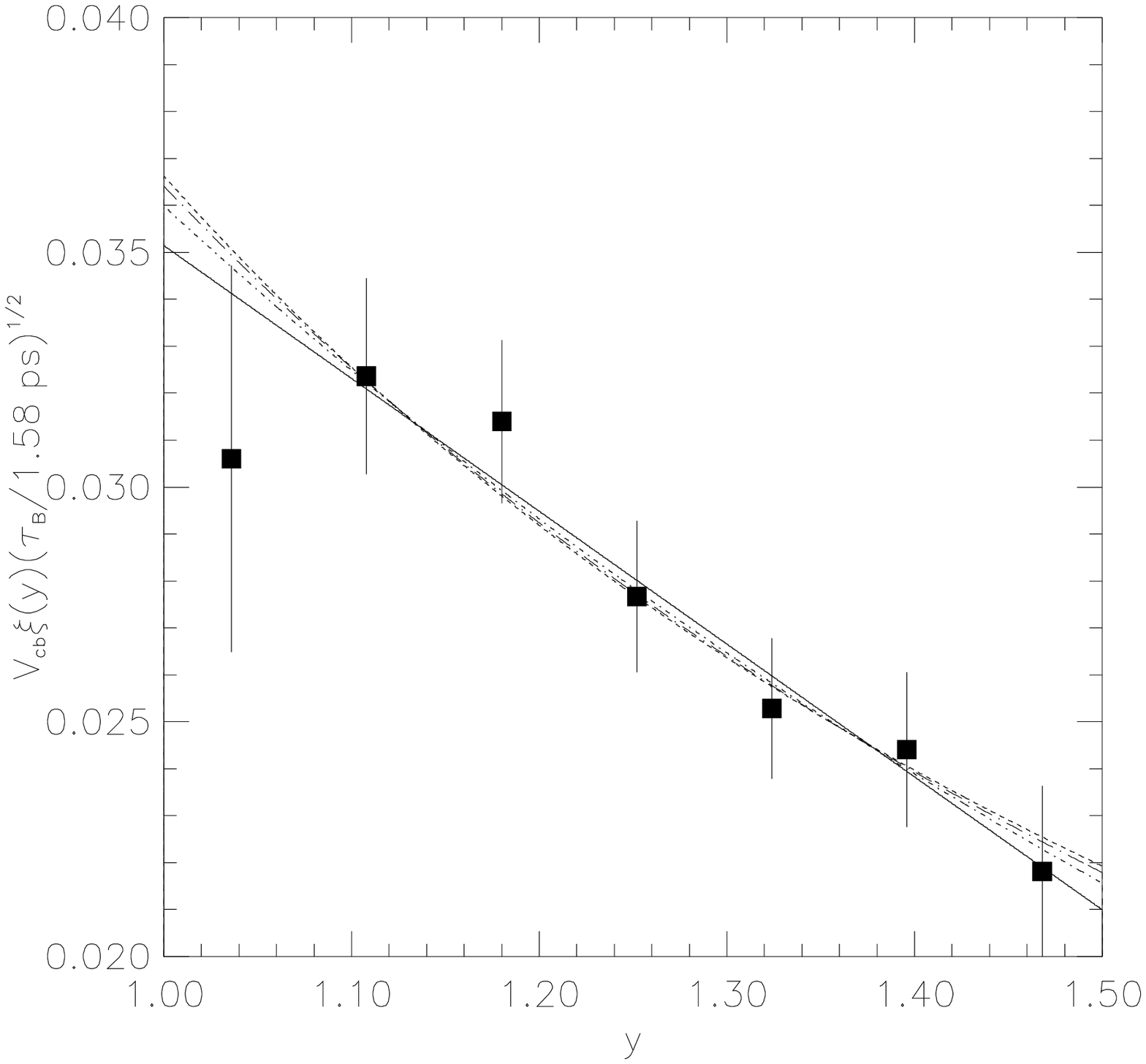,height=5.8in,bbllx=0bp,bblly=0bp,bburx=600bp,bbury=700bp,clip=}}
\vspace{-2.7cm}
\fcaption{\label{vcbxif} Fits to the CLEO data with different shapes. The curves
 are linear (solid), Neubert-Reickert (NR) exponential (dashed),
 pole (long dash-dot) and exponential (dot-dashed).} \end{figure}

Although the shape of the function is not specified in HQET general
considerations lead to the expectation that the slope is positive: there is
a pole in the amplitude as $y\to -1$ and $\xi(y)\to 0$ as $y$ increases. 
Shapes for $\xi(y)$ are suggested by quark models. I have fit the CLEO data
to different model functions as shown in Fig.~\ref{vcbxif}. The results are shown
in Table~\ref{table:vcbslope}.

\begin{table}[th]\centering\tcaption{Values of $|V_{cb}|\xi(1)$ for different fit shapes of CLEO II data }
\label{table:vcbslope}
\vspace*{2mm}
\begin{tabular}{llcc}\hline\hline
$\xi(y)$ & name& $\rho$ & $|V_{cb}|\xi(1)\times 10^3$  \\\hline
$1-\rho^2(y-1)$&linear&0.90$\pm$0.07&0.0351$\pm$0.0018$\pm$0.0018\\
${2\over y+1}exp\left[-(2\rho^2-1){y-1\over y+1}\right]$&NR
exp&0.90$\pm$0.12& 0.0366$\pm$0.0024$\pm$0.0018\\
$\left({2\over y+1}\right)^{2\rho^2}$&pole&1.07$\pm$0.11&
0.0364$\pm$0.0023$\pm$0.0018\\
$exp\left[-\rho^2(y-1)\right]$&exp&1.01$\pm$0.10&0.0360$\pm$0.0022$\pm$0.0018 \\
\hline\hline
\end{tabular}\end{table}
 
These shapes give larger values of $|V_{cb}|\xi(1)|$ than the linear fit
by (5$\pm$3)\%. I call this a model dependent error. The value then obtained
for $|V_{cb}|\xi(1)|$ is $(36.3\pm 1.6\pm 1.0)\times 10^{-3}$, and 
\begin{equation}
|V_{cb}|=0.0397\pm 0.0021\pm0.0017~~.
\end{equation}

\subsubsection{$|V_{cb}|$ using inclusive semileptonic decays}

\begin{figure}[htb]
\vspace{-1.6cm}
\centerline{\psfig{figure=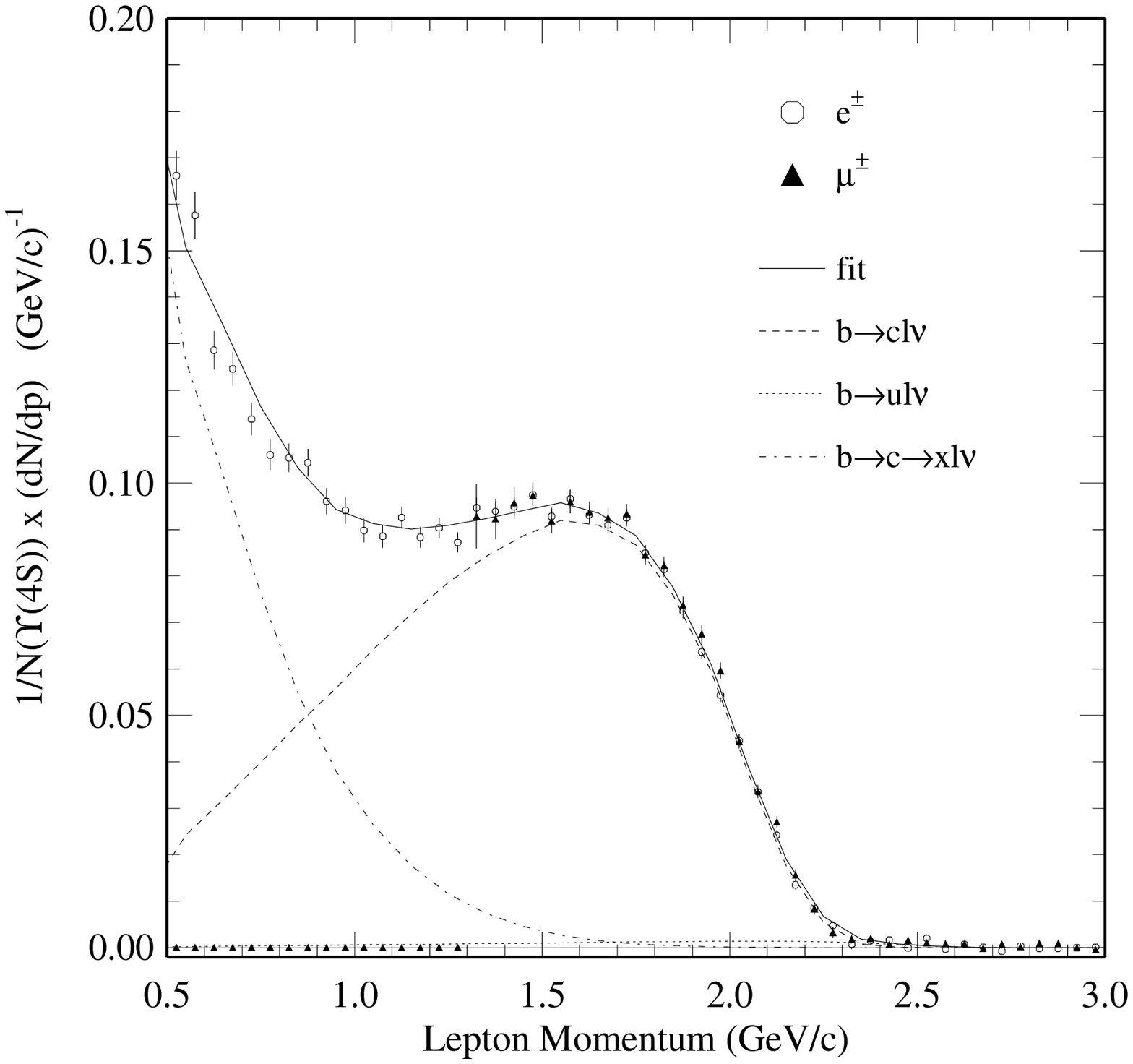,height=5.4in,bbllx=0bp,bblly=0bp,bburx=600bp,bbury=700bp,clip=}}
\vspace{-2.7cm}
\fcaption{\label{lepf} Fit to the CLEO inclusive lepton spectrum with the
 ACM model.} \end{figure}

The inclusive semileptonic branching ratio can also be used to measure
$V_{cb}$. While ${\cal B}(B\to X e^- {\bar \nu})$ this has traditionally been
done by measuring the inclusive lepton momentum spectrum using only single
lepton data, recently dilepton data have been used.
The inclusive lepton spectrum from the latest CLEO II data\cite{lepspec}
 is shown in Fig.~\ref{lepf}.
 Both electrons and muons are shown.
Leptons which arise from the continuum have been statistically subtracted
using the below resonance sample.
The peak at low momentum is due to the decay chain
${\bar B}\to DX, ~D\to Y\ell^+\nu$. The data are fit to two shapes whose
normalizations are allowed to float. The first shape is taken
from models of $B$ decay while the second comes from the measured shape of
leptons from $D$ mesons produced nearly at rest at the $\psi''$,
which is then smeared using the
measured momentum distribution of $D's$ produced in $B$ decay.
CLEO finds ${\cal B}_{sl}$ of 10.5$\pm$0.2\% and 11.1$\pm$0.3\% in the
ACM\cite{ACM} and ISGW$^*$ models, respectively.\cite{lepspec}
The ACM model will be described below. The ISGW$^*$ model is a variant
of the ISGW\cite{ISGW} model. The  ISGW model
 includes  all the exclusive single hadron modes, $D$, $D^*$, and $D^{**}$
 which contains several components. 
 CLEO lets the normalization of the $D^{**}$ components float in the fit,
and calls this model ISGW$^*$.

Next, I discuss how to use dilepton events to eliminate the secondary leptons
at low momentum. Consider the sign of the lepton charges for the four leptons
in the following decay sequence: $\Upsilon(4S)\to B^-B^+;~B^-\to D\ell_1^-{\bar
\nu},~B^+\to {\bar D}\ell_3^+\nu; ~D\to Y\ell^+_2\nu,~{\bar D}\to
Y'\ell^-_4{\bar\nu}.$ If a high momentum negative lepton ($\ell_1^-$) is found,
then if the second lepton is also negative it must come from the cascade decay
of the $B^+$ (i.e. it must be $\ell_4^-$). On the other hand the second lepton
being positive shows that it must be either the primary lepton from the
opposite $B^+$, ($\ell_3^+$), or the cascade from the same $B^-$, ($\ell_2^+$).
However the cascades from the same $B^-$ can be greatly reduced by insisting
that the cosine of the opening angle between the two leptons be greater than
zero as they tend to be aligned. The same arguments are applicable to
$\Upsilon(4S)\to B^o\bar{B}^o$, except that an additional correction must be
made to account for $B\bar B$ mixing.

\begin{figure}[htb]
\vspace{-5.6cm}
\centerline{\psfig{figure=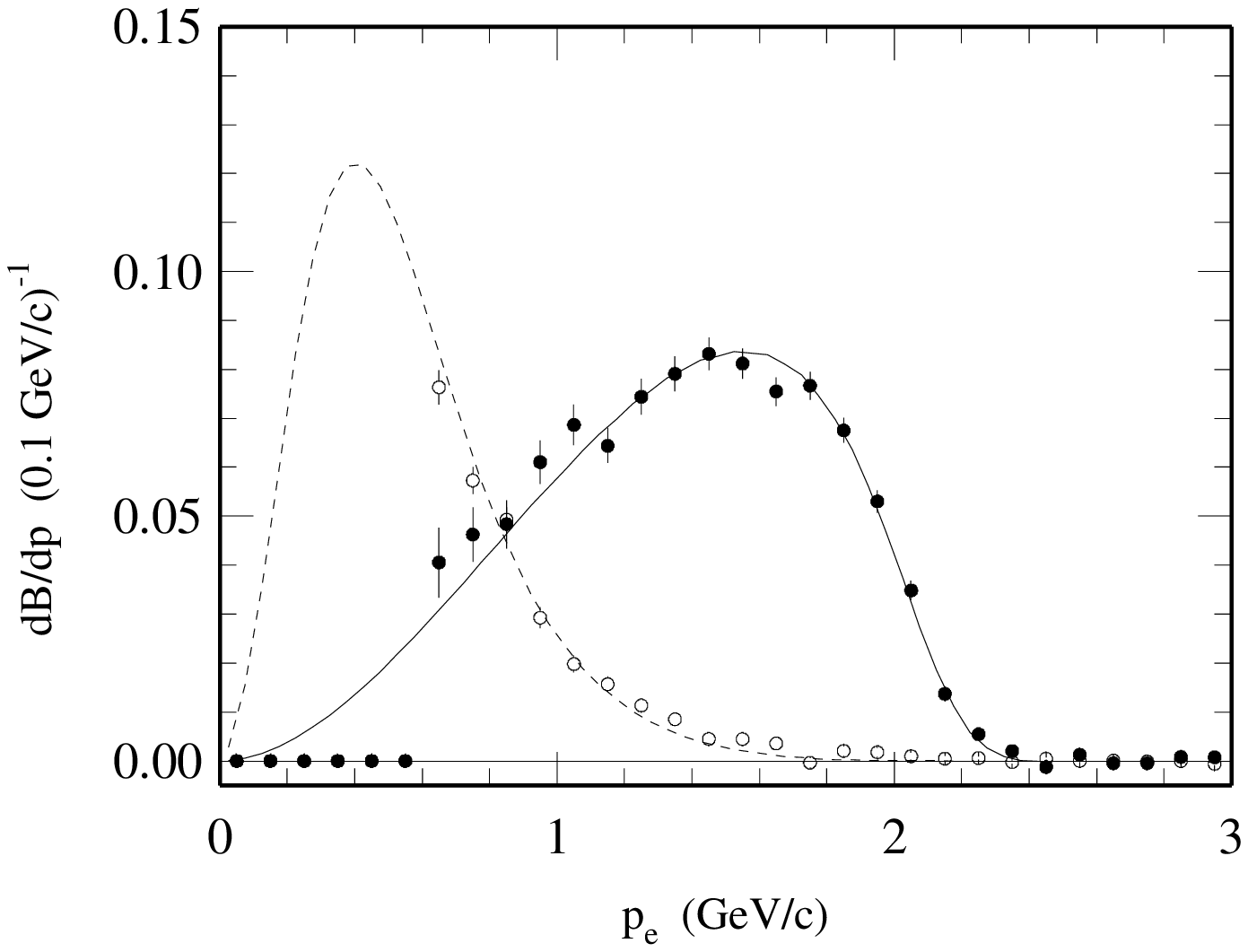,height=6.0in,bbllx=0bp,bblly=0bp,bburx=600bp,bbury=700bp,clip=}}
\vspace{-3cm}
\fcaption{\label{dilepf} The lepton momentum spectrum in dilepton events
from CLEO. The solid points are for opposite sign leptons, while the open
circles indicate like sign lepton pairs. The fit is to the ACM model.} \end{figure}

The CLEO II data are shown in Fig.~\ref{dilepf}. The data fit nicely to either
the ACM or ISGW$^*$ model. They find that the semileptonic branching ratio, ${\cal
B}_{sl},$ equals $(10.36\pm 0.17\pm 0.40)\%$ with a negligible dependence on the
model.\cite{Cdilep} This result confirms that the $B$
model shapes are appropriate down to lepton momenta of 0.6 GeV/c.
ARGUS\cite{ARGdil} did
the first analysis using this technique and found  ${\cal B}_{sl}=(9.6\pm 0.5\pm
0.4)\%$.

\renewcommand{\topfraction}{1.}
\renewcommand{\bottomfraction}{1.}
\setcounter{bottomnumber}{2}
\renewcommand{\textfraction}{0}

The next topic is to measure $V_{cb}$ using the inclusive lepton spectrum.
	Consider $\Gamma_{sl}\equiv \Gamma (B\to X e^- {\bar \nu})$ in the
simplest parton model:
\begin{equation}
\Gamma_{sl}={G^2_F m_b^5\over
192\pi^3}\left(p_c|V_{cb}|^2+p_u|V_{ub}|^2\right)\eta_{QCD},
\end{equation}
  where the $p$'s are phase space factors, and  the QCD correction, $\eta_{QCD} =
1-2\alpha_s/3\pi.$
   Since  $|V_{ub}|< <|V_{cb}|$, we ignore the 2nd term. To use the semileptonic
width to extract $|V_{cb}|$ using this expression requires a knowledge of $m_b^5$,
which is poorly understood. A way around this dilemma was found by
Altarelli \etal \cite{ACM}
They make two important corrections to the simple
parton model. First they treat the spectator quark in the $B$ meson as a
quasi-free particle with a Gaussian spectrum of Fermi-momentum, $p$:
\begin{equation}
f(p)={4p^2 \over \sqrt{\pi}p_f^3}\exp (-p^2/p_f^2).
\end{equation}
The average value, $p_f$,
is a free parameter in the model. Secondly, they include the effects
of gluon radiation from the quarks, which lowers the spectrum at high
lepton momentum. The semileptonic width is given explicitly as:
\begin{equation}
{d\Gamma (B\to D X\ell^-\bar{\nu})\over dx}=
{{m_b^5G_F^2V_{cb}^2}\over{96\pi^3}}\cdot\left[\Phi(x,\epsilon)
-G(x,\epsilon)\right],\label{eq:Altar}
\end{equation}
where $x=2E_{\ell}/m_b$, $E_{\ell}$ being the lepton energy, $\epsilon=m_c/m_b$,
$G(x,\epsilon)$ is a complicated gluon radiation function and
\begin{equation}
\Phi(x,\epsilon)={x^2(1-\epsilon^2-x)^2 \over (1-x)^3}
\left[(1-x)(3-2x)+(3-x)\epsilon^2\right].
\end{equation}
 Each value of the Fermi-momentum, $p$, leads to a different value of $m_b$ and
hence a different distribution for ${d\Gamma \over dx}$ which must
be convoluted with Eq. (\ref{eq:Altar}) to find the total theoretical lepton
momentum spectrum. The relationship
between $m_b$ and $p$ is just given by kinematics
\begin{equation}
m_b^2=m_B^2+m_{sp}^2-2m_B\sqrt{(p^2+m_{sp}^2)}.
\end{equation}
Here $m_B$ is the known value of the $B$ meson mass
of 5.280 GeV and $m_{sp}$
is the spectator quark mass.
 A fit to the shape of the lepton energy spectrum
then is needed to determine the free parameters $p_f$, $\epsilon$ and $m_{sp}$.
In turns out that one can fix $m_{sp}$ and any latent dependence is absorbed
by the other two.
So a fit to the data will determine ${\cal B}_{sl}$, $p_f$ and
$\epsilon$. In this way Altarelli et al. remove the explicit
dependence of the $m_b^5$ term in the total decay rate.
The ISGW and ISGW$^*$ models are also used.
The resulting values are given in Table~\ref{table:inclvcb}.
\begin{table}[th]\centering\tcaption{$V_{cb}$  Values from Inclusive leptons}
\label{table:inclvcb}
\vspace*{2mm}
\begin{tabular}{ccc}\hline\hline
Model & Experiment  & $V_{cb}$\\ \hline
ACM&CLEO I & 0.042$\pm$0.002$\pm$0.004\\
ACM&ARGUS & 0.039$\pm$0.001$\pm$0.003\\
ACM&CLEO II & 0.040$\pm$0.001$\pm$0.004\\
ISGW&CLEO I & 0.039$\pm$0.002$\pm$0.004\\
ISGW&ARGUS& 0.039$\pm$0.001$\pm$0.005\\
ISGW&CLEO II& 0.040$\pm$0.001$\pm$0.004\\
ISGW$^*$&CLEO I & 0.037$\pm$0.002$\pm$0.004\\
ISGW$^*$&CLEO II & 0.040$\pm$0.002$\pm$0.004\\ \hline\hline
\end{tabular}\end{table}

The representative value of $|V_{cb}|$ found from this analysis alone is
\begin{equation}
|V_{cb}|=0.039\pm0.001\pm0.004~~.
\end{equation}
 There are determinations of the inclusive
$B$ semileptonic branching ratio from LEP. These measurements average over
more $B$ species that at the $\Upsilon(4S)$. Since the lifetimes of some
of these, especially the $\Lambda_b$ appears to be shorter than for the
ground state mesons, the semileptonic branching ratio measured at LEP should
be lower than that measured on the $\Upsilon(4S)$, yet it is somewhat
higher.\cite{tomasz} Since the measurement at LEP is far more complicated, I have chosen to
leave out these results.

\begin{figure}[hbt]
\vspace{-0.8cm}
\centerline{\psfig{figure=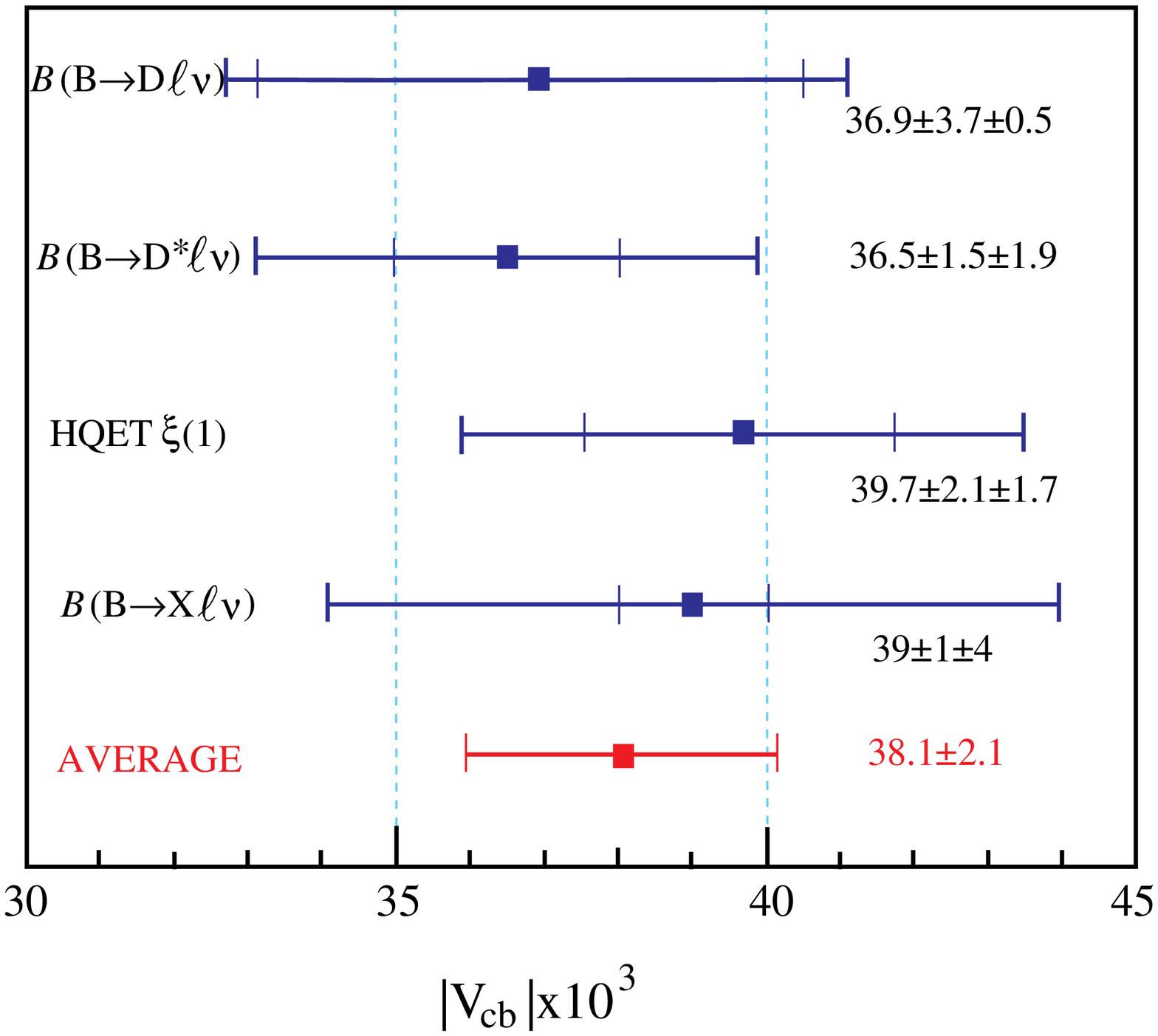,height=4.0in,bbllx=0bp,bblly=80bp,bburx=600bp,bbury=600bp,clip=}}
\vspace{-.2cm}
\fcaption{\label{Vcb}Results of four different methods used to evaluate 
$V_{cb}$, and the resulting average. The horizontal lines show the values,
the statistical errors out to the thin vertical lines, and the systematic errors
added on linearly out to the thick vertical lines. } \end{figure}

The results of using all four methods to find $V_{cb}$ are shown in
Fig.~\ref{Vcb}. It is remarkable that all four separate methods give such
consistent results. Advocates for any particular method can choose among these
results. I have chosen to average them. The errors are handled by adding
the statistical and systematic errors on each method and then adding the
different methods in quadrature. This should give a generous estimate of the
final error. The average value of $V_{cb}$ is 0.0381$\pm$0.0021, which gives
a value for the CKM parameter 
\begin{equation}
A = 0.784\pm 0.043 ~~.
\end{equation}

\subsection{The CKM element $V_{ub}$}

\begin{figure}[hbt]
\vspace{-1.5cm}
\centerline{\psfig{figure=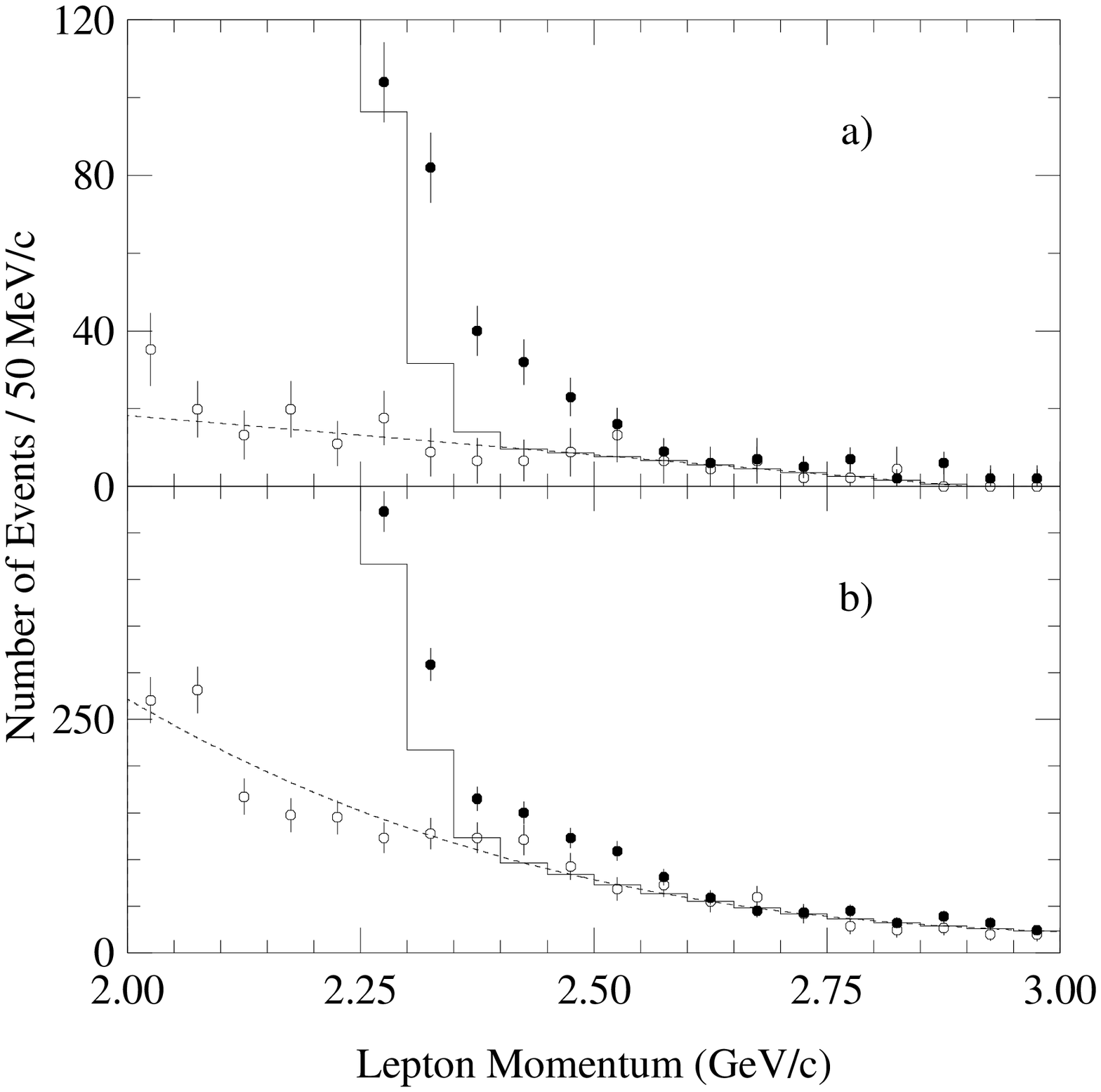,height=5.0in,bbllx=0bp,bblly=0bp,bburx=600bp,bbury=700bp,clip=}}
\vspace{-2.2cm}
\fcaption{\label{btouf}Lepton yield versus momentum from CLEO II
 for the ``strict" cut sample, $R_2 < 0.2$,
$P_{miss} > 1$ GeV/c and  the lepton and missing
momentum direction point into opposite hemispheres,
(a) and the $R_2 < 0.3$ sample (b). The filled points are from data
taken on the peak of the $\Upsilon(4S)$, while the open points are continuum
data scaled appropriately. The dashed curves are fits to the continuum
data, while the solid histograms are predictions of the sum of
$b\to c\ell\nu$ and continuum lepton production.} \end{figure}

The first evidence of a non-zero value of $V_{ub}$ was obtained by CLEO I
who saw a non-zero excess beyond the endpoint allowed for $B\to D \ell\nu$
transitions.\cite{cleo_btou} This result was quickly confirmed by
ARGUS.\cite{argus_btou}
The latest evidence from
CLEO II\cite{Cbtounu} is shown in Fig.~\ref{btouf}. $R_2$ is the second
Fox-Wolfram event shape variable,\cite{foxwolf} which tends to zero for
spherical events, such as $\Upsilon (4S)$ decays and to one for jet-like
events. $P_{miss}$ is the missing momentum in the event.

\begin{figure}[hbt]
\vspace{-0.8cm}
\centerline{\psfig{figure=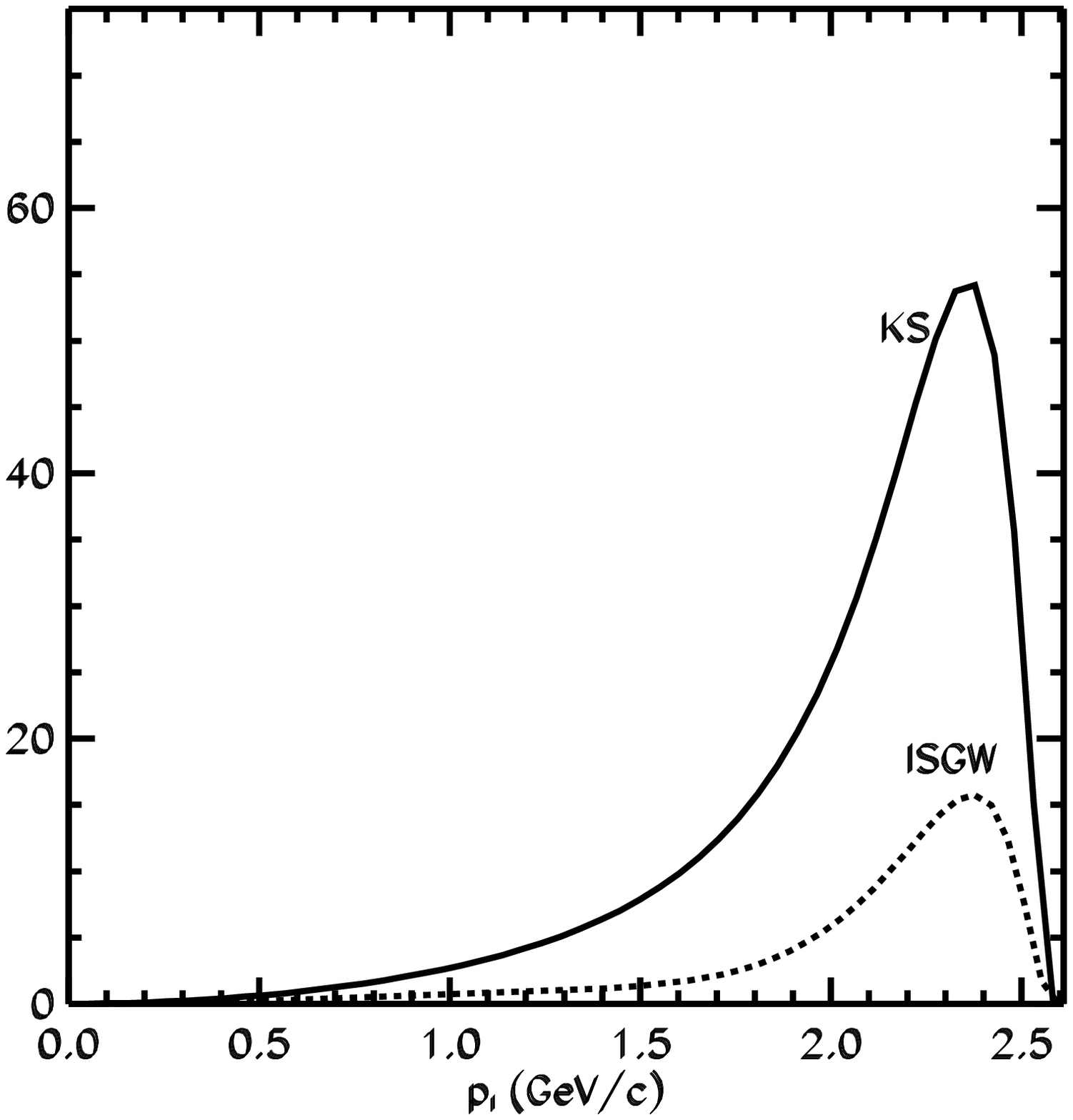,height=3.5in,bbllx=0bp,bblly=0bp,bburx=600bp,bbury=700bp,clip=}}
\vspace{-2.cm}
\fcaption{\label{isw_ks} Lepton momentum spectra, for $B\to\rho\ell\nu$
in the KS and the original ISGW model.} \end{figure}

\begin{figure}[hbt]
\vspace{-1.5cm}
\centerline{\psfig{figure=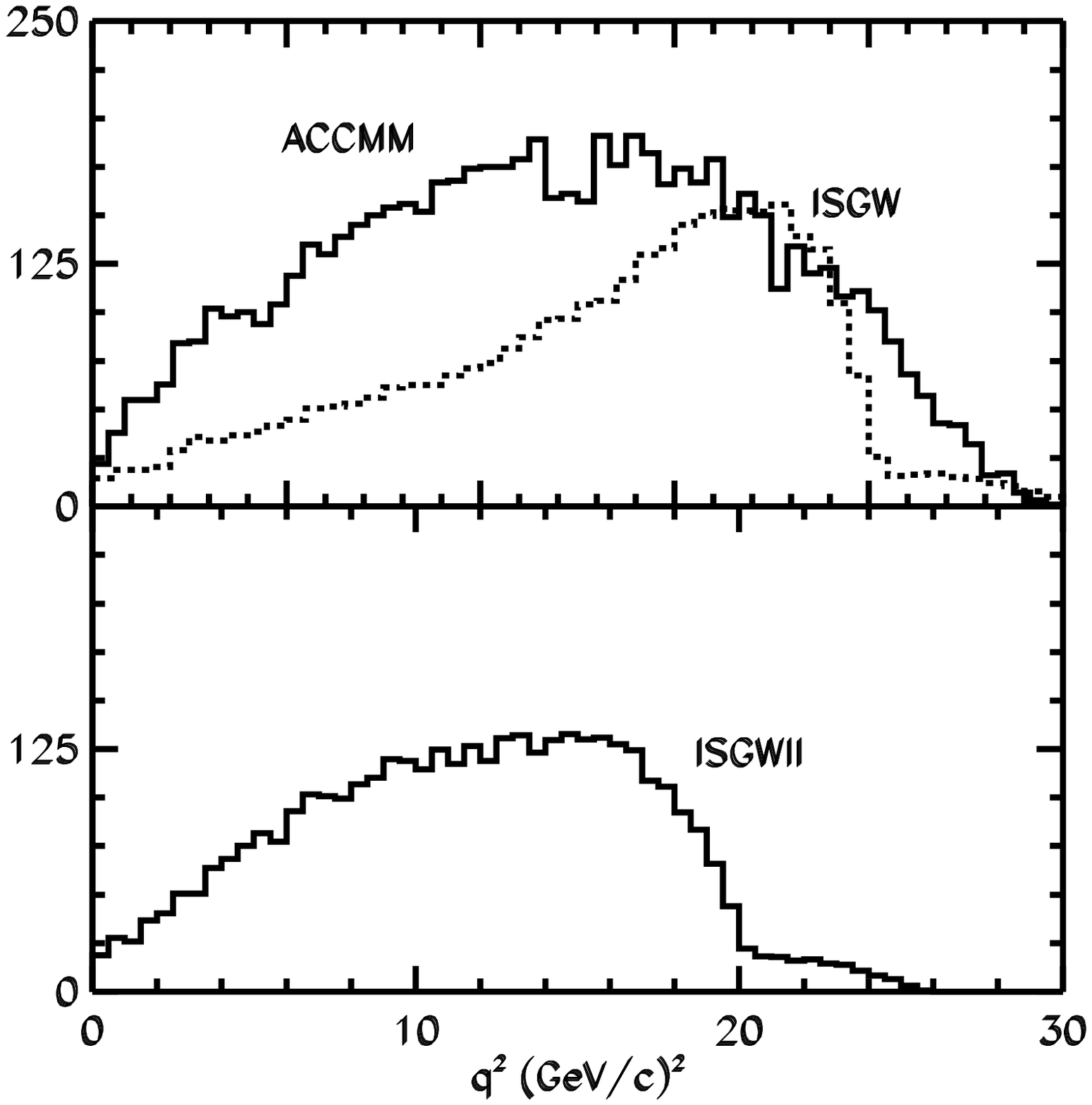,height=4.5in,bbllx=0bp,bblly=0bp,bburx=600bp,bbury=700bp,clip=}}
\vspace{-2.2cm}
\fcaption{\label{qsq} $q^2$ distribution, for charmless semileptonic
$b$ decays in the model of Altarelli \etal (ACCMM) and the orginal ISGW model
shown on top, and the new ISGW model shown on the bottom. The areas reflect
the predicted widths, but the vertical scale is arbitrary. The high $q^2$ tails
on the ISGW models arise from the $\pi\ell\nu$ final state.} \end{figure}

The branching ratios are small. CLEO  finds that the rate in the lepton momentum
interval $2.6>p_{\ell}>2.4$ GeV/c, ${\cal B}_u(p)$, is $(1.5\pm 0.2\pm 0.2)
\times 10^{-4}.$
To extract $V_{ub }$ from this measurement we need  to use theoretical models.
It is convenient to define: $\Gamma(b\to u\ell\nu)=\gamma_u |V_{ub}|^2$,
and $\Gamma(b\to c\ell\nu)=\gamma_c |V_{cb}|^2$. In addition,
$f_u(p)$ is the fraction of the spectrum predicted in the end point region by
different models, and ${\cal B}_{sl}$ is the semileptonic branching ratio. Then:
\begin{equation}
{|V_{ub}|^2 \over |V_{cb}|^2} = {{\cal B}_u(p)\over {\cal B}_{sl}}\cdot {\gamma_c
\over f_u(p)\gamma_u}.
\end{equation}

These models disagree as to which final states populate the endpoint region. Most
models agree roughly on values of $\gamma_c$. However, models differ greatly in
the value of the product $\gamma_u\cdot f_u(p)$. There are two important
reasons for these differences. First of all, different authors disagree as to
the importance of the specific exclusive final states such as $\pi\ell\nu$,
$\rho\ell\nu$ in the lepton endpoint region. For example, the Altarelli et al.
model doesn't consider individual final states and thus can be seriously
misleading if the endpoint region is dominated by only one or two final states.
In fact, several inventors of exclusive models have claimed that the endpoint
is dominated by only a few final states.\cite{ISGW}$^,$\cite{WSB} Secondly, even among
the exclusive form-factor models there are large differences in the absolute
decay rate predictions. This is illustrated in Fig.~\ref{isw_ks}. The differences
in the exclusive models are much larger in $b\to u$ transitions than in $b\to
c$ transitions because the $q^2$ range is much larger.

Artuso has explicitly shown that the $q^2$ distributions were very different
in the ACM and original ISGW model.\cite{artu1} However, the new ISGW II model agrees
much better with ACM (see Fig.~\ref{qsq}).\cite{artu2}

Measurement of exclusive charmless semileptonic decays can put constraints
on the models and therefore restrict the model dependence. In principle, the
ratio of rates for $\pi\ell\nu$ and $\rho\ell\nu$ can be measured as well
as the $q^2$ dependence of the form-factors. However, measurement of these
rates is difficult. CLEO has recently succeeded in measuring the branching
ratios.\cite{pilnuex} 

A neutrino reconstruction technique is used. The neutrino energy and 
momentum is determined by evaluating the missing momentum and energy in the 
entire event:
\begin{eqnarray}
E_{miss}&=&2E_{beam}-\sum_{i}E_i \\
\overrightarrow{p}\!_{miss} & = & \sum_{i}\overrightarrow{p}\!_i~~.
\end{eqnarray}

\begin{figure}[htbp]
\vspace{-0.6cm}
\centerline{\psfig{figure=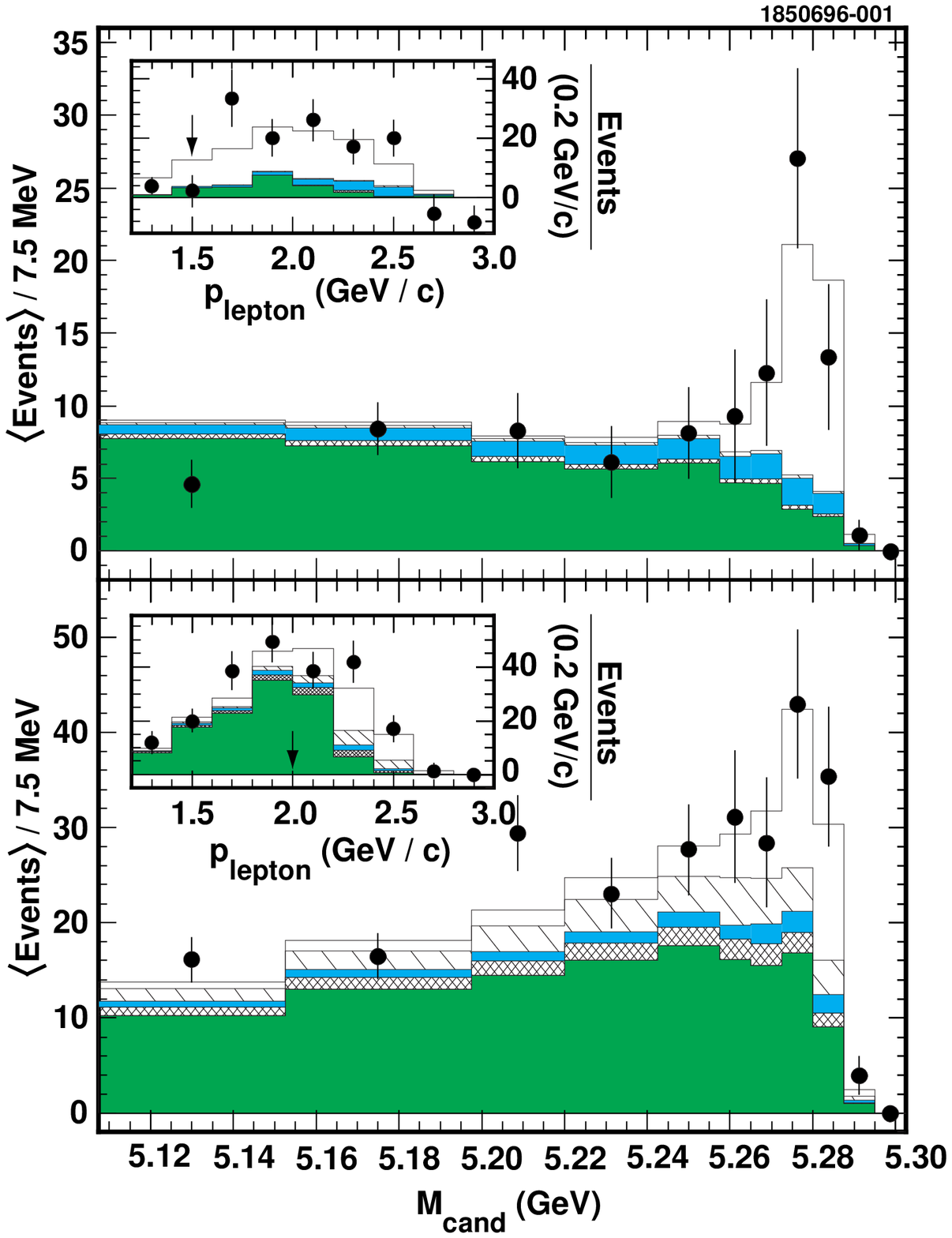,height=6.5in,bbllx=0bp,bblly=0bp,bburx=600bp,bbury=730bp,clip=}}
\vspace{-3.8cm}
\fcaption{\label{xlnu} The $B$ candidate mass distributions, $M_{cand}$,
for the sum of the scalar $\pi^+\ell\nu$ and $\pi^o\ell\nu$ (top) and the
vector modes ($\rho$ and $\omega$) (bottom). The points are the data after
continuum and fake background subtractions. The unshaded histogram is
the signal, while the dark shaded shows the $b\to c X$ background estimate,
the cross-hatched, estimated $b\to u\ell\nu$ feedown. For the $\pi$ (vector)
modes, the light-shaded and hatched histograms are $\pi\to\pi$
(vector$\to$vector) and vector$\to\pi$ ($\pi\to$vector) crossfeed,
respectively. The insets show the lepton momentum spectra for the events
in the $B$ mass peak (the arrows indicate the momentum cuts).}
 \end{figure}

 Criteria are imposed to guard against events with
false large missing energies. First, the net charge is required to be zero.
Secondly, events with two identified leptons (implying two neutrinos) are
rejected. Leptons are required to have momenta greater than 1.5 GeV/c in
the case of $\pi\ell\nu$ and greater than 2.0 GeV/c in the case of
$\rho\ell\nu$. In addition, the candidate neutrino mass is calculated as
\begin{equation}
M^2_{\nu}=E^2_{miss}-\overrightarrow{p}^2\!\!_{miss}~~.
\end{equation}
Candidate events containing a neutrino are kept if 
$M^2_{\nu}/2E_{miss}< 300 $ {\rm MeV}. Then the semileptonic $B$ decay
candidates ($\pi^o,~\pi^+,~\rho^o,~\omega^o,~\rho^+)\ell\nu$ are reconstructed
using the neutrino four-vector found from the missing energy 
measurement.\cite{neut} 
The beam constrained invariant mass, $M_{cand}$ is defined as
\begin{equation}
M^2_{cand}=E^2_{beam}-\left(\overrightarrow{p}\!_{\nu}+\overrightarrow{p}\!_{\ell}
+\overrightarrow{p}\!_{(\pi{\rm ~or~}\rho})\right)^2,
\end{equation}
and with the use of the neutrino four-vector is essentially the same as any 
other full $B$ reconstruction analysis done at the $\Upsilon(4S)$.
The $M_{cand}$ distributions are shown in Fig.~\ref{xlnu}.

\begin{figure}[htb]
\centerline{\psfig{figure=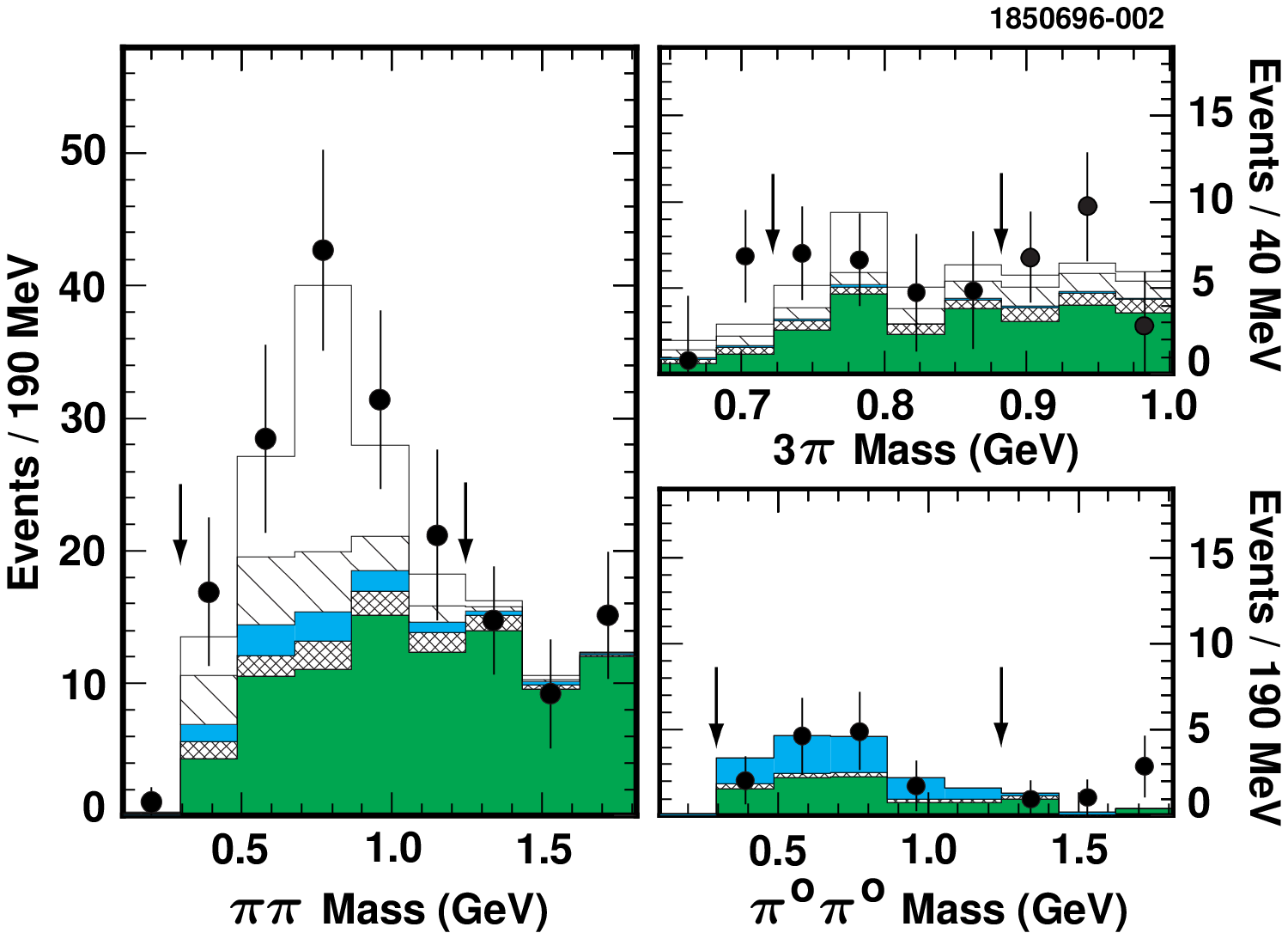,height=7.in,bbllx=0bp,bblly=0bp,bburx=600bp,bbury=700bp,clip=}}
\vspace{-9.1cm}
\fcaption{\label{rholnu} Mass distributions for $\pi^+\pi^-$ plus
$\pi^+\pi^o$ (left), $3\pi$ (upper right) and $\pi^o\pi^o$ (lower right),
for events which are candidates $B\to x\ell\nu$ decays which satisfy
all the other $B$ candidate cuts including a cut on the $B$ mass.The
shading is the same as on the previous figure. The arrows indicate the mass
range used in the analysis.}
 \end{figure}

It is often difficult to prove that a $\pi\pi$ system indeed is dominantly
resonant $\rho$. CLEO attempts to show $\rho$ dominance by plotting the
$\pi^+\pi^-$ and $\pi^+\pi^o$ summed mass spectrum in Fig.~\ref{rholnu}. They also
show a test case of $\pi^o\pi^o\ell\nu$, which cannot be $\rho$, since
$\rho^o$ cannot decay to $\pi^o\pi^o$. There is an enhancement in the
$\pi^+\pi^-$ plus $\pi^+\pi^o$ sum, while the $\pi^o\pi^o$ shows a relatively flat spectrum
that is explained by background. The 3$\pi$ spectrum shows little evidence
of resonant $\omega$, however. More data is needed to settle this issue.
CLEO proceeds by assuming they are seeing purely resonant decays in the
vector channel.

The measured branching ratio is model dependent due to different form-factor
dependences on $q^2$ and lepton momentum. Therefore, CLEO reports
different branching ratios for a selection of models. The ratio of
$\rho\ell\nu /\pi\ell\nu$ is also given, see Table~\ref{table:vubex}, and
compared to model predictions; the errors are non-Gaussian, but the
KS model has only a 0.5\% likelihood of being consistent with the data.

\begin{figure}[htb]
\centerline{\psfig{figure=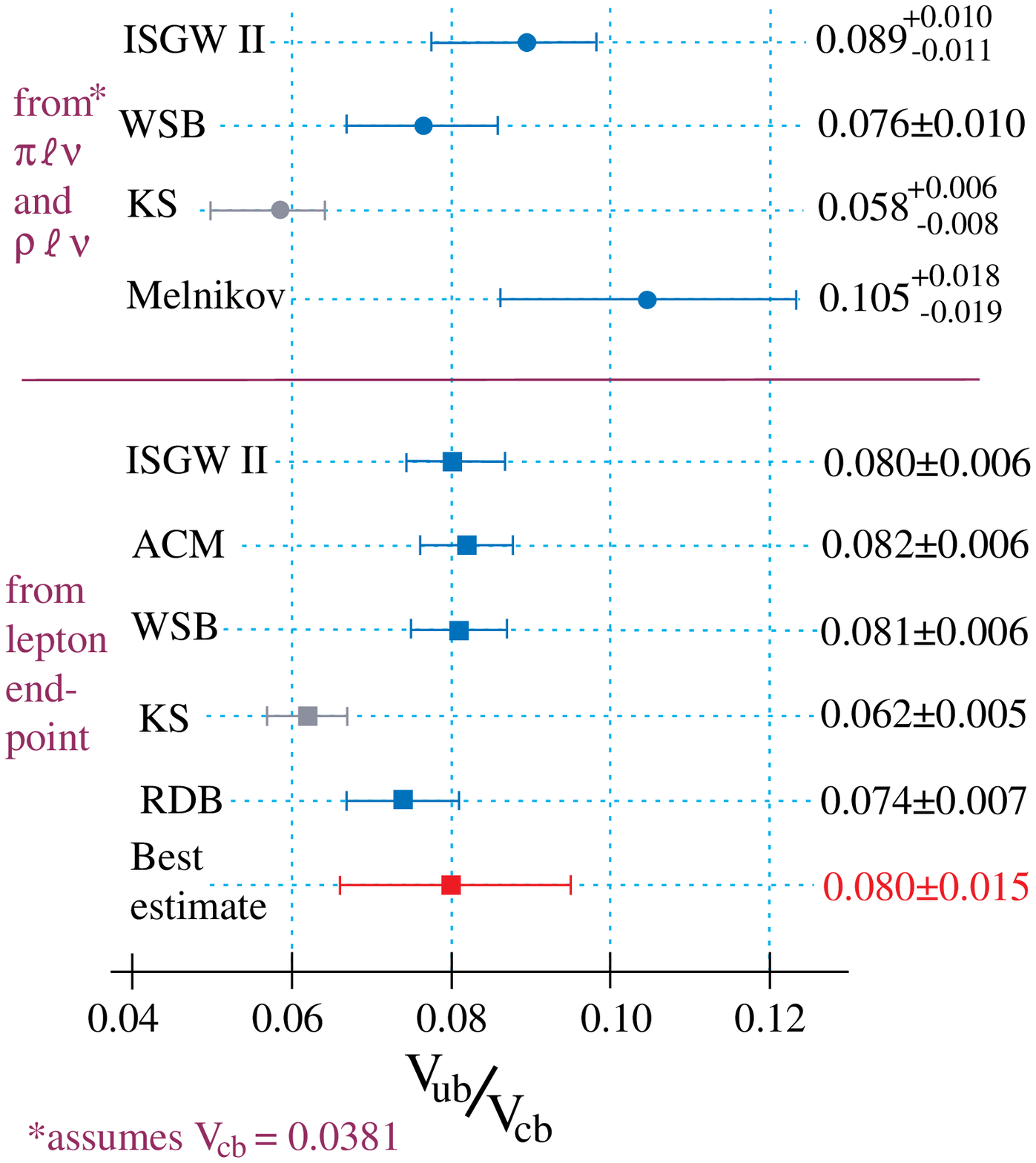,height=7.0in,bbllx=0bp,bblly=0bp,bburx=600bp,bbury=750bp,clip=}}
\vspace{-5.2cm}
\fcaption{\label{Vub_cb} Values of $V_{ub}/V_{cb}$ obtained from the
exclusive $\pi\ell\nu$ and $\rho\ell\nu$ analyses combined and taking
$V_{cb}$ = 0.0381, and results from the inclusive endpoint analysis. The
best estimate combining all models except KS is also given.}
 \end{figure}

The values of $V_{ub}$ obtained from both the exclusive  and the
inclusive analyses are summarized 
in Fig.~\ref{Vub_cb}. For the inclusive analysis, results from CLEO I 
and ARGUS  have been included in the average.\cite{vubold} Since the KS model predicts the wrong pseudoscalar/vector
ratio, it is excluded from the average. The ISGW model has been dropped in
favor of the ISGW II model. The range of model predictions is now narrowed
compared to former analyses. However, the model variations still dominate
the error. A conservative estimate gives
\begin{equation}
\left|V_{ub}\over V_{cb}\right| = 0.080\pm 0.015 ~~,
\end{equation} 
which provides a constraint
\begin{equation}
{\left(1\over \lambda^2\right)}{\left|V_{ub}\over V_{cb}\right|}^2 = 
\left(\rho^2+\eta^2\right)= (0.36\pm 0.07)^2~~.
\end{equation} 

\begin{table}[th]\centering\tcaption{ Results from exclusive semileptonic $b\to u$ transistions }
\label{table:vubex}
\vspace*{2mm}
\small
\begin{tabular}{ccccc}\hline\hline
Model & ${\cal B}(B\to\pi\ell\nu)$ &${\cal B}(B\to\rho\ell\nu)$ & 
$\Gamma(\rho)/\Gamma(\pi)$ & $\Gamma(\rho)/\Gamma(\pi)$\\ 
&$\times 10^4$&$\times10^4$ &&predicted\\\hline 
ISGW II & $2.0\pm 0.5\pm 0.3$ &  $2.2\pm 0.4^{+0.4}_{-0.6}$ & 
$1.1^{+0.5+0.2}_{-0.3-0.3}$ & 1.47\\
WSB & $1.8\pm 0.5\pm 0.3$ & $2.8\pm 0.5^{+0.5}_{-0.8}$ &
$1.6^{+0.7+0.3}_{-0.5-0.4}$ & 3.51\\
KS &$1.9\pm 0.5\pm 0.3$ & $1.9\pm 0.3^{+0.4}_{-0.5}$  &
{\bf $1.0^{+0.5+0.2}_{-0.3-0.3}$} &  4.55\\
Melikhov$^{\dagger}$ &  $1.8\pm 0.4\pm 0.3\pm 0.2$ & 
$2.8\pm 0.5^{+0.5}_{-0.8}\pm 0.4$ & $1.6^{+0.7+0.3}_{-0.5-0.4}\pm 0.11$ 
& 1.53$\pm$0.15\\\hline
\multicolumn{5}{l}{$\dagger$ The 3rd error arises from uncertainties in 
  the estimated form-factors}   
\end{tabular}\end{table}\normalsize
  
\subsection{$B_d^o-\bar{B}_d^o$ Mixing}

\begin{figure}[thb]
\vspace{-1.6cm}
\centerline{\psfig{figure=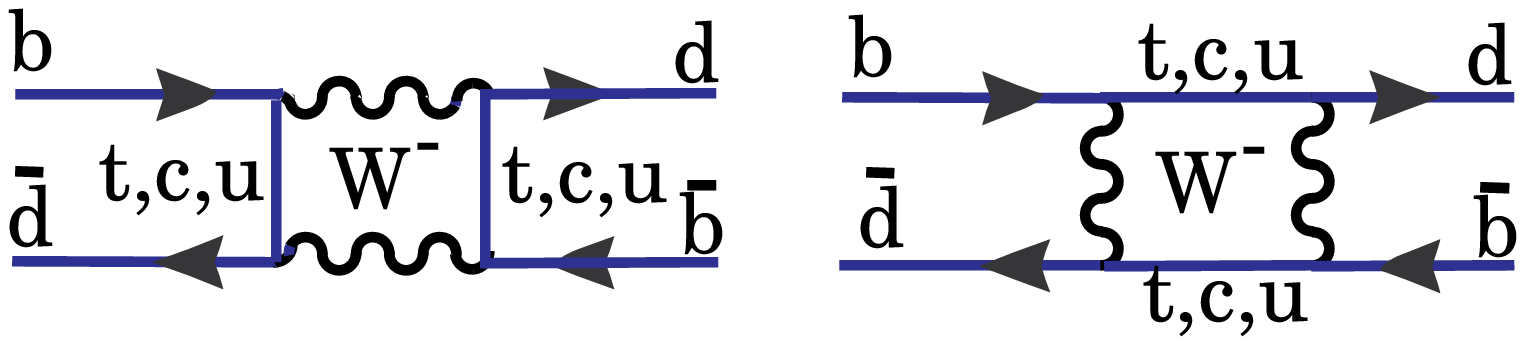,height=2.0in,bbllx=0bp,bblly=540bp,bburx=600bp,bbury=700bp,clip=}}
\vspace{-0.6cm}
\fcaption{\label{bmix}The two diagrams for $B_d$ mixing.} \end{figure}

Neutral $B$ mesons can transform to 
their anti-particles before they decay. The 
diagrams for this process are shown in Fig.~\ref{bmix}. Although $u$, $c$ and
$t$ quark exchanges are all shown, the $t$ quark plays a dominant role mainly
due to its mass, as the amplitude of this process is proportional to the mass
of the exchanged fermion.
 (We will discuss the phenomenon of
mixing in more detail in section 3.2).

 The probability of mixing is given 
by\cite{mixeq}
\begin{equation}
x\equiv \frac{\Delta m}{\Gamma}={G_F^2\over 6\pi^2}B_Bf_B^2m_B\tau_B|V^*_{tb}V_{td}|^2m_t^2
F{\left(m^2_t\over M^2_W\right)}\eta_{QCD},
\end{equation}
where $B_B$ is a parameter related to the probability of the $d$ and $\bar{b}$ 
quarks 
forming a hadron and must be estimated theoretically, $F$ is a known function 
which 
increases approximately as $m^2_t$, and $\eta_{QCD}$ is a QCD
correction, with value about 0.8. By far the largest uncertainty arises from the 
unknown decay constant, $f_B$. $B_d$ mixing was first discovered by the ARGUS 
experiment.\cite{Argmix} (There was a previous measurement by UA1 indicating
mixing for a mixture of $B_d^o$ and $B_s^o$.\cite{UA1mix} At the time it was quite a surprise, since $m_t$ was thought to be in
the 30 GeV range. Since
\begin{equation} 
 |V^*_{tb}V_{td}|^2\propto |(1-\rho-i\eta)|^2=(\rho-1)^2+\eta^2,
\end{equation}
measuring mixing gives a circle centered at (1,0) in the $\rho - \eta$ plane.

\begin{figure}[thb]
\vspace{-0.3cm}
\centerline{\psfig{figure=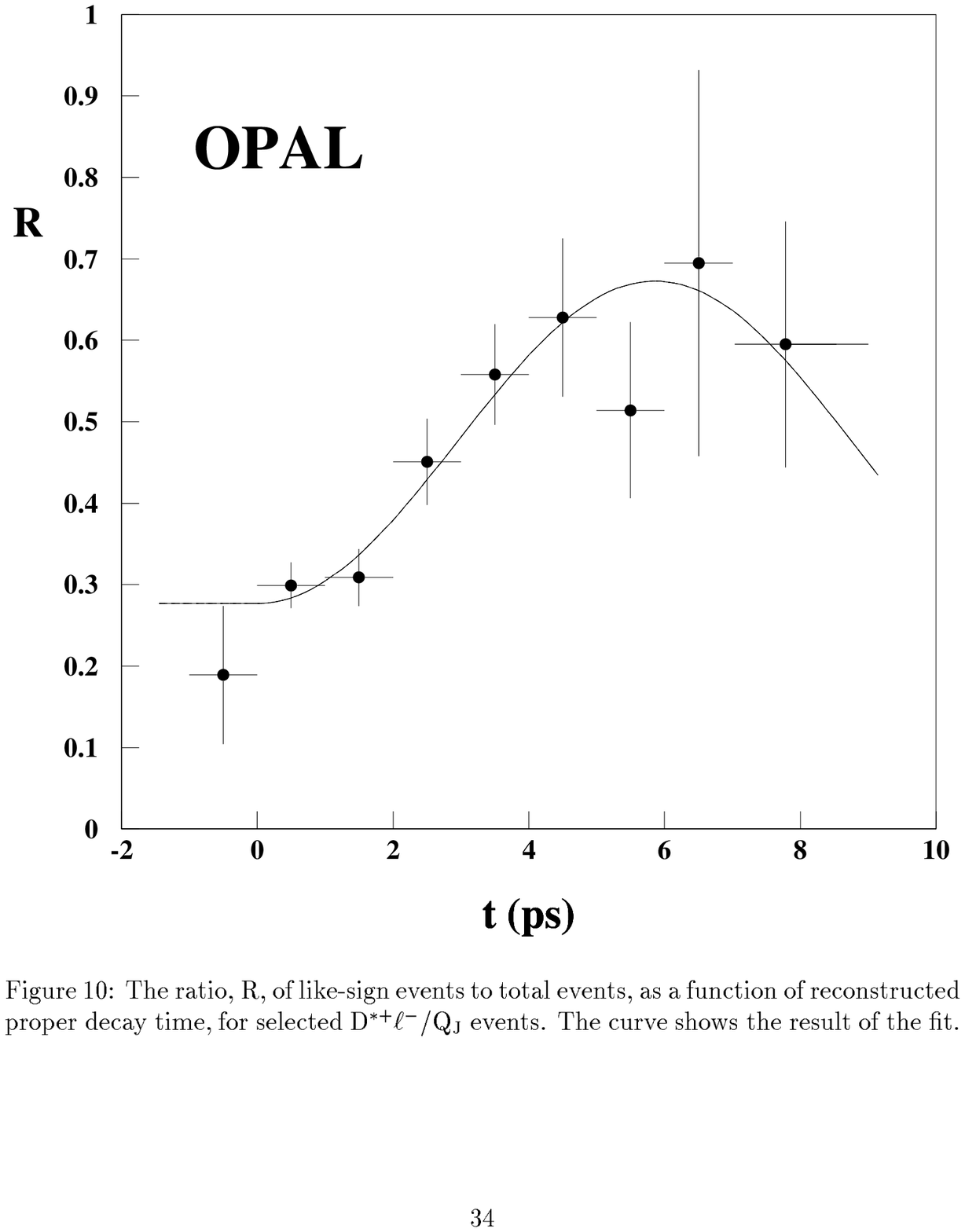,height=4.0in,bbllx=0bp,bblly=186bp,bburx=600bp,bbury=650bp,clip=}}
\vspace{.2cm}
\fcaption{\label{opal_mix}
The ratio, R, of like-sign to total events as a function of proper decay
time, for selected $B\to D^{*+}X\ell^-\bar{\nu}$ events. The jet charge in the
opposite hemisphere is used to determine the sign correlation. The curve is the result of 
a fit to the mixing parameter.}
\end{figure}

The best recent mixing measurements have been done at LEP, where the time-
dependent oscillations have been measured. The OPAL data\cite{OPALmix} is shown
in Fig.~\ref{opal_mix}. Averaging over all LEP experiments 
x=0.728$\pm$0.025.\cite{LEPosc}

\subsection{Rare $B$ Decays}
\begin{figure}[thbp]
\vspace{-2cm}
\centerline{\psfig{figure=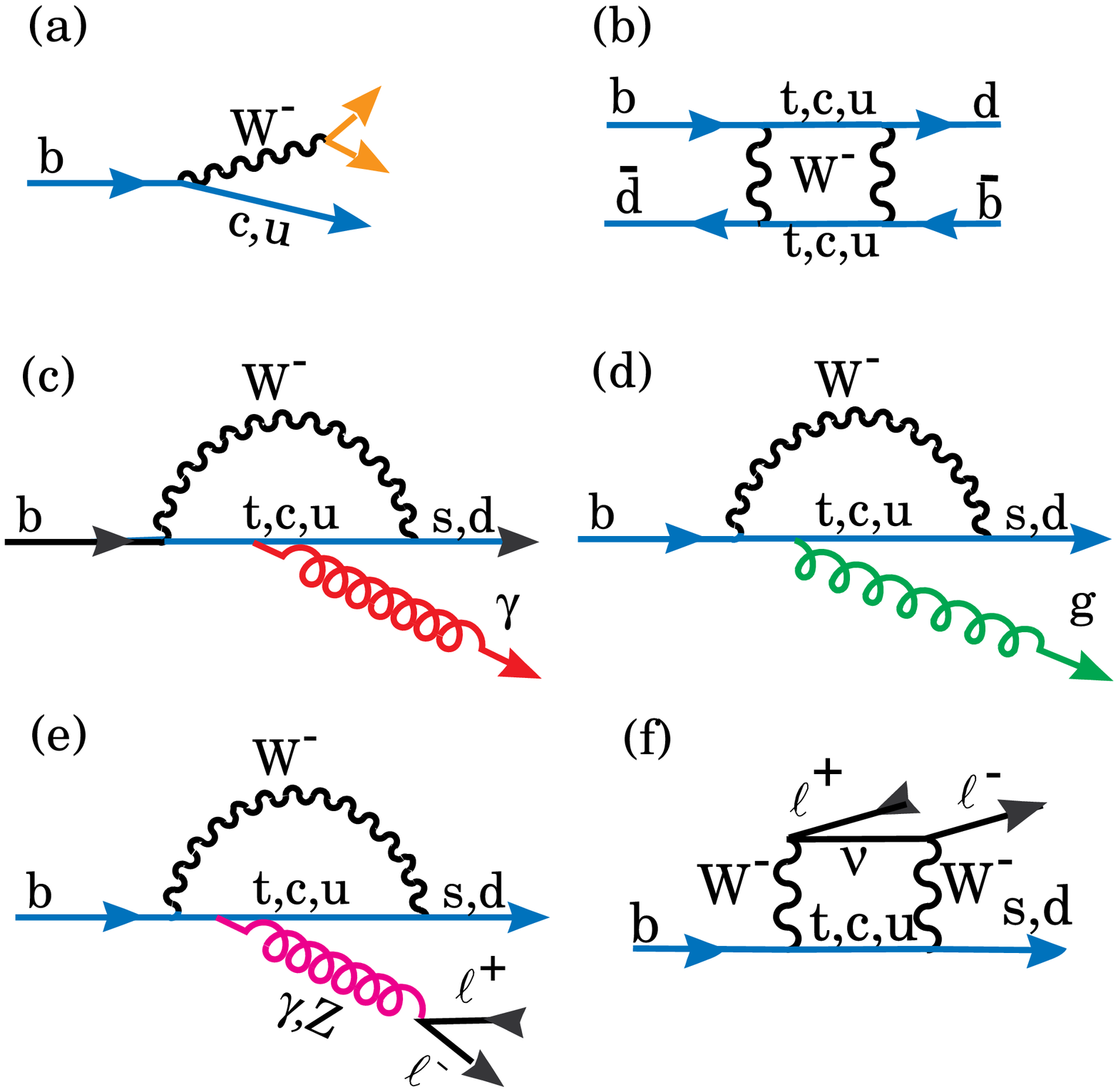,height=6.0in,bbllx=0bp,bblly=160bp,bburx=600bp,bbury=750bp,clip=}}
\vspace{-.3cm}
\fcaption{\label{rarebdecay} A compendium of rare $b$ decay diagrams. (a) The
spectator diagram, rare when $b\to u$; (b) one of the mixing diagrams;
(c) a radiative penguin diagram; (d) a gluonic penguin diagram;
(e) and (f) are dilepton penguin diagrams.} \end{figure}

The term ``rare $B$ decays" is loosely defined. The spectator process shown in 
Fig.~\ref{rarebdecay}(a) is included since $b\to u$ doesn't occur very often ($\approx$1\%),
and the mixing process which occurs often($\approx$17\%) is included since it involves two gauge bosons
(the so called box diagram Fig.~\ref{rarebdecay}(b)). Other loop or box diagrams are shown 
in 
Fig.~\ref{rarebdecay}(d-f).

CLEO found the first unambiguous loop process, the one shown in
Fig.~\ref{rarebdecay}(c).\cite{ksgamma} These  decays  involving a loop diagram are sometimes called
``penguins," an indefensible if amusing term that was injected into the
literature as a result of a bet. For the  Standard Model to be correct these
decays must exist. In fact, penguins are expected  to play an important role in
kaon decay, but there are no unique penguin final  states in kaon decay. Since
penguins are expected to be quite small in charm decay,  it is  only in $B$
decay that penguins can clearly be discerned.

CLEO first found the exclusive final state $B\to K^*\gamma$. An updated value 
for 
the branching ratio is\cite{ksgup}
\begin{equation}
{\cal B}(B\to K^*\gamma)=  (4.2\pm 0.8\pm 0.6)\times 10^{-5}~~.
\end{equation}

\begin{figure}[thb]
\centerline{\psfig{figure=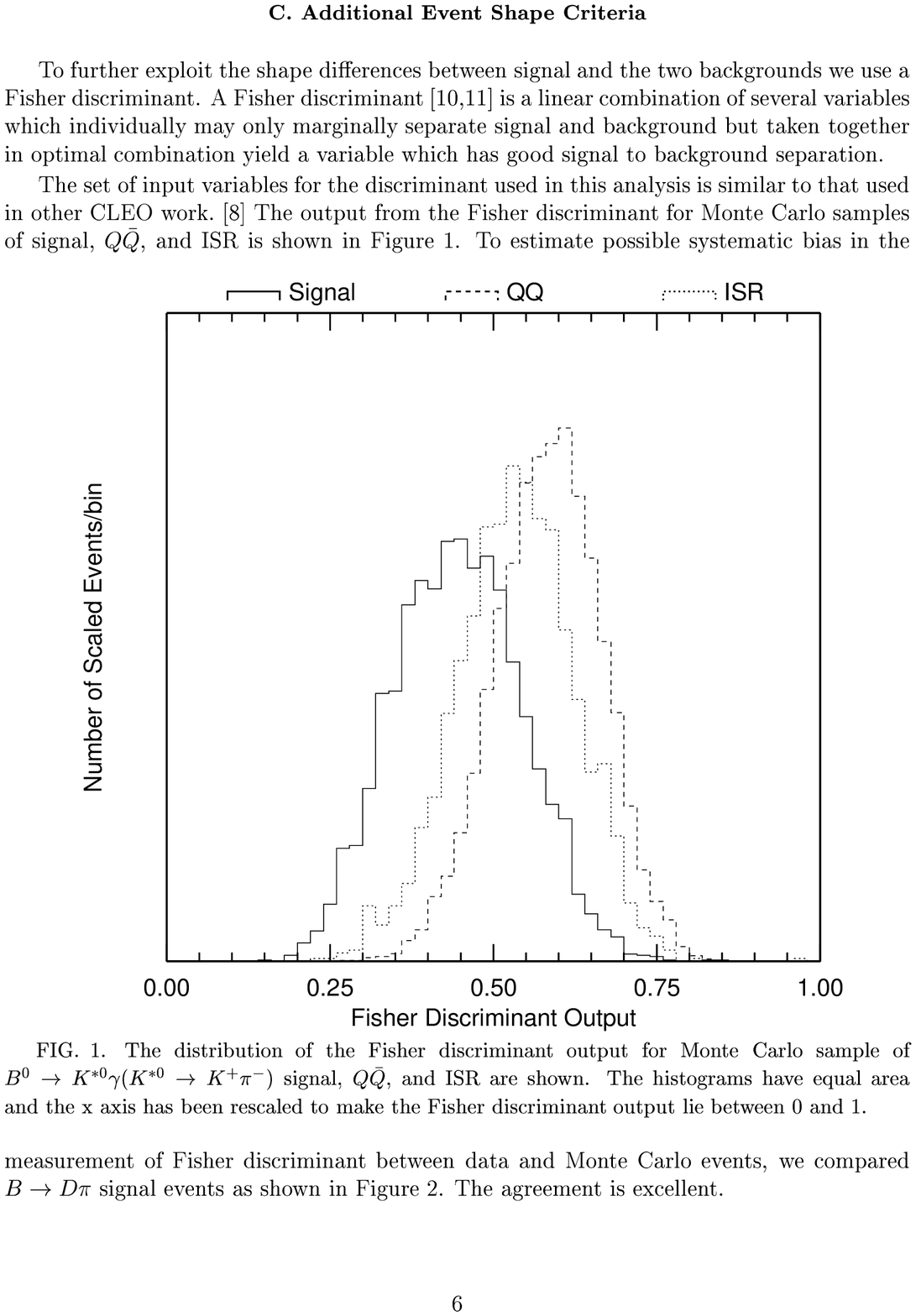,height=3.3in,bbllx=0bp,bblly=184bp,bburx=600bp,bbury=575bp,clip=}}
\vspace{.1cm}
\fcaption{\label{fischer} The distribution of the Fischer discriminant output
for Monte Carlo samples of $B^o\to K^{*o}\gamma(K^{*o}\to K^+\pi^-)$ signal, 
$Q\bar{Q}$ and ISR backgrounds. The histograms have equal area and the
x axis has been rescaled to make the Fischer discriminant output lie 
between 0 and 1.} \end{figure}

This analysis uses the standard $B$ reconstruction technique, summarized in
equation (\ref{eq:mcon}) used at the $\Upsilon (4S)$, combined with some additional 
background suppression cuts. These are separated into trying to insure that one
is dealing with a real $K^*$ and trying to supress background leading to hard 
photons.  The latter comes from initial state radiation (ISR), where one of the
beams radiates  a photon and then subsequently annihilates and from continuum 
quark-antiquark production ($Q\bar{Q}$). Suppression of ISR and $Q\bar{Q}$ is
accomplished by  combining event shape variables into a Fischer discriminant. A
Fischer discriminant\cite{Fischer} is a linear combination of several
variables which individually may have  poor separation between signal and
background, but when taken together yield  acceptable background rejection, the
correlations between the variables helping. The Fischer output distribution for
Monte Carlo simulations of signal, ISR and $Q\bar{Q}$ backgrounds are shown in
Fig.~\ref{fischer}.

\begin{figure}[thb]
\centerline{\psfig{figure=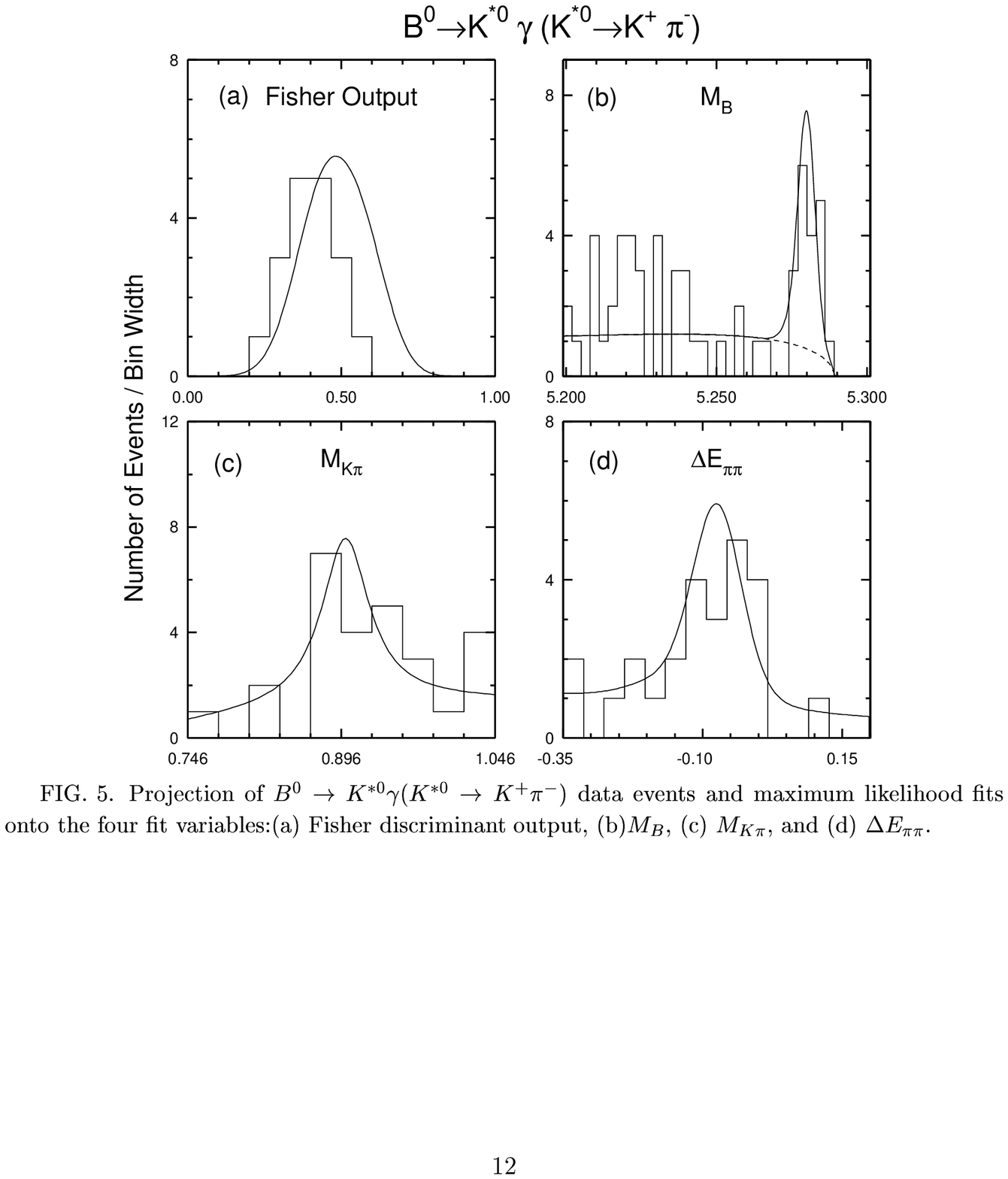,height=3.8in,bbllx=0bp,bblly=227bp,bburx=600bp,bbury=600bp,clip=}}
\vspace{.1cm}
\fcaption{\label{ks_proj} Projections of
 $B^o\to K^{*o}\gamma(K^{*o}\to K^+\pi^-)$ data events (histograms) and maximum
  likelihood
 fits (curves) onto the four fit variables: (a) Fischer discriminant output, (b) $M_B$,
 (c)$M_{K\pi}$ and (d) $\Delta E_{\pi\pi}$, which is the difference between
 the candidate $B$ energy and the beam energy assuming both charged tracks
 are pions.} \end{figure}

The branching ratio is extracted by making a maximum likelihood fit to four 
distributions, $M_B$, $\Delta E$, the $K\pi$ invariant mass $m(K\pi)$, and the Fischer discriminant. To 
illustrate what the signal shapes look like, projection plots are made by
applying  restrictive selection criteria on three of the four likelihood
variables and projecting  the  remaining events onto the axis of the fourth
variable. This is shown for the $K^{*o}\to K^- \pi^+$ mode in Fig.~\ref{ks_proj}.

\begin{figure}[thbp]
{\psfig{figure=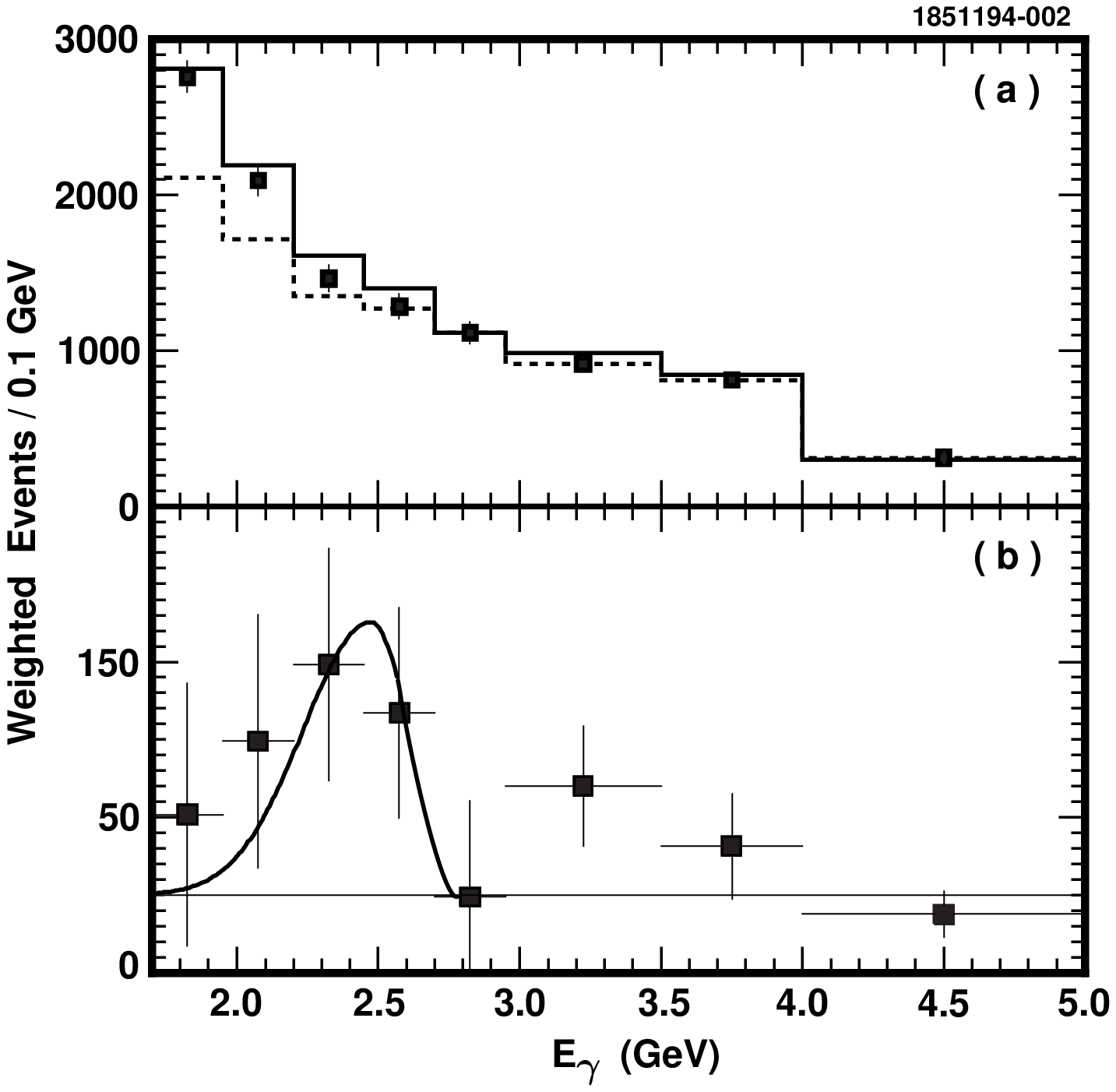,bbllx=50bp,bblly=0bp,bburx=600bp,bbury=700bp,width=2.95in,clip=}}
\vspace{-4cm}\vspace{-2.2in}\hspace{3in}
{\psfig{figure=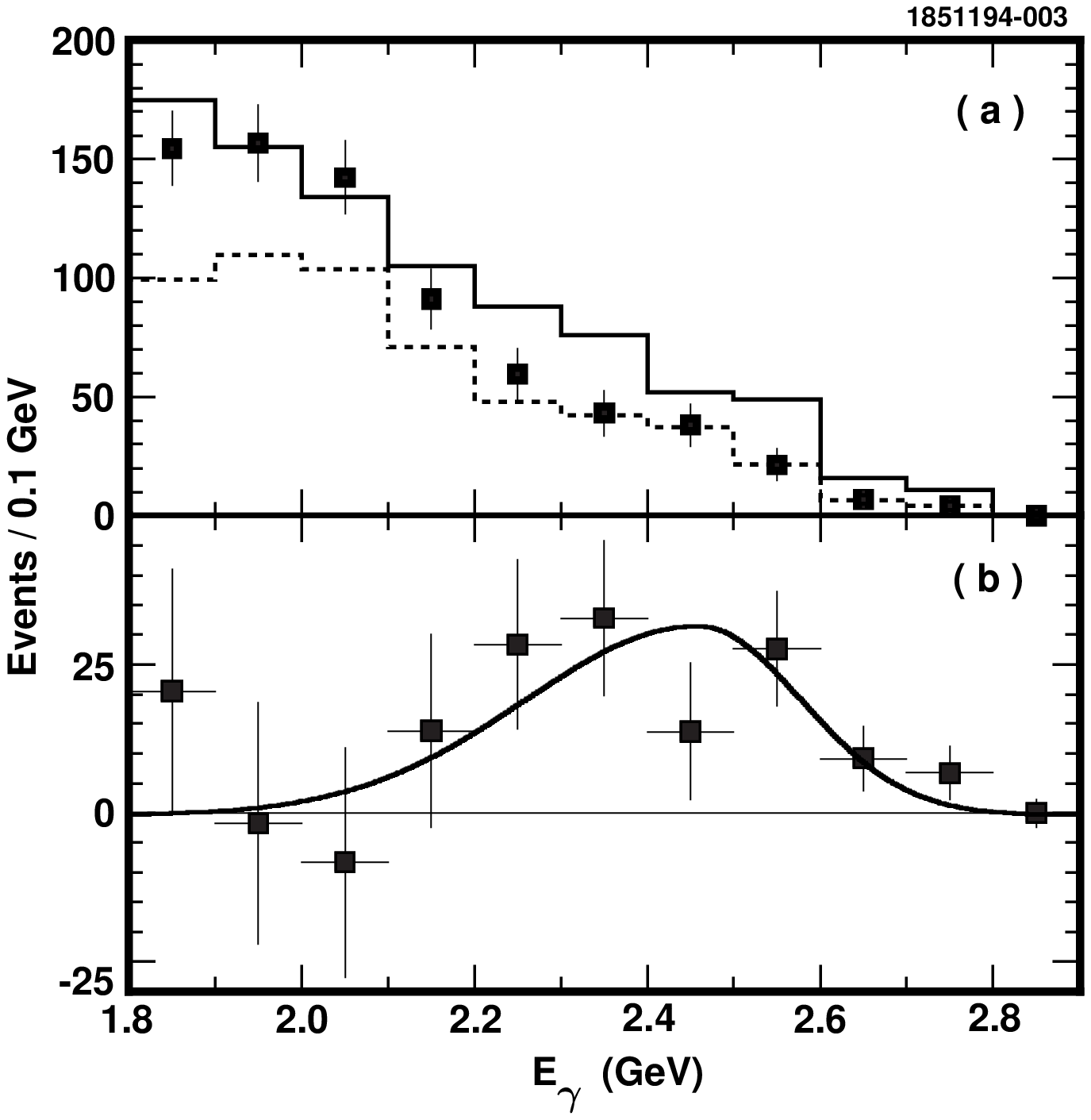,bbllx=0bp,bblly=0bp,bburx=550bp,bbury=700bp,width=2.95in,clip=}}
\vspace{-4.2cm}
\fcaption{\label{inclgamn}Photon energy spectra from the neural net analysis
shown on the left side, and from the $B$ reconstruction analysis, shown on
the right side. In (a) 
 the on resonance date are  the solid lines, the scaled off resonance data
are the dashed lines, and the sum of  backgrounds from off resonance data and
$b\to c$ Monte Carlo are shown as the  square points with error bars. In (b)
 the backgrounds have been subtracted to show the net signal for $b\to
s\gamma$; the solid lines are fits of the signal  using a spectator model
prediction.}  \end{figure}

The extraction of the inclusive rate for $b\to s\gamma$ is more difficult.
There are  two separate CLEO analyses.\cite{btosg}  The first one measures the
inclusive photon  spectrum  from $B$ decay near the maximum momentum end,
similar to what is done to extract an inclusive $b\to X\ell\nu$ signal, but
with the additional problem that the expected branching ratio is much lower.
The main problem is to reduce the ISR  and $Q\bar{Q}$  backgrounds. Here 
instead of using a Fischer discriminant, a set of event shape variables and
energies  formed in a series of cones parallel and antiparallel to the
candidate photon  direction  are fed into a neural net trained on Monte Carlo.
The result is shown in Fig.~\ref{inclgamn}(leftside).

 The second technique constructs the inclusive rate
by summing up the possible  exclusive final states. Since the photons are
expected to be at high momentum, and  therefore take away up to half the $B$'s
rest energy, the  number of hadrons in the final state is quite limited. The
analysis looks for the  final states $B\to K ~{\rm n}\pi\gamma $ where n is
allowed to be a maximum of  4, but  only one  can be a $\pi^o$. Only one entry
per event is allowed. Here background  reduction is accomplished by using the
full power of the exclusive $B$  reconstruction analysis. The resulting
$\gamma$ energy spectrum is shown on the right side of Fig.~\ref{inclgamn}.

The branching ratios found are $(1.88\pm 0.74)\times 10^{-4}$ and
$(2.75\pm 0.67)\times 10^{-4}$ for the neural net and $B$ reconstruction
analyses, respectively. The average of the two results, taking into account the
correlations between the two techniques is
\begin{equation}
{\cal B}(b\to s\gamma)= (2.3\pm 0.5\pm 0.4)\times 10^{-4}~~.
\end{equation}

The theoretical prediction for the branching ratio is given by\cite{bsgc7}
\begin{equation}
{{\Gamma(b\to s\gamma)} \over {\Gamma(b\to c\ell\nu)}} = 
\left|{V_{ts}^*V_{tb} \over V_{cb}}\right|
{\alpha\over 6\pi g(m_c/m_b)}|C_7^{eff}(\mu)|^2,
\end{equation}
where $g(m_c/m_b)$ is a known function. While $C_7$ is calculated 
perturbatively 
at  $\mu$ equal to the $W$ mass, the evolution to $b$ mass scale causes 
$\approx$25\% uncertainty in the prediction, since the proper point could be
$m_b/2$ or $2m_b$. In the leading log approximation the theoretical prediction
is ${\cal B}(b\to s\gamma)= (2.8\pm 0.8)\times 10^{-4}$,\cite{bsgc7} while an
incomplete next to leading order 
calculation, gives 
$\sim 1.9\times 10^{-4}$.\cite{Ciuchini} A recently completed next to leading
order calculation gives $3.3\times 10^{-4}$.\cite{btosgnll} In all
cases the data are consistent with the prediction.

\begin{figure}[thbp]
\vspace{-.7cm}
\centerline{\psfig{figure=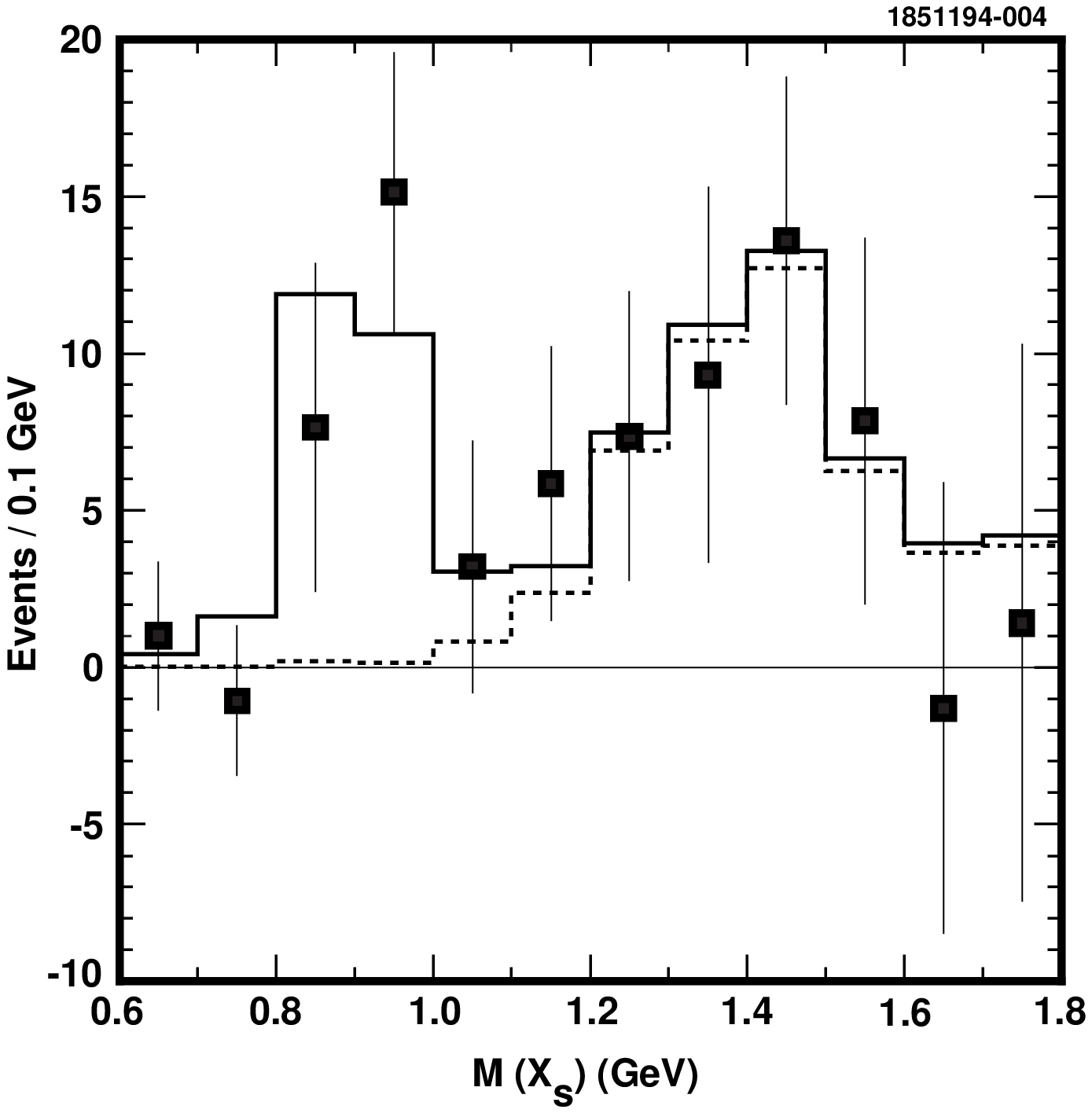,height=4.5in,bbllx=0bp,bblly=0bp,bburx=600bp,bbury=750bp,clip=}}
\vspace{-4.3cm}
\fcaption{\label{xs_mass}The apparent $K~{\rm n}\pi$ mass distribution for the
$B$  reconstruction analysis. The points are the background subtracted data,
not efficiency corrected,  the solid histogram is fit to the data using 
several $K^*$ resonance as input to a  Monte  Carlo simulation, while the
dotted histogram shows all the fit components but the $K^*(890)$.}
 \end{figure}

The second analysis also produces the mass spectrum of the $K~n\pi$ system,
shown in Fig.~\ref{xs_mass}. A clear $K^*(890)$ component is observed. 
The best way to
measure the fraction of $K^*(890)$ is to divide the exclusive result by the average 
inclusive result. This number can test theoretical models, but mostly we are testing
the prediction of the exclusive rate which is the far more difficult 
calculation than the inclusive rate.

The CLEO result is\cite{ksgup}
\begin{equation}
{\Gamma(B\to K^*\gamma) \over \Gamma(b\to s\gamma)} = 0.181\pm 0.068~~.
\end{equation}
Model predictions vary between 4 and 40\%.\cite{playstone}

\begin{figure}[htb]
\vspace{-.3cm}
\centerline{\psfig{figure=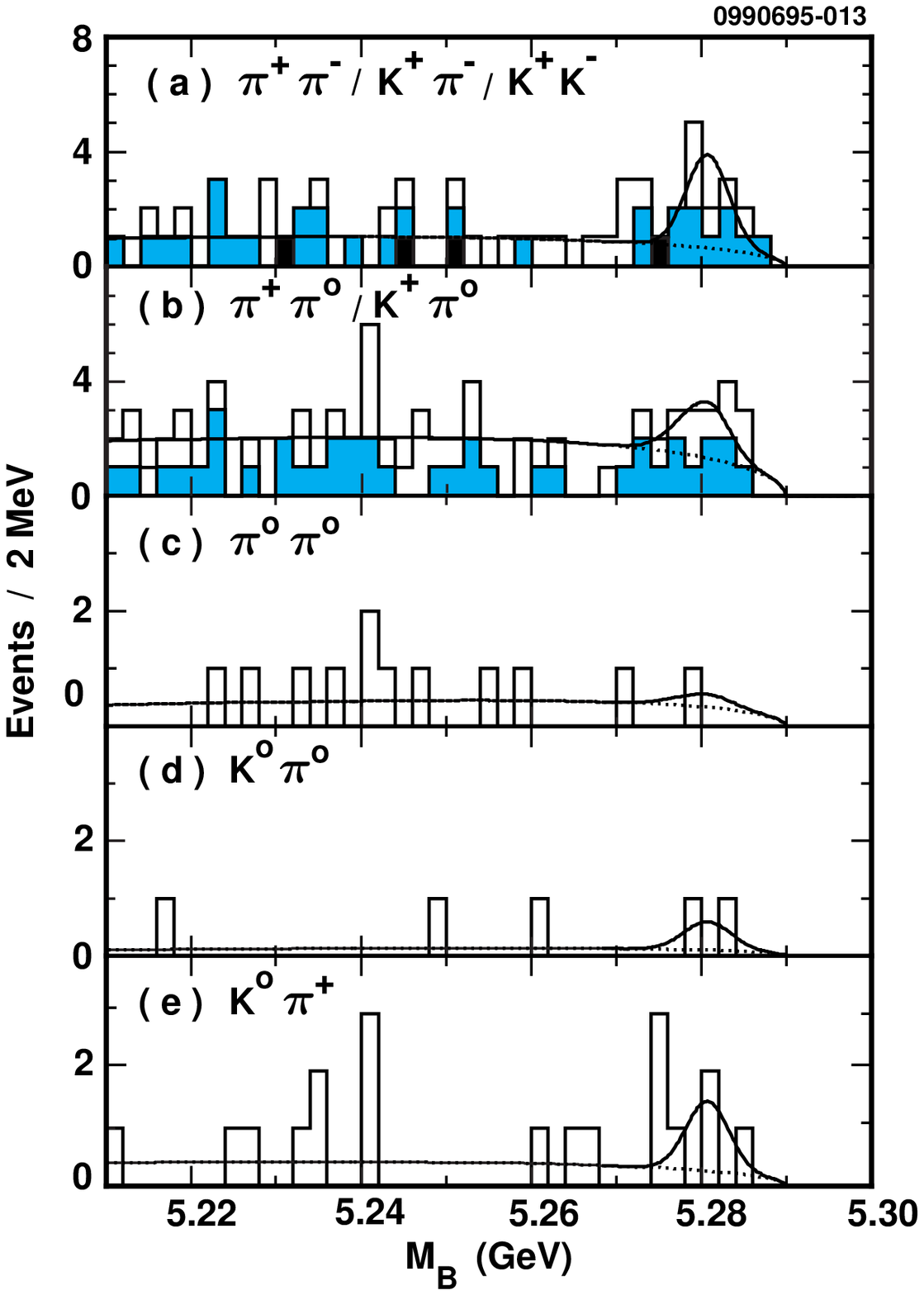,height=6.0in,bbllx=0bp,bblly=0bp,bburx=600bp,bbury=700bp,clip=}}
\vspace{-4.2cm}
\fcaption{\label{kpi_mass}$M_B$ plots for (a) $B^o\to \pi^+\pi^-$ (unshaded),
 $B^o\to K^+\pi^-$, and $B^o\to K^+ K^-$ (black) (b) 
 $B^+\to \pi^+\pi^o$ (unshaded) and $B^+\to K^+\pi^o$ (grey), (c)
 $B^o\to \pi^o\pi^o$, (d) $B^o\to K^o\pi^o$, and (e)
 $B^+\to K^o\pi^+$. The projection of the total likelihood fit (solid curve)
 and the continuum background component (dotted curve) are overlaid.}
\end{figure}

Rare hadronic final states have also been measured. CLEO reported a signal in the 
sum of  $K^{\pm}\pi^{\mp}$ and $\pi^+\pi^-$ final states.\cite{CleoKpi} The particle 
identification 
could not uniquely separate high momentum kaons and pions. While the $K\pi$
mode results from a penguin diagram the $\pi\pi$ mode results mainly from
a $b\to u$ spectator diagram.
 The reconstructed $B$ mass plot is shown in Fig.~\ref{kpi_mass}, along with the results of 
several other searches from an updated analysis,\cite{Kpiup} based
on 2.4 fb${-1}$ of integrated luminosity on the $\Upsilon (4S)$. Here a best guess
 is made as to which final state is present. The
resulting rate is
\begin{equation}
{\cal B}(B^o\to K^{\pm}\pi^{\mp}+\pi^+\pi^-)=(1.8^{+0.6+0.2}_{-0.5-0.3}\pm 0.2)
\times 10^{-5}~~.
\end{equation}

\begin{figure}[htbp]
\vspace{-.15cm}
\centerline{\psfig{figure=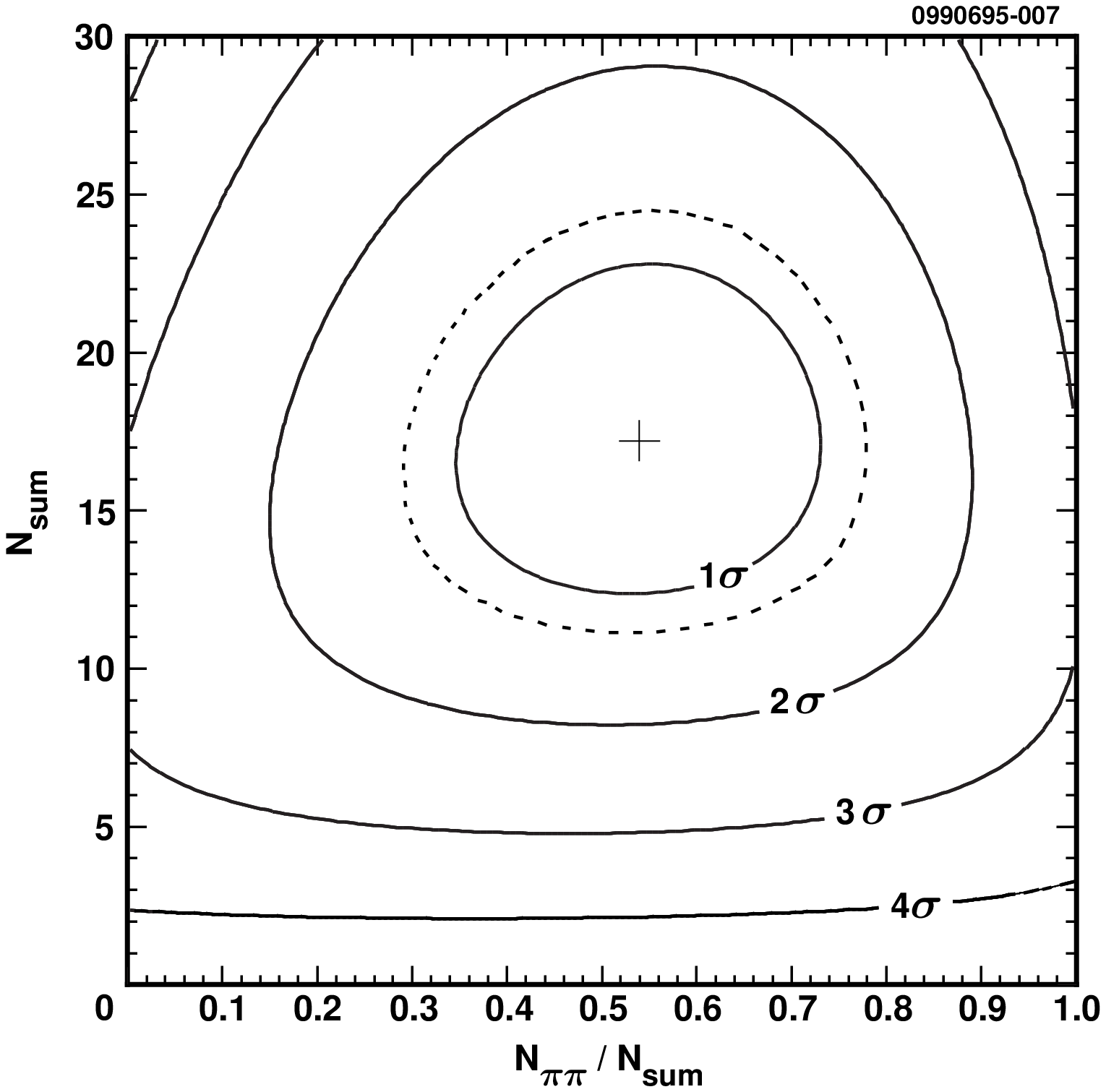,height=4.0in,bbllx=0bp,bblly=0bp,bburx=600bp,bbury=700bp,clip=}}
\vspace{-3.5cm}
\fcaption{\label{dipifrac} The central value (+) of the likelihood fit to 
$N_{sum}\equiv 
N_{\pi\pi}+N_{K\pi}$ and the fraction $N_{\pi\pi}/N_{sum}$. The solid curves are 
the
$n\sigma$ contours and the dotted curve is the 1.28$\sigma$ contour.}
\end{figure}

An attempt to separate the kaon and pion components using the small difference 
in reconstructed energy and whatever particle identification power exists leads
to the dipion fraction shown in Fig.~\ref{dipifrac}. The best current guess is
that  approximately half of the rate is due to $\pi^+\pi^-$.

\begin{figure}[hbtp]
\vspace{-1.3cm}
\centerline{\psfig{figure=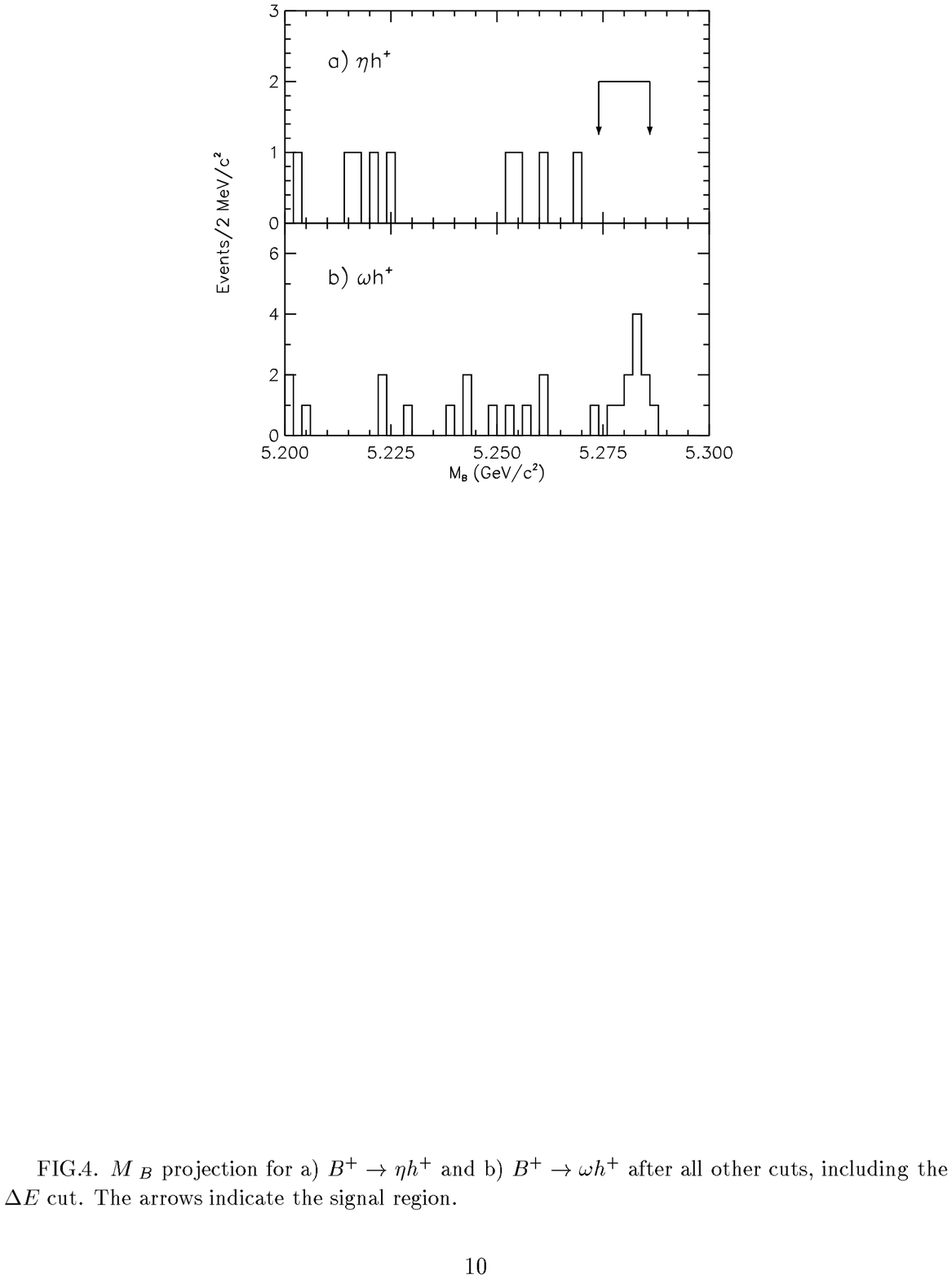,height=3.5in,bbllx=0bp,bblly=400bp,bburx=600bp,bbury=700bp,clip=}}
\vspace{-.6cm}
\fcaption{\label{omega_pi} The $M_B$ projection for a) $B^+\to \eta h^+$ and
b) $B^+\to \omega h^+$ after all other cuts, including the $\Delta E$ cut.
The arrows indicate the signal region.}
\end{figure}

CLEO also has found a signal in the sum of $\omega\pi^+$ and $\omega K^+$
decays.\cite{omega-pi} The $B$ mass plot is shown in Fig.~\ref{omega_pi}. The signal is 10 events observed on a 
background of 2$\pm$0.3 events. The branching ratio is
\begin{equation}
{\cal B}(B^+\to \omega\pi^+ +\omega K^+)=(2.8\pm 1.0\pm 0.5)
\times 10^{-5}~~.
\end{equation}

\begin{figure}[htbp]
\vspace{-.2cm}
\centerline{\psfig{figure=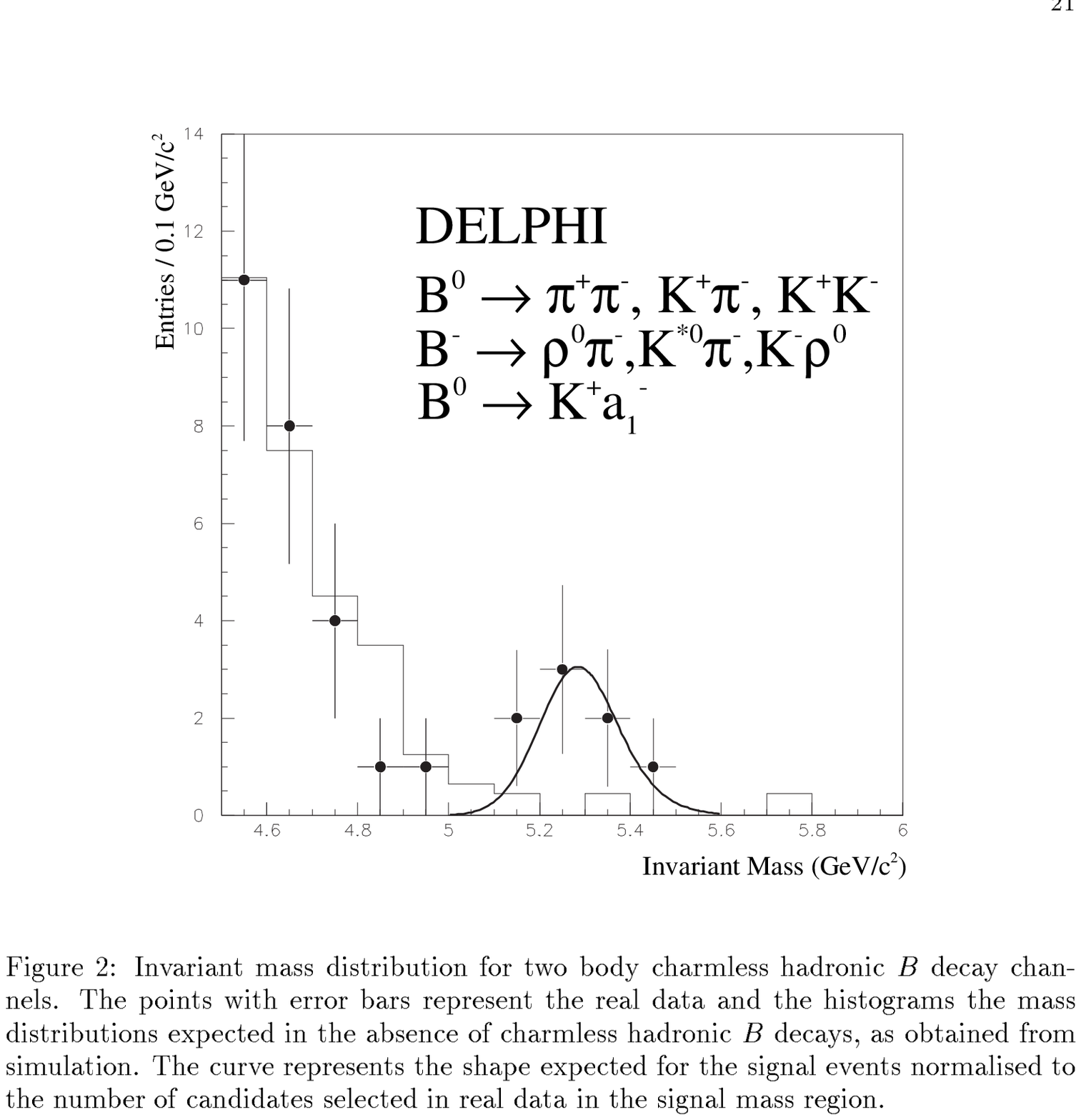,height=3.3in,bbllx=0bp,bblly=400bp,bburx=600bp,bbury=750bp,clip=}}
\vspace{-.2cm}
\fcaption{\label{delph_rare}Invariant mass distribution for two-body charmless
hadronic $B$ decays. The points with error bars represent the real data and the
histograms the mass distributions expected in the absence of such decays as
obtained from simulation. The curve represents the shape expected for the
signal events normalized to the number of candidates selected in real data in
the signal mass region.}
\end{figure}

\begin{figure}[htbp]
\vspace{-1.2cm}
{\psfig{figure=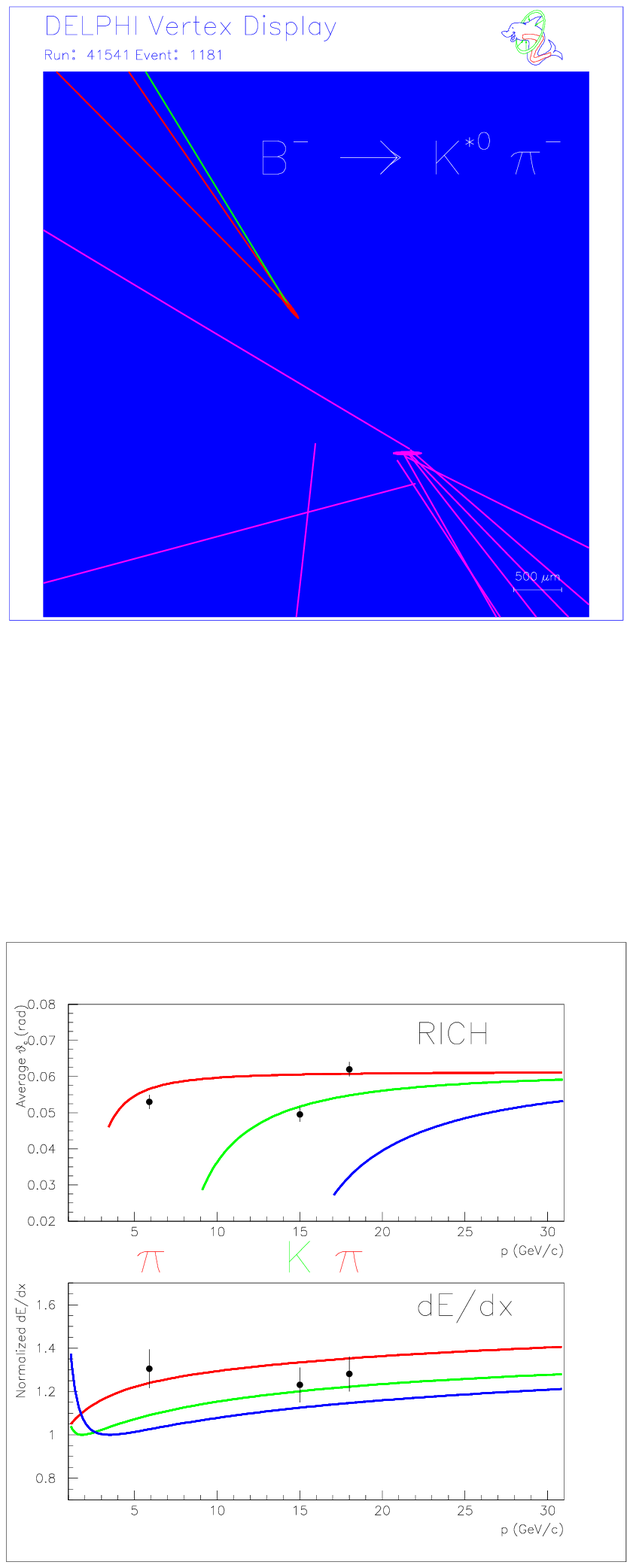,width=2.95in,bbllx=130bp,bblly=450bp,bburx=400bp,bbury=750bp,clip=}}
\vspace{-3.05in}\hspace{3in}
{\psfig{figure=delph_rare_evt.ps,width=2.95in,bbllx=130bp,bblly=100bp,bburx=400bp,bbury=370bp,clip=}}
\vspace{-.5cm}
\fcaption{\label{delph_rare_evt}The candidate $B^-\to K^{*o}\pi^-$ decay: a magnified view of the 
extrapolated tracks at the vertex is displayed above. The primary and secondary 
vertices are indicated by error ellipses corresponding to 3$\sigma$ regions. The plot 
below summarizes the hadron identification properties. The lines represent the 
expected response to pions (upper), kaons (middle) and protons (lower), and the points with error bars the 
measured values for the reconstructed $B$ decay products.}
\end{figure}

DELPHI also reports a signal of 11 ``rare" events over a background of 1 event.
The invariant mass plot is shown in Fig.~\ref{delph_rare}.\cite{delph_r}
One of these events appears to be uniquely identified as a $K^{*o}\pi^-$ final
state  and  this then is an unambiguous hadronic penguin decay. The evidence
is shown in Fig~\ref{delph_rare_evt}.

\section{Importance of  Further Study of $B$ Decays}
\subsection{Tests of the Standard Model via the CKM triangle}
The unitarity of the CKM matrix\footnote{Unitarity implies that any pair of 
rows or 
columns are orthogonal.} allows us to construct six relationships. The most useful 
turns out to be
\begin{equation}
V_{ud}V_{td}^*+V_{us}V_{ts}^*+V_{ub}V_{tb}^*=0~~.
\end{equation}
To a good approximation
\begin{equation}
V_{ud} \approx V_{tb}^*\approx 1 {\rm ~~ and~~}V_{ts}^*\approx -V_{cb},
\end{equation}
then
\begin{equation}
{V_{ub}\over V_{cb}} + {V_{td}^*\over V_{cb}} - V_{us} = 0~~.
\end{equation}
Since $V_{us}=\lambda$, we can define a triangle with sides
\begin{eqnarray}
1 & & \\ 
\left|{V_{td}\over A\lambda^3 }\right| &=&{1\over \lambda}\sqrt{\left(\rho-
1\right)^2+\eta^2}
={1\over \lambda} \left|{V_{td}\over V_{ts}}\right|\\
\left|{V_{ub}\over A\lambda^3}\right|  &=&{1\over \lambda}\sqrt{\rho^2+\eta^2}
={1\over \lambda} \left|{V_{ub}\over V_{cb}}\right|.
\end{eqnarray}

\begin{figure}[htbp]
\vspace{-.8cm}
\centerline{\psfig{figure=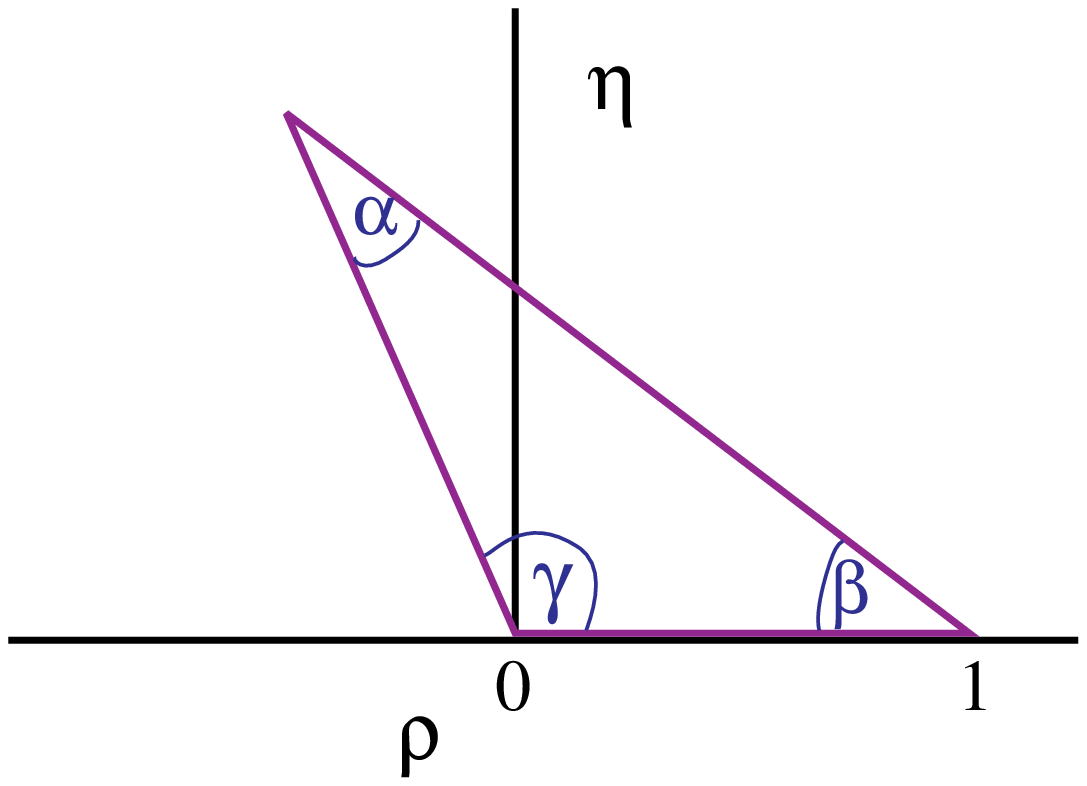,height=3in,bbllx=0bp,bblly=450bp,bburx=600bp,bbury=750bp,clip=}}
\vspace{-.6cm}
\fcaption{\label{ckm_tri}The CKM triangle shown in the $\rho-\eta$ plane. The
left side is determined by $|V_{ub}/V_{cb}|$ and the right side can be
determined using mixing in the neutral $B$ system. The angles can be found
by making measurements of CP violation in $B$ decays.}
\end{figure}

The CKM triangle is depicted in Fig.~\ref{ckm_tri}. 
We know two sides already:
the base is  defined as unity and the left side is determined by the
measurements of  $|V_{ub}/V_{cb}|$. The right side can be determined using
mixing measurements in  the neutral $B$ system. We will see, however, that
there is a large error due to  the  uncertainty in $f_B$. Later we will discuss
other measurements that can access this side. The figure also shows the angles
as $\alpha,~\beta$, and $\gamma$. These angles can be determined by measuring
CP violation in the $B$ system. First we  discuss CP violation in the $K^o_L$
system which also provides constraints on $\rho$ and $\eta$.

{\bf To test the Standard Model we can measure all three sides and all three angles. 
If we see consistency between all of these measurements we have defined the 
parameters 
of the Standard Model. If we see inconsistency, the breakdown can point us beyond 
the Standard Model.}

\subsection{CP Violation}

The fact that the CKM matrix is complex allows CP violation. This is not only 
true  for three generations of quark doublets, but for any number greater than
two. Now  let  us explain what we mean by CP violation. C is a quantum
mechanical operator  that  changes particle to antiparticle, while P switches
left to right, i.e. $x\to-x$. Thus  under  a P operation,
$\overrightarrow{p}\to- \overrightarrow{p}$ since $t$ is unaffected.

\begin{figure}[htb]
\vspace{-.2cm}
\centerline{\psfig{figure=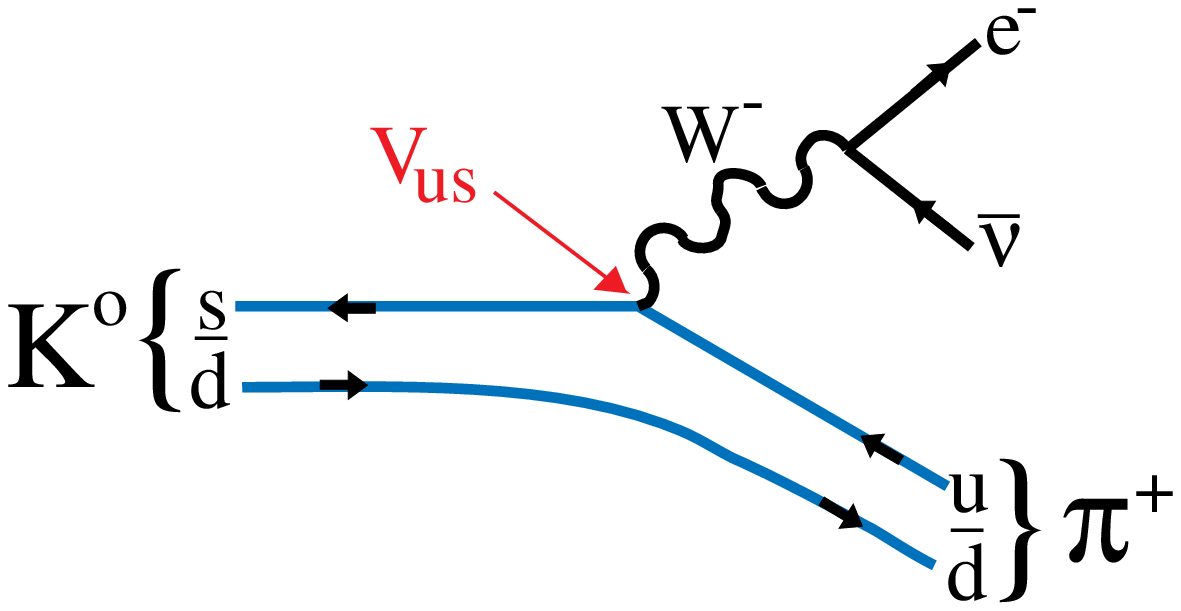,height=2.5in,bbllx=0bp,bblly=480bp,bburx=600bp,bbury=700bp,clip=}}
\vspace{-.15cm}
\fcaption{\label{k0_decay}Semileptonic decay of a $s$ quark contained in
a $K^o$ meson.}
\end{figure}

\begin{figure}[htb]
\vspace{-1.2cm}
\centerline{\psfig{figure=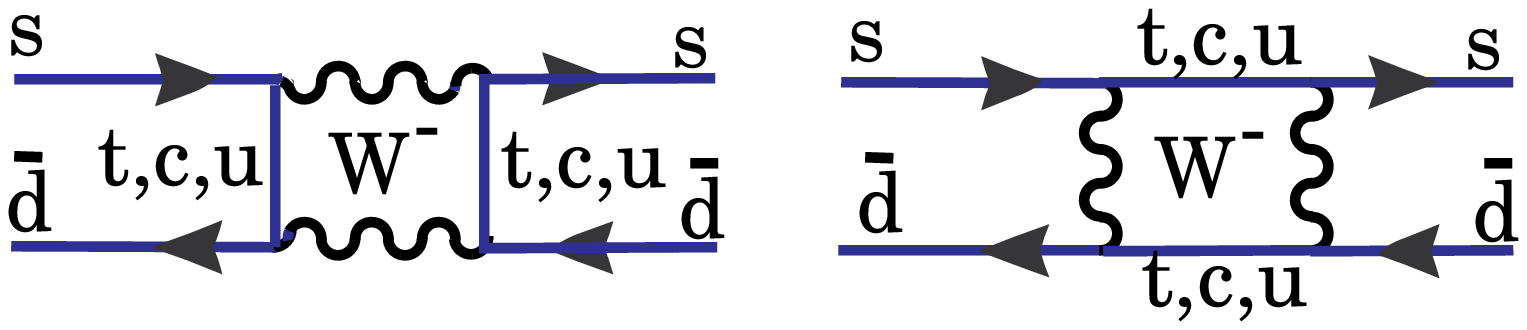,height=2.in,bbllx=0bp,bblly=540bp,bburx=600bp,bbury=700bp,clip=}}
\vspace{-.15cm}
\fcaption{\label{kmix} $K^o-\bar{K}^o$ mixing diagrams.}
\end{figure}

Examples of CP violation have been found in the $K^o$ system. Let us examine 
one 
such measurement. Consider the $K^o$ to be composed of long lived and short 
lived 
components having equal weight, so the wave function is
\begin{equation}
\big| K^o\big> = {1\over \sqrt{2}}\left(\big|K_S \big> + \big|K_L \big>\right)~~.
\end{equation}
In the case of neutral kaons there is a large difference in lifetimes between
the short lived and long lived components. The lifetimes are $9\times 10^{-11}$ sec
and $5\times 10^{-8}$ sec. Suppose we set up a detector far away from the $K^o$
production target. Then after the $K_S$ decay away we have only a $K_L$
beam. We find both
\begin{equation}
K_L\to e^+\nu_e\pi^-  {~\rm and~}
K_L\to e^-\bar{\nu}_e\pi^+
\end{equation}
are present. Now the initial state was a $K^o$, which contains an $\bar{s}$
quark and can only decay semileptonically into the $e^+\nu_e\pi^- $ final state
as shown in  Fig.~\ref{k0_decay}.  Thus we have found evidence that both $K^o$
and $\overline{K}\!^o$ are present. This phenomenon, $K^o\Leftrightarrow 
\overline{K}\!^o$ is called mixing and can be depicted by the diagram shown in 
Fig.~\ref{kmix}, much like the diagram for $B^o\- \overline{B}\!^o$ mixing.
However, here  the  $c$-quark loop has the largest amplitude, unlike the $B$
case, where the $t$-quark  is  dominant. (This is because the CKM couplings are
so much larger, i.e. $V_{cs}$ and  $V_{cd}$ $\gg$ $V_{ts}$ and $V_{td}$ and
this compensates for the decrease due to $(m_c/m_t)^2$.) There are also
hadronic intermediate states which contribute to the real part of the mixing
amplitude, such as $K^o\to\pi\pi\to \bar{K}^o$.


An example of CP violation is the measured rate asymmetry in our 
$K^o_L$
detector\cite{PDG} 
\begin{equation}
\delta=2Re(\epsilon)={{\#(K_L\to e^+\nu_e\pi^-)-\#(K_L\to e^-\bar{\nu}_e\pi^+)} \over
{ \#(K_L\to e^+\nu_e\pi^-)+\#(K_L\to e^-\bar{\nu}_e\pi^+) } }= 3.3\times10^{-3}~~.
\end{equation}

\begin{figure}[htb]
\vspace{-2cm}
\centerline{\psfig{figure=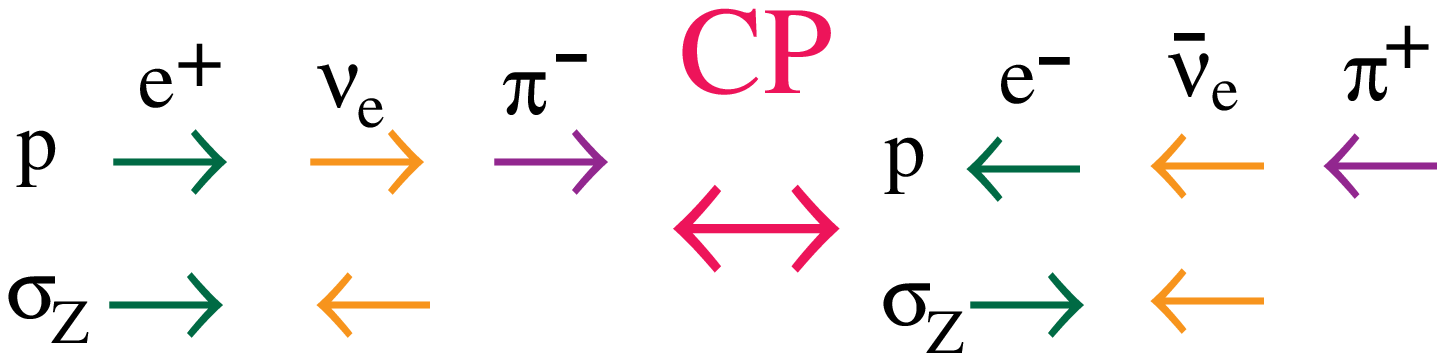,height=2.7in,bbllx=0bp,bblly=400bp,bburx=600bp,bbury=600bp,clip=}}
\vspace{-.4cm}
\fcaption{\label{KLCP}The momentum and spin orientations of the two final
states in semileptonic $K^o_L$ decay, showing that they are mapped into
one another by a CP transformation.}
\end{figure}

Let us look at why this violates CP. In Fig.~\ref{KLCP} the momentum and spin vectors
for  the two final states are shown. The CP operation transforms the
$e^+\nu_e\pi^-$  to the $e^-\bar{\nu}_e\pi^+$ final state and vice-versa. Thus
CP invariance would  imply equal rates for the two processes, contrary to what
is observed. 

CP violation thus far has only been seen in the neutral kaon 
system.\footnote{The other observed example of CP violation is the decay 
$K^o_L\to  \pi\pi$.} If we can find CP violation in the $B$ system we could see
if the CKM  model works or perhaps go beyond the model. Speculation has it that
CP violation  is  responsible for the baryon-antibaryon asymmetry in our
section of the Universe. If  so,  to understand the mechanism of CP violation
is critical in our conjectures of why  we  exist.\cite{langacker}

\begin{figure}[htbp]
\vspace{-1.2cm}
\centerline{\psfig{figure=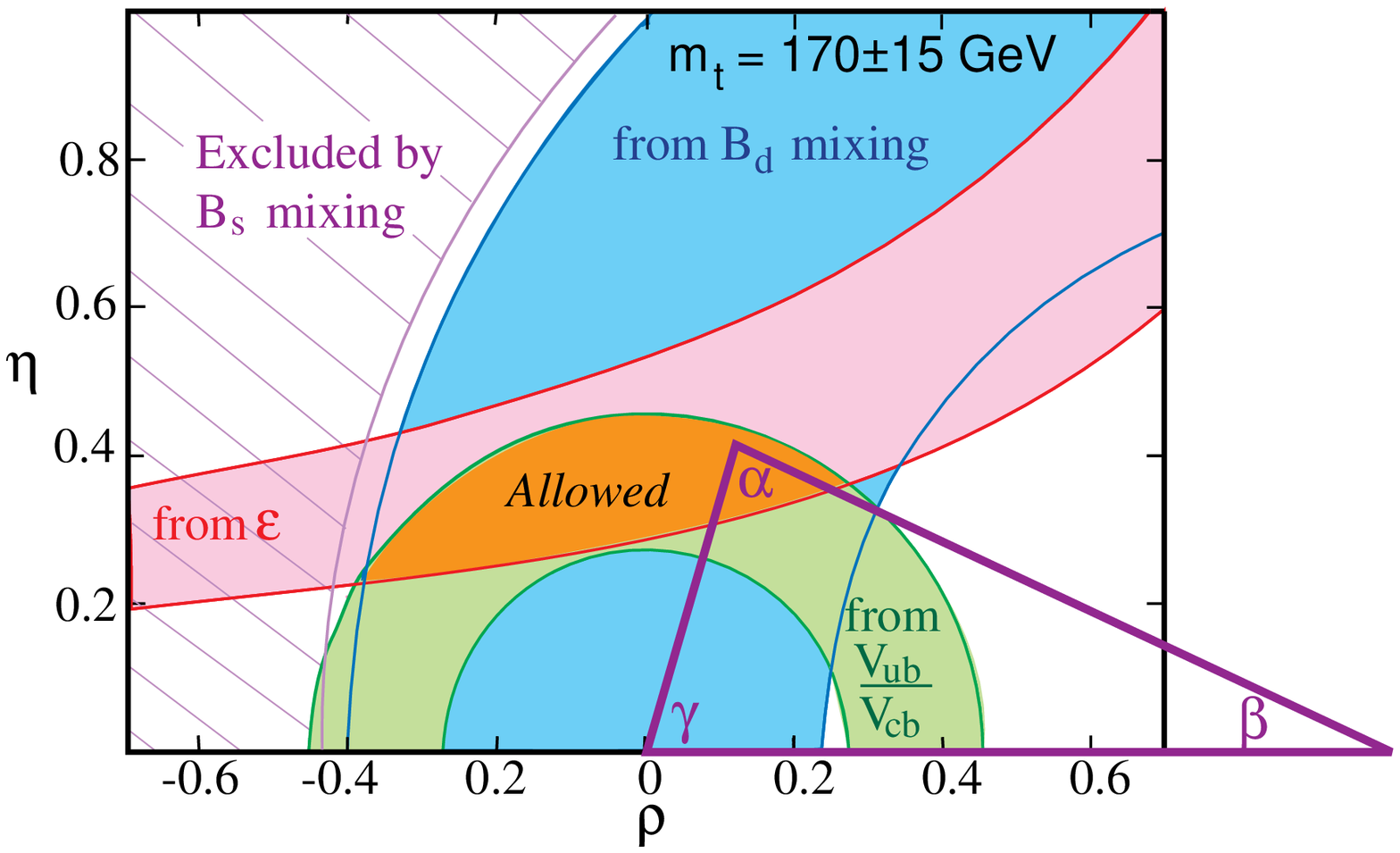,height=6.0in,bbllx=0bp,bblly=200bp,bburx=600bp,bbury=700bp,clip=}}
\vspace{-4.2cm}
\fcaption{\label{ckm_fig}The regions in $\rho-\eta$ space (shaded) consistent
with measurements of CP violation in $K_L^o$ decay ($\epsilon$), $V_{ub}/V_{cb}$
in semileptonic $B$ decay, $B_d^o$ mixing, and the excluded region from
limits on $B_s^o$ mixing. The allowed region is defined by the overlap of
the 3 permitted areas, and is where the apex of the  CKM triangle  sits.}
\end{figure}

\begin{figure}[htbp]
\vspace{-2.3cm}
\centerline{\psfig{figure=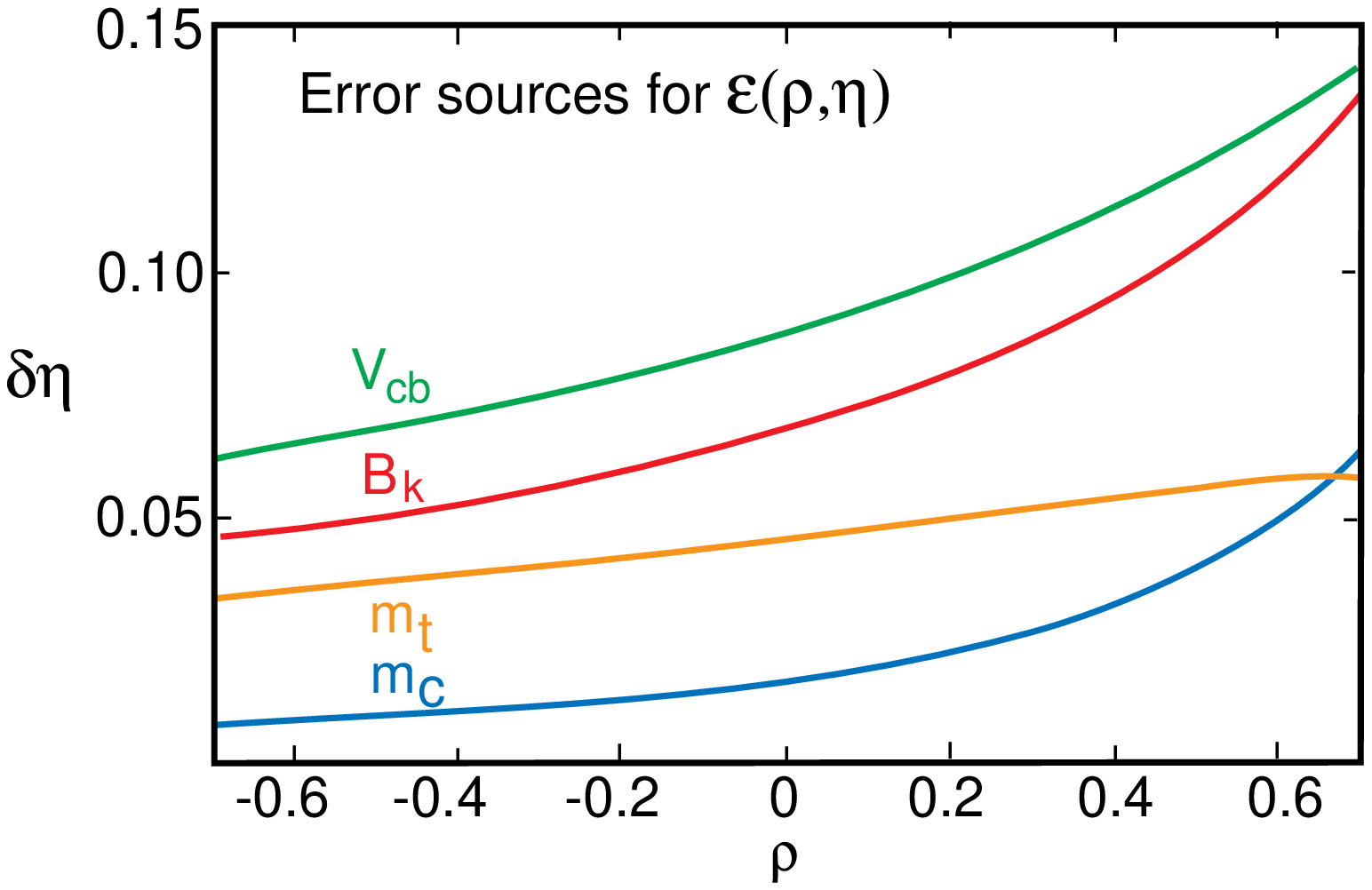,height=4.0in,bbllx=0bp,bblly=300bp,bburx=600bp,bbury=700bp,clip=}}
\vspace{-.55cm}
\fcaption{\label{err_eps}Error sources in units of $\delta\eta$ on the value
of $\eta$ as a function of $\rho$ provided by the CP violation constraint
in $K^o_L$ decay.}
\end{figure}

There is a constraint on $\rho$ and $\eta$ given by the $K_L^o$ CP violation 
measurement ($\epsilon$), given by\cite{buras}
\begin{equation}
\eta\left[(1-\rho)A^2(1.4\pm 0.2)+0.35\right]A^2{B_K \over 0.75}=(0.30\pm 0.06),
\end{equation}
where the errors arise from uncertainties on $m_t$ and $m_c$. The constraints on 
$\rho$ versus $\eta$ from the $V_{ub}/V_{cb}$ measurement, $\epsilon$ and $B$ 
mixing are shown in Fig.~\ref{ckm_fig}. The width of the $B$ mixing band is caused mainly 
by 
the uncertainty on $f_B$, taken here as $240> f_B > 160$ MeV. The width of the 
$\epsilon$ band is caused by errors in $A$, $m_t$, $m_c$ and $B_K$. The size of 
these error sources is shown in Fig.~\ref{err_eps}. The largest error still comes from the 
measurement of $V_{cb}$, with the theoretical estimate of $B_K$ being a close 
second. The errors on $m_t$ and $m_c$ are less important. 

\subsection{Ways of Measuring CP violation in $B$ Decays}
\subsubsection{CP Violation in Charged $B$ Decays}
The theoretical basis of the study of CP violation in $B$ decays was given
in series of papers by Carter and Sanda and Bigi and Sanda.\cite{sanda}
We start with charged $B$ decays.
Consider the final states $f^{\pm}$ which can be reached by two distinct weak 
processes ${\cal A}$ and ${\cal B}$. Then the strong  ($s$) and weak ($w$) parts are
\begin{equation}
{\cal A}=a_se^{i\theta_s}a_we^{i\theta_w},~~{\cal B}=b_se^{i\delta_s}b_we^{i\delta_
w}
~~.
\end{equation}
Under the CP operation the strong phases remain constant but the weak phases 
change sign, so
\begin{equation}
\overline{\cal A}=a_se^{i\theta_s}a_we^{-i\theta_w},~~
\overline{\cal B}=b_se^{i\delta_s}b_we^{-i\delta_w}~~.
\end{equation}
The rate difference is
\begin{eqnarray}
\Gamma-\overline{\Gamma}&=&|{\cal A}+{\cal B}|^2-
|\overline{\cal A}+\overline{{\cal B}}|^2 \\
&=&2a_sa_wb_sb_w\sin(\delta_s-\theta_s)\sin(\delta_w-\theta_w)~~.
\end{eqnarray}
A weak phase difference is guaranteed in the appropriate decay mode (different 
CKM 
phases), but the strong phase difference is not; it is very difficult to predict
the magnitude of  strong 
phase differences.

\begin{figure}[htbp]
\vspace{-1.1cm}
\centerline{\psfig{figure=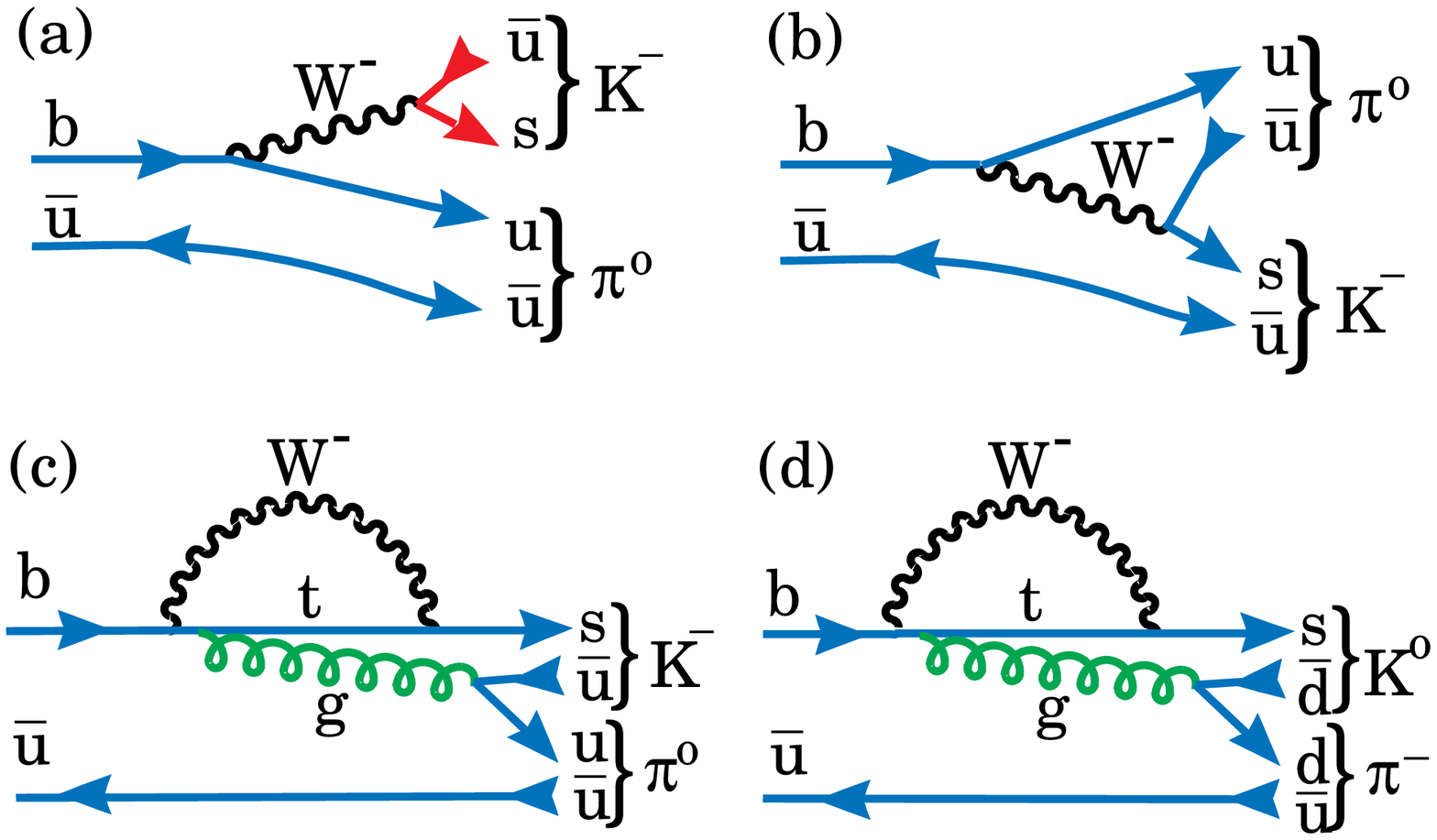,height=3.9in,bbllx=0bp,bblly=350bp,bburx=600bp,bbury=750bp,clip=}}
\vspace{-.7cm}
\fcaption{\label{kpi} Diagrams for $B^-\to K^-\pi^o$, (a) and (b) are tree level
 diagrams where (b) is color suppressed; (c) is a  penguin diagram. (d) shows
  $B^-\to K^o\pi^-$, which cannot be produced via a tree diagram.}
\end{figure}

 As an example consider the possibility of observing CP violation 
by measuring a rate difference between $B^-\to K^-\pi^o$ and $B^+\to K^+\pi^o$.
The $K^-\pi^o$ final state can be reached either by tree or penguin diagrams
as shown in Fig.~\ref{kpi}. 
The tree diagram has an imaginary part coming from the $V_{ub}$ coupling, while 
the penguin term does not, thus insuring a weak phase difference. This type of CP 
violation is called ``direct." Note also that the process  $B^-\to K^o\pi^-$
can only be produced by the penguin diagram in Fig.~\ref{kpi}(d). Therefore,
we do not expect a rate difference between  $B^-\to K^o\pi^-$ and
$B^+\to K^o\pi^+$. 

\subsubsection{Formalism in neutral $B$ decays}

Consider the operations of C and P:
\begin{eqnarray}
&C|B(\overrightarrow{p})\big>=|\overline{B}(\overrightarrow{p})\big>,~~~~~~~~~
&C|\overline{B}(\overrightarrow{p})\big>=|{B}(\overrightarrow{p})\big> \\
&P|B(\overrightarrow{p})\big>=-|{B}(-\overrightarrow{p})\big>,~~~~~
&P|\overline{B}(\overrightarrow{p})\big>=-|{B}(-\overrightarrow{p})\big> \\
&CP|B(\overrightarrow{p})\big>=-|\overline{B}(-\overrightarrow{p})\big>,~~~
&CP|\overline{B}(\overrightarrow{p})\big>=-|{B}(-\overrightarrow{p})\big> ~~.
\end{eqnarray}
For neutral mesons we can construct the CP eigenstates
\begin{eqnarray}
\big|B^o_1\big>&=&{1\over \sqrt{2}}\left(\big|B^o\big>-
\big|\overline{B}^o\big>\right)~~,\\
\big|B^o_2\big>&=&{1\over 
\sqrt{2}}\left(\big|B^o\big>+\big|\overline{B}^o\big>\right)~~,
\end{eqnarray}
where 
\begin{eqnarray}
CP\big|B^o_1\big>&=&\big|B^o_1\big>~~, \\
CP\big|B^o_2\big>&=&-\big|B^o_2\big>~~.
\end{eqnarray}
Since $B^o$ and $\overline{B}^o$ can mix, the mass eigenstates are a superposition of
$a\big|B^o\big> + b\big|\overline{B}^o\big>$ which obey the Schrodinger equation
\begin{equation}
i{d\over dt}\left(\begin{array}{c}a\\b\end{array}\right)=
H\left(\begin{array}{c}a\\b\end{array}\right)=
\left(M-{i\over 2}\Gamma\right)\left(\begin{array}{c}a\\b\end{array}\right).
\label{eq:schrod}
\end{equation}
If CP is not conserved then the eigenvectors, the mass eigenstates $\big|B_L\big>  $ 
and  $\big|B_H\big>$, are not the CP eigenstates but are 
\begin{equation}
\big|B_L\big> = p\big|B^o\big>+q\big|\overline{B}^o\big>,~~\big|B_H\big> = 
p\big|B^o\big>-q\big|\overline{B}^o\big>,
\end{equation}
where
\begin{equation}
p={1\over \sqrt{2}}{{1+\epsilon_B}\over {\sqrt{1+|\epsilon_B|^2}}},~~
q={1\over \sqrt{2}}{{1-\epsilon_B}\over {\sqrt{1+|\epsilon_B|^2}}}.
\end{equation}
CP is violated if $\epsilon_B\neq 0$, which occurs if $|q/p|\neq 1$.

The time dependence of the mass eigenstates is 
\begin{eqnarray}
\big|B_L(t)\big> &= &e^{-\Gamma_Lt/2}e^{im_Lt/2} \big|B_L(0)\big> \\
 \big|B_H(t)\big> &= &e^{-\Gamma_Ht/2}e^{im_Ht/2} \big|B_H(0)\big>,
\end{eqnarray}
leading to the time evolution of the flavor eigenstates as
\begin{eqnarray}
\big|B^o(t)\big>&=&e^{-\left(im+{\Gamma\over 2}\right)t}
\left(\cos{\Delta mt\over 2}\big|B^o(0)\big>+i{q\over p}\sin{\Delta mt\over 
2}\big|\overline{B}^o(0)\big>\right) \\
\big|\overline{B}^o(t)\big>&=&e^{-\left(im+{\Gamma\over 2}\right)t}
\left(i{p\over q}\sin{\Delta mt\over 2}\big|B^o(0)\big>+
\cos{\Delta mt\over 2}\big|\overline{B}^o(0)\big>\right),
\end{eqnarray}
where $m=(m_L+m_H)/2$, $\Delta m=m_H-m_L$ and 
$\Gamma=\Gamma_L\approx \Gamma_H.$ Note, that the probability of a 
$B^o$ decay as a function of $t$ is given by $\big<B^o(t)\big|B^o(t)\big>^*$, and is 
a pure exponential, $e^{-\Gamma t/2}$, in the absence of CP violation.

\subsubsection{Indirect CP violation in the neutral $B$ system}

As in the example described earlier for $K_L$ decay, we can look for the rate 
asymmetry
\begin{eqnarray}
a_{sl}&=&{{\Gamma\left(\overline{B}^o(t)\to X\ell^+{\nu}\right)-
\Gamma\left({B}^o(t)\to X\ell^-\overline{\nu}\right)}\over 
{\Gamma\left(\overline{B}^o(t)\to X\ell^+{\nu}\right)+
\Gamma\left({B}^o(t)\to X\ell^-\bar{\nu}\right)}}\\
&=&{{1-\left|{q\over p}\right|^4}\over {1+\left|{q\over p}\right|^4}}\approx 
O\left(10^{-2}\right).
\end{eqnarray}
These final states occur only through mixing as the direct decay occurs
only as  $B^o\to X\ell^+\nu$. To generate CP violation we need an interference
between two diagrams. In this case the two diagrams are the mixing diagram
with the $t$-quark and the mixing diagram with the $c$-quark quark. This is identical
to what happens in the $K^o_L$ case. This type of CP violation is called
``indirect." The small size of the expected asymmetry is caused by the off 
diagonal elements of the $\Gamma$ matrix in equation (\ref{eq:schrod}) being very small
compared to the off diagonal elements of the mass matrix, i.e. 
$\left|\Gamma_{12}/M_{12}\right|<<1$. This results from the nearly equal widths
of the $B_L^o$ and $B_H^o$.\cite{Khoze}

\subsubsection{CP violation for $B$ via interference of mixing and decays}

\begin{figure}[htb]
\vspace{-.2cm}
\centerline{\psfig{figure=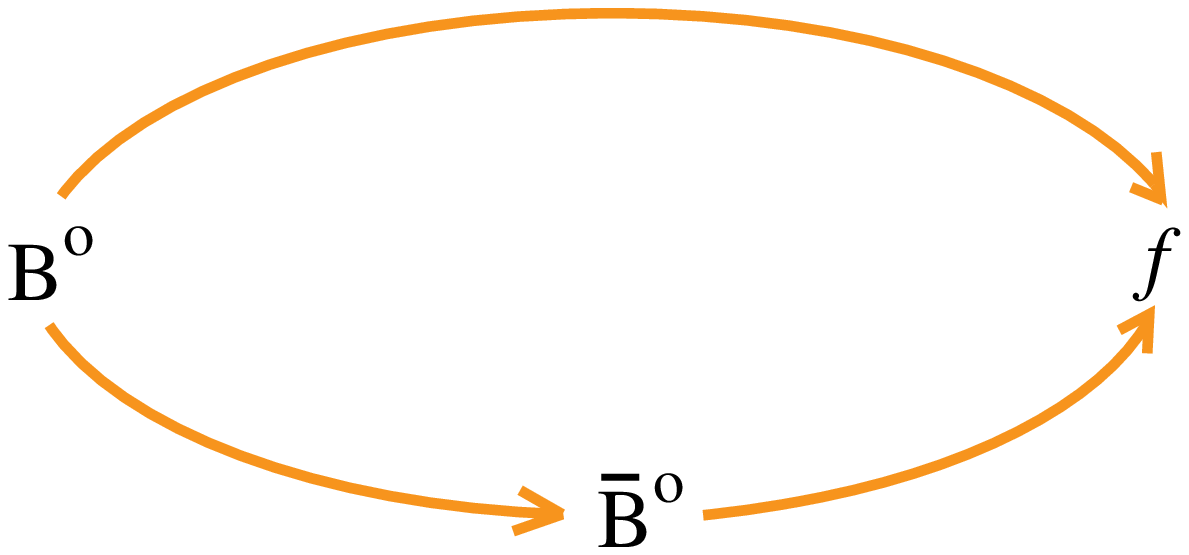,height=3.0in,bbllx=0bp,bblly=100bp,bburx=600bp,bbury=500bp,clip=}}
\vspace{-4.5cm}
\fcaption{\label{eigen_CP}Two interfering ways for a $B^o$ to decay into
a final state $f$.}
\end{figure}

Here we choose a final state $f$ which is accessible to both $B^o$ and $\overline{B}^o$ 
decays. The second amplitude necessary for interference is provided by mixing. 
Fig.~\ref{eigen_CP} shows the decay into $f$ either directly or indirectly via 
mixing.
It is 
necessary only that $f$ be accessible directly from either state, however if $f$ is a CP 
eigenstate the situation is far simpler. For  CP eigenstates
\begin{equation}
CP\big|f_{CP}\big>=\pm\big|f_{CP}\big>.
\end{equation}

It is useful to define the amplitudes
\begin{equation}
A=\big<f_{CP}\big|{\cal H}\big|B^o\big>,~~
\bar{A}=\big<f_{CP}\big|{\cal H}\big|\overline{B}^o\big>.
\end{equation}
If $\left|{\bar{A}\over A}\right|\neq 1$, then we have ``direct" CP violation in the 
decay 
amplitude, which we will discuss in detail later. Here CP can be violated by having
\begin{equation}
\lambda = {q\over p}\cdot {\bar{A}\over A}\neq 1,
\end{equation}
which requires only that $\lambda$  acquire a non-zero phase, i.e. $|\lambda|$ 
could be 
unity and CP violation can occur. 

A comment on neutral $B$ production at $e^+e^-$ colliders is in order. At the 
$\Upsilon 
(4S)$ resonance there is coherent production of $B^o\bar{B}^o$ pairs. This puts the 
$B$'s in a $C~=~-1$ state. In hadron colliders, or at $e^+e^-$ machines operating 
at the $Z^o$, the $B$'s are produced incoherently. For the rest of this article I will 
assume incoherent production except where explicitly noted.

The asymmetry, in this case, is defined as
\begin{equation}
a_{f_{CP}}={{\Gamma\left(B^o(t)\to f_{CP}\right)- 
\Gamma\left(\overline{B}^o(t)\to 
f_{CP}\right)}\over
{\Gamma\left(B^o(t)\to f_{CP}\right)+ \Gamma\left(\overline{B}^o(t)\to 
f_{CP}\right)}},
\end{equation}
which for $|q/p|=1$ gives
\begin{equation}
a_{f_{CP}}={{\left(1-|\lambda|^2\right)\cos\left(\Delta mt\right)-2{\rm Im}\lambda 
\sin(\Delta mt)}\over {1+|\lambda|^2}}.
\end{equation}
For the cases where there is only one decay amplitude $A$, $|\lambda |$ equals 1, 
and we have
\begin{equation}
a_{f_{CP}}=-{\rm Im}\lambda \sin(\Delta mt).
\end{equation}
Only the amplitude, ${\rm -Im}\lambda$ contains information about the level of CP 
violation, 
the sine term is determined only by $B_d$ mixing. In fact, the time integrated 
asymmetry is given by
\begin{equation}
a_{f_{CP}}=-{x \over {1+x^2}}{\rm Im}\lambda = -0.48 {\rm Im}\lambda ~~. \label{eq:aint}
\end{equation}
This is quite lucky as the maximum size of the coefficient is $-0.5$.

Let us now find out how ${\rm Im}\lambda$ relates to the CKM parameters. Recall 
$\lambda={q\over p}\cdot {\bar{A}\over A}$. The first term is the part that comes 
from mixing:
\begin{equation}
{q\over p}={{\left(V_{tb}^*V_{td}\right)^2}\over {\left|V_{tb}V_{td}\right|^2}}
={{\left(1-\rho-i\eta\right)^2}\over {\left(1-\rho+i\eta\right)\left(1-\rho-
i\eta\right)}}
=e^{-2i\beta}{\rm~~and}
\end{equation}
\begin{equation}
{\rm Im}{q\over p}= -{{2(1-\rho)\eta}\over {\left(1-\rho\right)^2+\eta^2}}=\sin(2\beta).
\end{equation}

\begin{figure}[htb]
\vspace{-.6cm}
\centerline{\psfig{figure=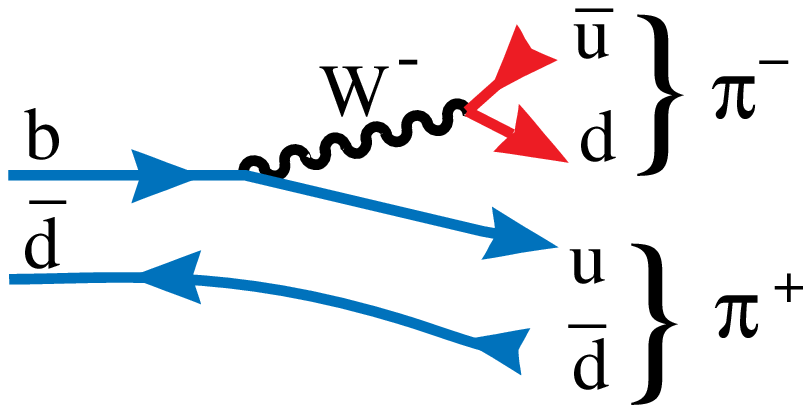,height=2.2in,bbllx=0bp,bblly=550bp,bburx=600bp,bbury=750bp,clip=}}
\vspace{-.5cm}
\fcaption{\label{pippim}Decay diagram at the tree level for $B^o\to\pi^+\pi^-$.}
\end{figure}

To evaluate the decay part we need to consider specific final states. For example, 
consider 
$f\equiv\pi^+\pi^-$. The simple spectator decay diagram is shown in 
Fig.~\ref{pippim}.
For the moment we will assume that this is the only diagram which contributes. 
Later I will 
show why this is not true. For this $b\to u\bar{u}d$ process we have
\begin{equation}
{\bar{A}\over A}={{\left(V_{ud}^*V_{ub}\right)^2}\over 
{\left|V_{ud}V_{ub}\right|^2}}={{(\rho-i\eta)^2}\over
 {(\rho-i\eta)(\rho+i\eta)}}=e^{-2i\gamma},
\end{equation}
and 
\begin{equation}
{\rm Im}(\lambda)={\rm Im}(e^{-2i\beta}e^{-2i\gamma})=
{\rm Im}(e^{2i\alpha})=\sin(2\alpha).
\end{equation}

\begin{figure}[htb]
\vspace{-1.2cm}
\centerline{\psfig{figure=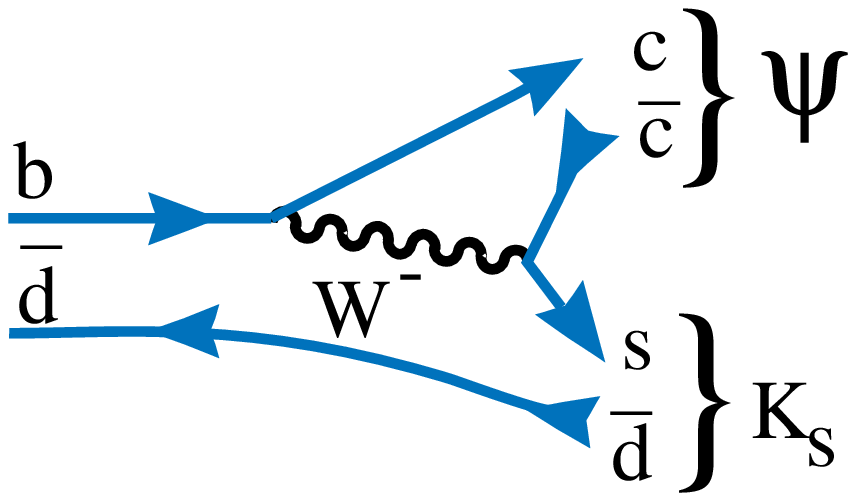,height=2.0in,bbllx=0bp,bblly=330bp,bburx=600bp,bbury=550bp,clip=}}
\vspace{-.05cm}
\fcaption{\label{psi_ks}Decay diagram at the tree level for $B^o\to\psi K_S$.}
\end{figure}

For our next example let's consider the final state $\psi K_S$. The decay diagram is 
shown in Fig.~\ref{psi_ks}. In this case we do not get a phase from the decay part because
\begin{equation}
{\bar{A}\over A} = {{\left(V_{cb}V_{cs}^*\right)^2}\over 
{\left|V_{cb}V_{cs}\right|^2}}
\end{equation}
is real.
In this case the final state is a state of negative $CP$, i.e. 
$CP\big|\psi K_S\big>=-\big|\psi K_S\big>$. This introduces an additional
minus sign in the result for ${\rm Im}\lambda$.
Before finishing discussion of this final state we need to consider in more detail 
the presence of the $K_S$ in the final state. Since neutral kaons can mix, we pick 
up another mixing phase (see Fig.~\ref{kmix}). This term creates a phase given by
\begin{equation}
\left({q\over p}\right)_K={{\left(V_{cd}^*V_{cs}\right)^2}\over 
{\left|V_{cd}V_{cs}\right|^2}},
\end{equation}
which is real. It necessary to include this term, however, since there are other 
formulations of the CKM matrix than Wolfenstein, which have the phase in a 
different location. It is important that the physics predictions not depend on the 
CKM convention.\footnote{Here we don't include CP violation in the neutral kaon 
since it is much smaller than what is expected in the $B$ decay.}

In summary, for the case of $f=\psi K_S$, ${\rm Im}\lambda=-\sin(2\beta)$.

\subsubsection{Comment on Penguin Amplitude}

\begin{figure}[htb]
\vspace{-3.2cm}
\centerline{\psfig{figure=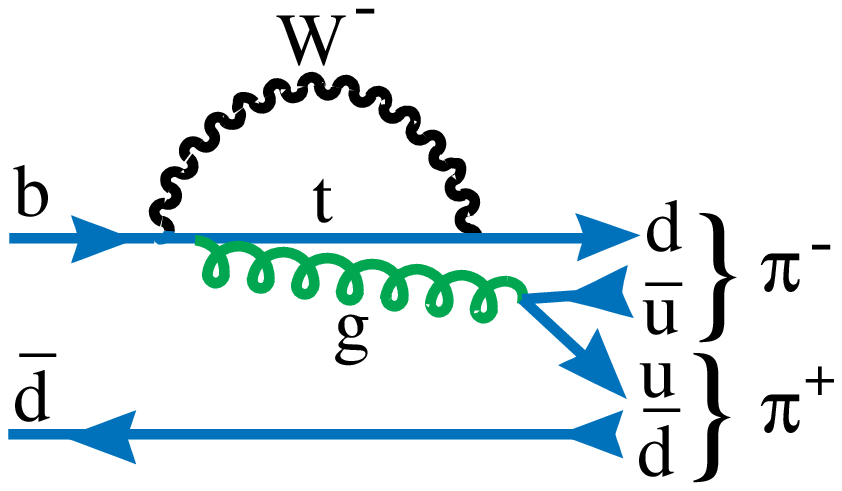,height=3.0in,bbllx=0bp,bblly=500bp,bburx=600bp,bbury=750bp,clip=}}
\vspace{-.05cm}
\fcaption{\label{pipi_penguin}Penguin diagram for $B^o\to\pi^+\pi^-$.}
\end{figure}

In principle all processes can have penguin components. One such diagram is
shown  in Fig.~\ref{pipi_penguin}. The $\pi^+\pi^-$ final state is expected to have a rather
large penguin  amplitude $\sim$10\% of the tree amplitude. Then $|\lambda |\neq
1$ and  $a_{\pi\pi}(t)$ develops a $\cos(\Delta mt)$ term. It turns out (see
Gronau\cite{GRL}), that $\sin(2\alpha)$ can be extracted using isospin
considerations and  measurements of the branching ratios for $B^+\to \pi^+\pi^o$
and $B^o\to  \pi^o\pi^o$.

In the $\psi K_S$ case, the penguin amplitude is expected to be small since a 
$c\bar{c}$ pair must be ``popped" from the vacuum. Even if the penguin decay 
amplitude were of significant size, the decay phase is the same as the tree level 
process, namely zero.

\subsubsection{What actually has to be measured?}

\begin{figure}[htb]
\vspace{-4.6cm}
\centerline{\psfig{figure=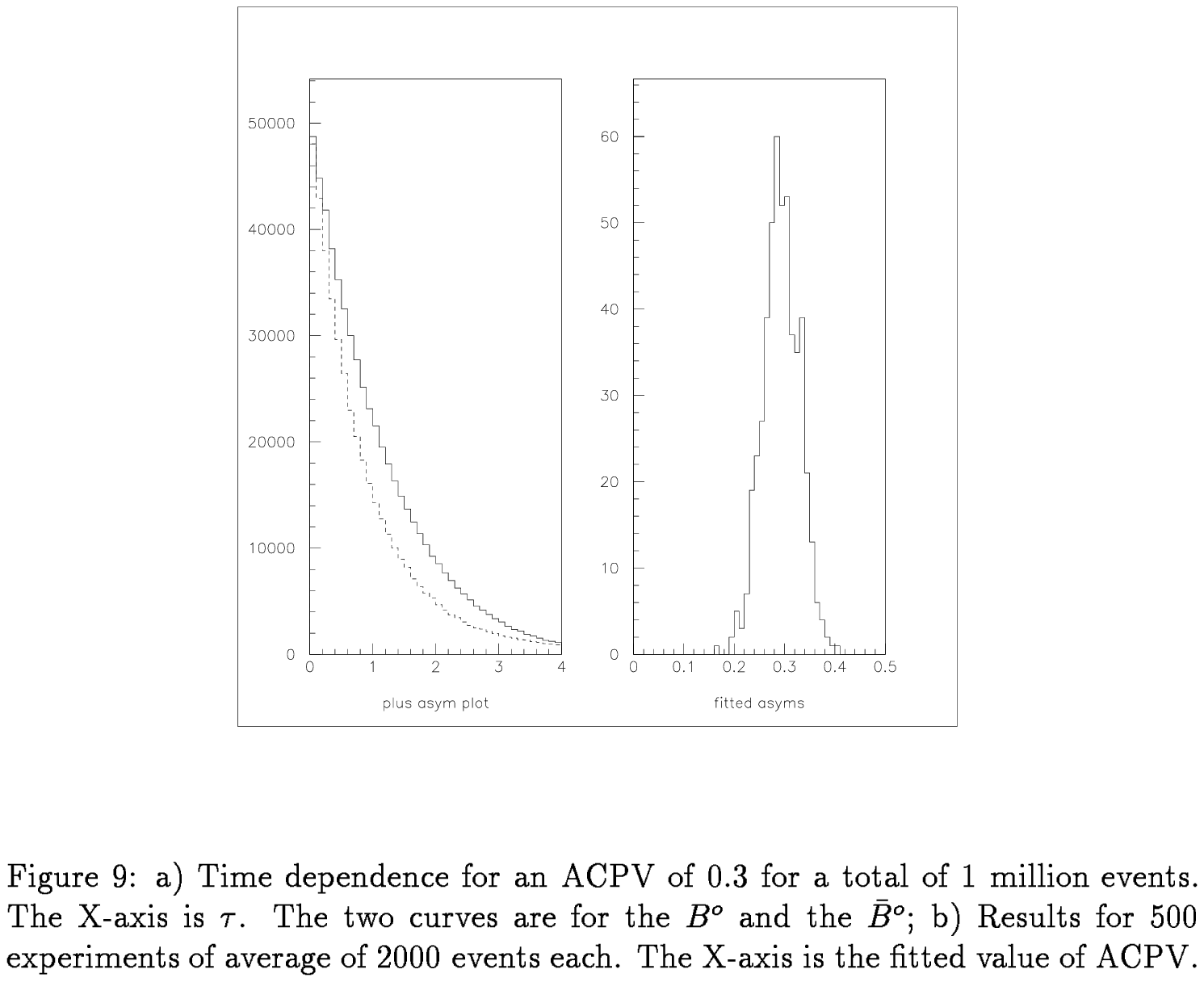,height=5.5in,bbllx=0bp,bblly=260bp,bburx=600bp,bbury=700bp,clip=}}
\vspace{-.15cm}
\fcaption{\label{fitted_asy}(a) Time dependence of $B^o$ and $\overline{B}^o$
decaying
into a CP eigenstate, for an asymmetry of 0.3 for a total
of 1 million events. The x-axis is proper time. In (b) the fitted asymmetry
results are shown for 500 ``experiments" of average of 2000 events each.}
\end{figure}

In charged $B$ decays we only have to measure  a branching ratio difference 
between $B^+$ and $B^-$ to see CP violation. For neutral $B$ decays  we must
find the flavor of the  other $b$-quark produced in the event (this is called
tagging), since we do not have any $B^o$ beams. We then measure a  rate
asymmetry
\begin{equation}
a_{asy}={{\#(f,\ell^+)-\#(f,\ell^-)}\over {\#(f,\ell^+)+\#(f,\ell^-)}},
\end{equation}
where $\ell^{\pm}$ indicates the charge of the lepton from the ``other" $b$
and  thus provides a flavor tag. In Fig.~\ref{fitted_asy}(a) the time dependence
for the $B^o$ and $\bar{B}^o$ are shown as a function of $t$ in the $B$ rest
frame for 500  experiments of an average of 2000 events each with an input
asymmetry of 0.3. In  Fig.~\ref{fitted_asy}(b) the fitted asymmetry is shown for
500 different ``experiments."

\subsection{Better Measurements of the sides of the CKM triangle}

One side of the triangle is determined by $|V_{ub}/V_{cb}|$. It appears that the best 
way to improve the values now is to measure the form-factors in the reactions 
$B\to\pi\ell\nu$ and $B\to\rho\ell\nu$. This will decrease the model dependence 
error, still the dominant errors, in the $V_{ub}$ determination. Lattice gauge 
model calculations are appearing and should be quite useful.

The other side of the triangle can determined by measuring $B_s$ mixing, using
the ratio
\begin{equation}
{x_s\over x_d}=\left({B_s\over B}\right)\left({f_{B_s}\over f_B}\right)^2
\left({\tau_{B_s}\over \tau_B}\right)\left|{V_{td}\over V_{ts}}\right|^2, \label{eq:bs}
\end{equation}
where
\begin{equation}
\left|{V_{td}\over V_{ts}}\right|^2=\lambda^2\left[(\rho-1)^2+\eta^2\right].
\end{equation}
The large uncertainty in using the $B_d$ mixing measurement to constrain 
$\rho$ and $\eta$ is largely removed as the ratio of the first three factors in 
equation (\ref{eq:bs}) is already known to 10\%.

As an alternative to measuring $x_s$, we can measure the ratio of the penguin 
decay rates 
\begin{equation}
{{\cal B}(B\to\rho\gamma)\over {\cal B}(B\to K^*\gamma)}=\xi\left|{V_{td}\over 
V_{ts}}\right|^2,
\end{equation}
where $\xi$ is a model dependent correction due to different form-factors.
Soni\cite{soni} has claimed that ``long distance" effects, basically other 
diagrams spoil this simple relationship.  This is unlikely for 
$\rho^o\gamma$ but possible for $\rho^+\gamma$.\footnote{One example is the 
$B^-$ decay which proceeds via $b\to u W^-$, where the $W^-\to \bar{u}d\to
\rho^-$ and the $u$ combines with the spectator $\bar{u}$ to form a photon.}
 If this occurs, however, then it 
is possible to find CP violation by looking for a difference in rate between 
$\rho^+\gamma$ and $\rho^-\gamma$.

The CLEO II data are already background limited. The limit quoted is\cite{ksgup}
\begin{equation}
{{\cal B}(B\to\rho\gamma)\over {\cal B}(B\to K^*\gamma)}< 0.19
\end{equation}
at 90\% confidence level.

\subsection{Rare decays as Probes beyond the Standard Model}

Rare decays have loops in the decay diagrams so they are sensitive to high mass 
gauge bosons and fermions. However, it must be kept in mind that any new effect 
must be consistent with already measured phenomena such as $B_d^o$ mixing and 
$b\to s\gamma$.

Let us now consider searches for other rare $b$ decay processes. The process $b\to 
s\ell^+\ell^-$ can result from the diagrams in Fig.~\ref{rarebdecay}(e or f). When 
searching for such decays, care must be taken to eliminate the mass region in the 
vicinity of the $\psi$ or $\psi '$ resonances, lest these more prolific processes, which 
are not rare decays, contaminate the sample. The result of searches are shown in 
Table~\ref{table:btoslplm}. 

\begin{table}[th]\centering\tcaption{Searches for $b\to s\ell^+\ell^-$ decays}
\label{table:btoslplm}
\vspace*{2mm}
\begin{tabular}{lclc}\hline\hline
$b$ decay mode & 90\% c.l. upper limit & Group &
 Ali et al. Prediction\cite{ali_rare}\\\hline
$s\mu^+\mu^- $& $50\times 10^{-6}$ & UA1\cite{ua1_rare} &\\
$K^{*o}\mu^+\mu^- $& $25\times 10^{-6}$ & CDF\cite{cdf_dilep} & $2.9\times 10^{-6}$\\
         & $23\times 10^{-6}$ & UA1\cite{ua1_rare} &\\
          & $31\times 10^{-6}$ & CLEO\cite{cleo_kll} &\\
$K^{*o}e^+e^- $& $16\times 10^{-6}$ & CLEO\cite{cleo_kll}& $5.6\times 10^{-6}$\\
 $K^{-}\mu^+\mu^- $& $9\times 10^{-6}$ & CLEO\cite{cleo_kll}& $0.6\times 10^{-6}$\\
          & $10\times 10^{-6}$ & CDF\cite{cdf_dilep}&\\
$K^{-}e^+e^- $& $12\times 10^{-6}$ & CLEO\cite{cleo_kll}& $0.6\times 10^{-6}$\\
\hline\hline
\end{tabular}\end{table}

\begin{figure}[htb]
\vspace{-.6cm}
\centerline{\psfig{figure=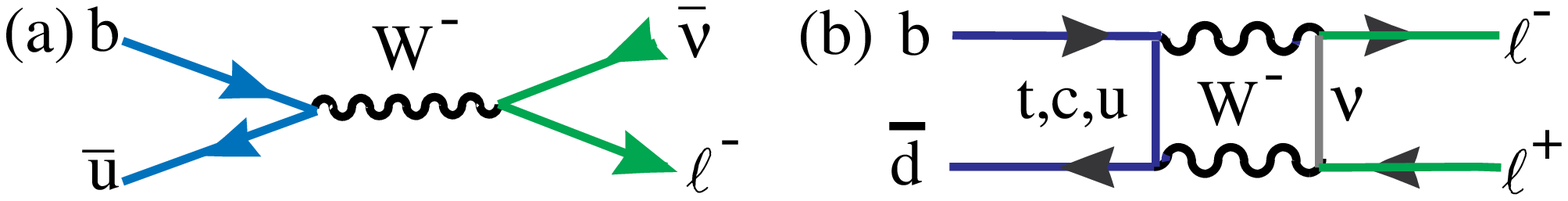,height=1.5in,bbllx=0bp,bblly=600bp,bburx=600bp,bbury=750bp,clip=}}
\vspace{-.45cm}
\fcaption{\label{b_dilep}Decay diagrams resulting in dilepton final states.
(a) is an annihilation diagram, and (b) is a box diagram.}
\end{figure}

$B$'s can also decay into dilepton final states. The Standard Model diagrams are 
shown in Fig.~\ref{b_dilep}. In (a) the decay rate is proportional to $|V_{ub}f_B|^2$. The 
diagram in (b) is much larger for $B_s$ than $B_d$, again the factor of 
$|V_{ts}/V_{td}|^2$. Results of searches are given in Table~\ref{table:dilep}.

\begin{table}[th]\centering\tcaption{Upper limits on
 $b\to $ dilepton decays (@90\% c.l.)}
\label{table:dilep}
\vspace*{2mm}
\footnotesize
\begin{tabular}{lcccccc}\hline\hline
&\multicolumn{2}{c}{${\cal B}(B^o\to\ell^+\ell^-)$}&
${\cal B}(B_s\to\ell^+\ell^-)$&\multicolumn{3}{c}
{${\cal B}(B^-\to\ell^-\bar{\nu})$}\\
&$e^+e^-$&$\mu^+\mu^-$&$\mu^+\mu^-$&$e^-\bar{\nu}$&
$\mu^-\bar{\nu}$&$\tau^-\bar{\nu}$\\\hline
SM$^{\dagger}$&$2\times 10^{-15}$&$8\times 10^{-11}$&$2\times 10^{-9}$&
$ 10^{-15}$&$ 10^{-8}$&$ 10^{-5}$\\
UA1\cite{ua1_rare}&&$8.3\times 10^{-6}$&&&&\\
CLEO\cite{cleo_dilep}&$5.9\times 10^{-6}$&$5.9\times 10^{-6}$&&
$1.5\times 10^{-5}$&$ 2.1\times 10^{-5}$&$ 2.2\times 10^{-3}$\\
CDF\cite{cdf_dilep}&&$1.6\times 10^{-6}$&$8.4\times 10^{-6}$&&&\\
ALEPH\cite{aleph_dilep}&&&&&&$1.8\times 10^{-3}$\\
\hline\hline
\multicolumn{7}{l}{$^{\dagger}$SM is the Standard Model
prediction.\cite{SM_pred}}\\
\end{tabular}\end{table}
\normalsize

\section{Future Experiments}

\subsection{$e^+e^-$ machines operating at the $\Upsilon(4S)$}

Recall that only $B$ meson pairs are produced at the
$\Upsilon(4S)$ as shown in Fig.~\ref{epemtoBB}.
Since each $B$ has about 30 MeV of kinetic energy, it moves on the average only 30 
$\mu$m before it decays. Another important consequence is that the decay 
products mix together and do not appear in distinct jets. To measure the important 
time difference required in CP violation experiments via mixing, it is necessary
to to give the $B$'s a Lorentz boost which can be accomplished by using
asymmetric beam energies.\cite{pierre}

Let me amplify on this last statement. The asymmetry I presented 
\begin{equation} 
a_{f_{CP}}=-{\rm Im}\lambda \sin(\Delta mt),   \label{eq:acp}
\end{equation}  
is calculated for incoherent production of the $B^o$ and
another $b$ quark ($t$ is  the time from production of the $B^o$ until it
decays). In $e^+e^-$ production  the $B$'s can be produced in a coherent state.
At the $\Upsilon(4S)$ $C=-1$, while  at higher energies, where $B^*\bar{B}$
($B^*\to B\gamma$) is produced, $C=+1$. For coherent production equation
(\ref{eq:acp}) gets modified to 
\begin{equation} 
a_{f_{CP~C=\pm }}=-{\rm Im}\lambda
\sin\left(\Delta m(t\pm t')\right),   \label{eq:acp2} 
\end{equation}  
where $t$
refers to the decay time of $f_{CP}$ and $t'$ the decay time of the  tagging
$B$. In principle, $a_{f_{CP}}$ can be measured by taking a time integral.  For
incoherent production this works fine (see equation (\ref{eq:aint})).  Here,
however, the  integral over the $C=-1$ case gives exactly zero, necessitating
the time dependent measurement. The integral over the $C=+1$ case, does not give zero,
but the measured  cross-section for $B^*\bar{B}$ is about 1/7 that of the
$\Upsilon (4S)$.\cite{fivesxc}$^,$\cite{fivesls}

The one serious disadvantage of the $\Upsilon(4S)$  machines is that the 
cross-section is only 1 nb, so at a peak luminosity of $3\times 10^{33}$, we expect
only 60 million $B^o$'s/year. For example, for a rare process with a branching
ratio of  $5\times 10^{-6}$ and a ``typical" efficiency of 20\%, we get only 60
events/year.

It is also important to note that there will not be much more $B$ physics from 
LEP. The data sample has been collected and there are no current plans to get 
another large sample of $Z^o$ decays to add to the brilliant $b$ physics already 
done.

The CESR machine will be upgraded to produce a luminosity in excess of $2\times
10^{33}$cm$^{-2}$s$^{-1}$, albeit with symmetric energy beams.  Both the KEK 
laboratory in Japan and SLAC in Stanford, Cal. will construct asymmetric energy 
machines with planned luminosities in excess of $3\times 
10^{33}$cm$^{-2}$s$^{-1}$. 

The advantages of such machines are that the $b$ cross-section is 1/4 of the total, 
and the relatively clean enviornment and low interaction rates allow for superb 
photon detection using CsI crystal calorimeters\cite{csi} and for planned particle 
identification systems which should provided excellent $\pi /K $ 
separation.\cite{rich}

\subsection{Hadron machines}

\begin{figure}[htb]
\vspace{-.2cm}
\centerline{\psfig{figure=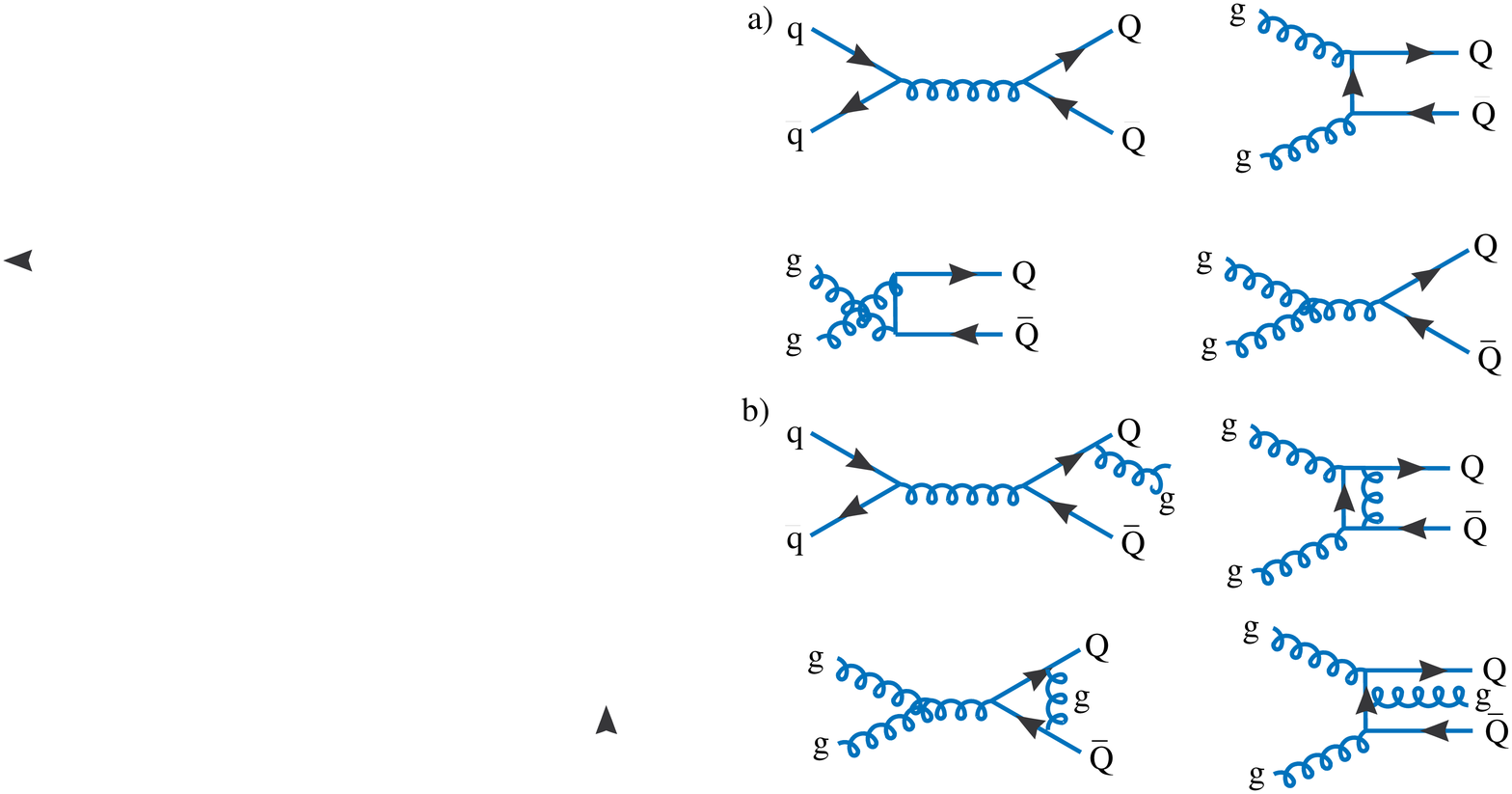,height=3.0in,bbllx=0bp,bblly=170bp,bburx=600bp,bbury=760bp,clip=}}
\vspace{-.05cm}
\fcaption{\label{hadron_b_prod} Feynman diagrams for heavy quark production
in hadronic collisions (a) of order $\alpha_s^2$, and (b) some diagrams
of order $\alpha_s^3$.}
\end{figure}

\begin{figure}[htb]
\vspace{-.8cm}
\centerline{\psfig{figure=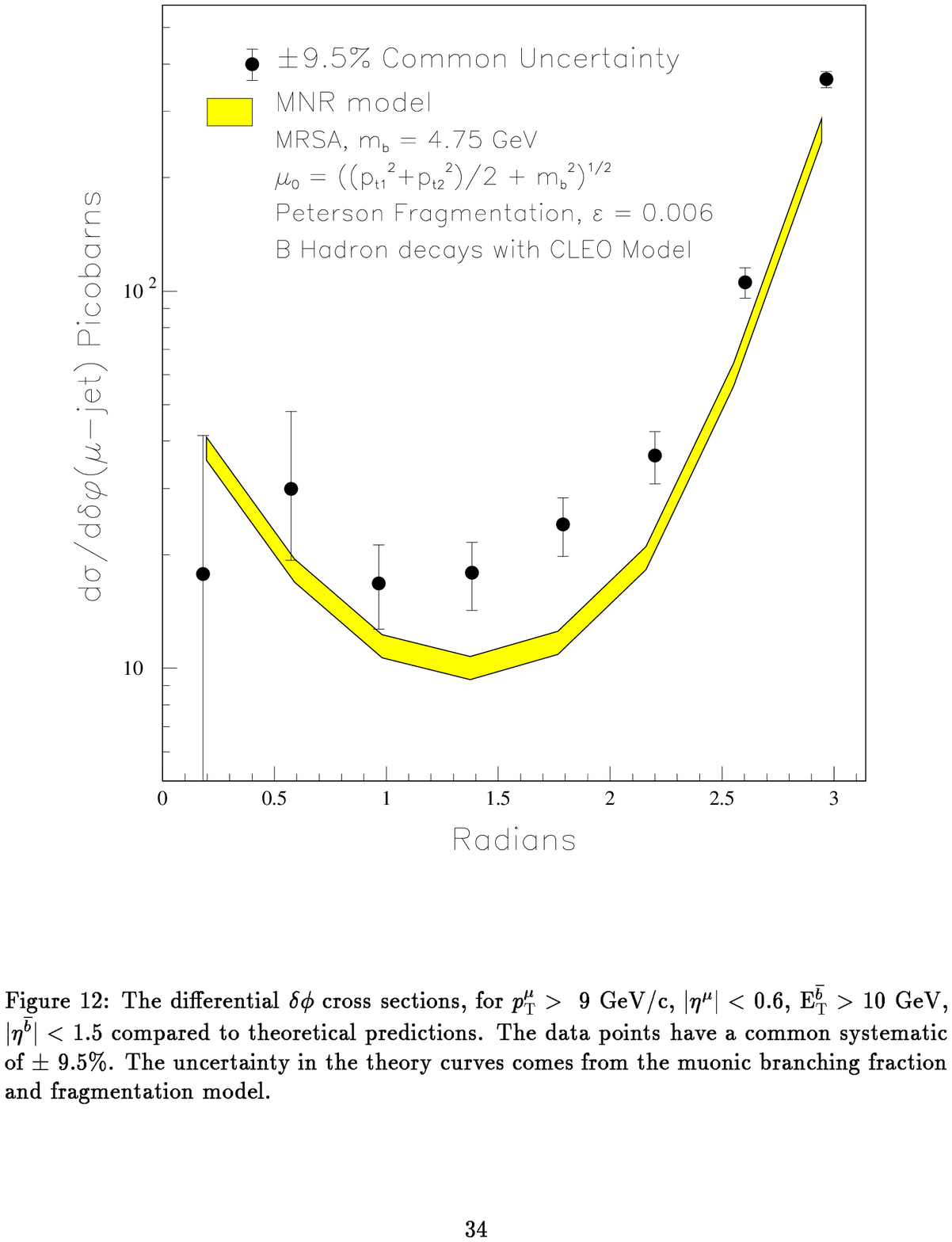,height=3.5in,bbllx=0bp,bblly=200bp,bburx=600bp,bbury=700bp,clip=}}
\vspace{-.3cm}
\fcaption{\label{cdf_dphi}The differential $\delta\phi$ cross-sections for
$p^{\mu}_T> 9 $ GeV/c, $\left|\eta^{\mu}\right|<$0.6, E$^{\bar{b}}_T>$10 GeV,
$\left|\eta^{\bar{b}}\right|<1.5$ compared with theoretical predictions. The
data points have a common systematic uncertainty of $\pm$9.5\%. The uncertainty
in the theory curve arises from the error on the muonic branching ratio and
the uncertainty in the fragmentation model.}
\end{figure}

Let us first discuss the characteristics of hadronic $b$ production.
Hadronic $b$ production mechanisms are shown in
Fig.~\ref{hadron_b_prod}.\cite{hbprod}
 The relative 
contribution of the terms proportional to $\alpha_s^2$ and those proportional
to $\alpha_s^3$ is not well known. This is an important issue since the correlations 
in rapidity, $\eta$ and in azimuthal angle between the $b$-quark and the 
$\bar{b}$-quark depends on the production mechanism. It is generally thought that 
$\left|\eta_b-\eta_{\bar{b}}\right|< 2$. In Fig.~\ref{cdf_dphi} I show the 
azimuthal opening angle distribution between a muon from a $b$ quark decay and the
$\bar{b}$ jet as measured by CDF\cite{cdf_prod} and 
compare with the MNR predictions.\cite{MNR} The model does a good job in
representing the shape which shows a strong back-to-back correlation. The
normalization is about a factor of two higher in the data than the theory,
which is generally
true of CDF $b$ cross-section measurements.\cite{cdf_bx} 
In hadron colliders all $B$ species are produced at the same time.

\begin{figure}[hbt]
\vspace{-.3cm}
\centerline{\psfig{figure=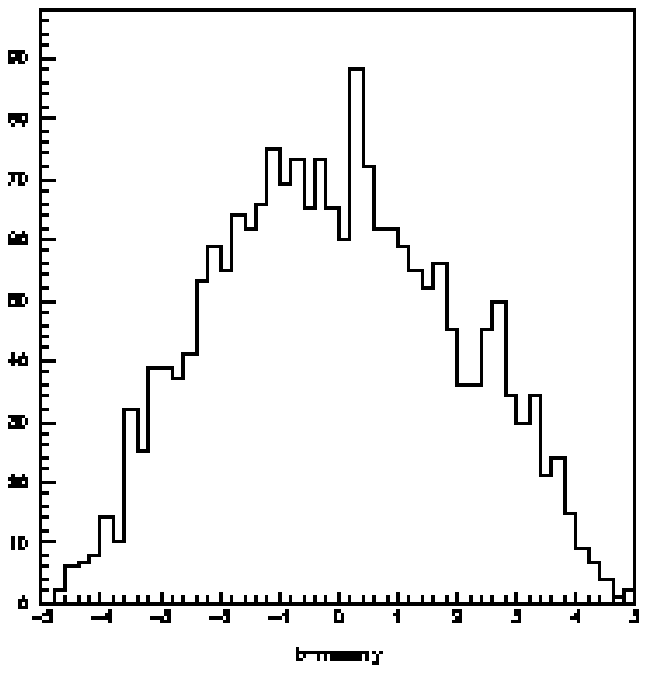,height=2.8in,bbllx=0bp,bblly=305bp,bburx=600bp,bbury=490bp,clip=}}
\fcaption{\label{byield_vs_eta}The predicted distribution of $B$'s versus $\eta$
for 1.8 TeV p$\bar{p}$ collisions.}
\end{figure}

\begin{figure}[hbt]
\vspace{.1cm}
\centerline{\psfig{figure=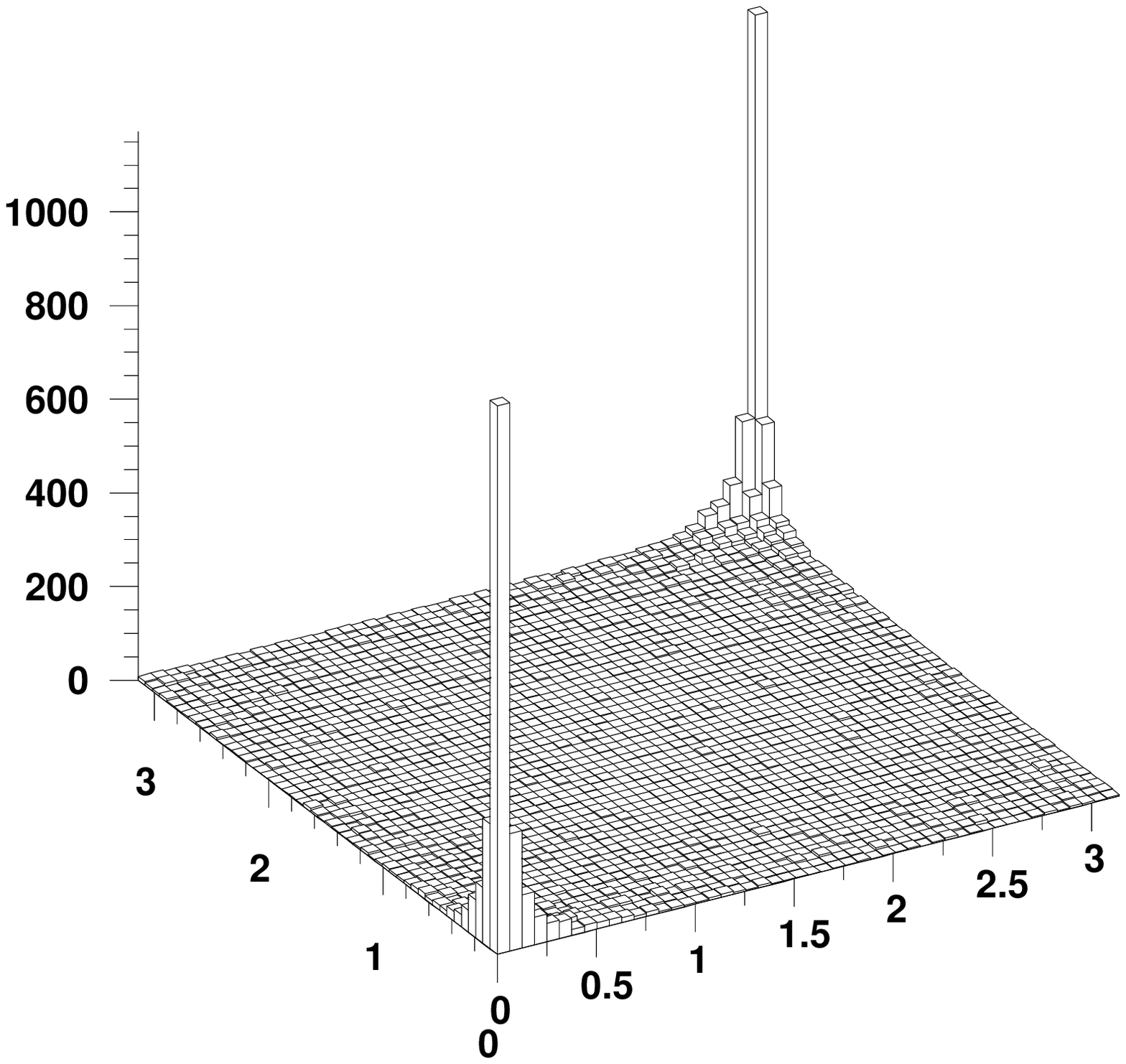,height=2.9in,bbllx=0bp,bblly=0bp,bburx=600bp,bbury=490bp,clip=}}
\vspace{-.5cm}
\fcaption{\label{bvb_ang}Production anglesof $B$ versus production angle
of the $\overline{B}$ in the laboratory (in radians) for the LHC collider
calculated using PYTHIA.} 
\end{figure}

The $B$ meson transverse momentum distribution is severely limited and peaks 
near the $B$ meson mass. The distribution in $\eta$, however is spread widely. In 
Fig.~\ref{byield_vs_eta} I show the predicted (Pythia) distribution at the
Tevatron collider. 
It should
 be realized that this 
distribution in $\eta$ reflects into a sharply peaked distribution in spatial angle
($\cos(\theta)$). The laboratory angular distributions of the $B$ and
$\overline{B}$ mesons expected at the LHC are shown in Fig.~\ref{bvb_ang}. Most
of the events are far forward with the $B$ and $\overline{B}$ being strongly
correlated.\cite{erhan}

Let us review some properties of current and proposed hadron $b$ collider 
experiments.\cite{beauty9} 
\begin{itemize}
\item The CDF and D0 detectors already exist at the Fermilab collider. 
The $b$ cross-section is $\sim$50 $\mu$b, with the ratio
 $\sigma(b)/\sigma(total)=10^{-3}$. The luminosity is now close to $10^{31}$ and 
will increase with the advent of the main injector to $10^{32}$. However, the 
restrictive trigger limits the $b$ sample.
\item The HERA-$b$ experiment at DESY collides the HERA proton 
beam with fixed wire targets. The $b$ cross-section is only $\sim$6 nb with 
 $\sigma(b)/\sigma(total)=10^{-6}$. In order to produce enough $b$'s they plan on 
four interactions per crossing. The goal is to measure CP violation in the $\psi 
K_S$ decay mode and possibly investigate other modes that are accessible by 
triggering on dileptons. The experiment is now under construction.
\item The LHC-B experiment is being planned. At the LHC the $b$ cross-section 
is $\sim$300 $\mu$b, with the ratio
 $\sigma(b)/\sigma(total)=3\times 10^{-3}$. The experiment can run at a luminosity 
of $10^{32}$, $\approx$240 Billion $B^o$/year are produced.
\item Also at the LHC, the Atlas and CMS experiments will have some 
$B$ capabilities.
\item There is now a proposal for a dedicated $B$ collider experiment at 
Fermilab called BTEV. Here $\approx$60 Billion $B^o$/year are produced.
\end{itemize}

\subsection{Detector Considerations}

\begin{figure}[htb]
\vspace{.4cm}
\centerline{\psfig{figure=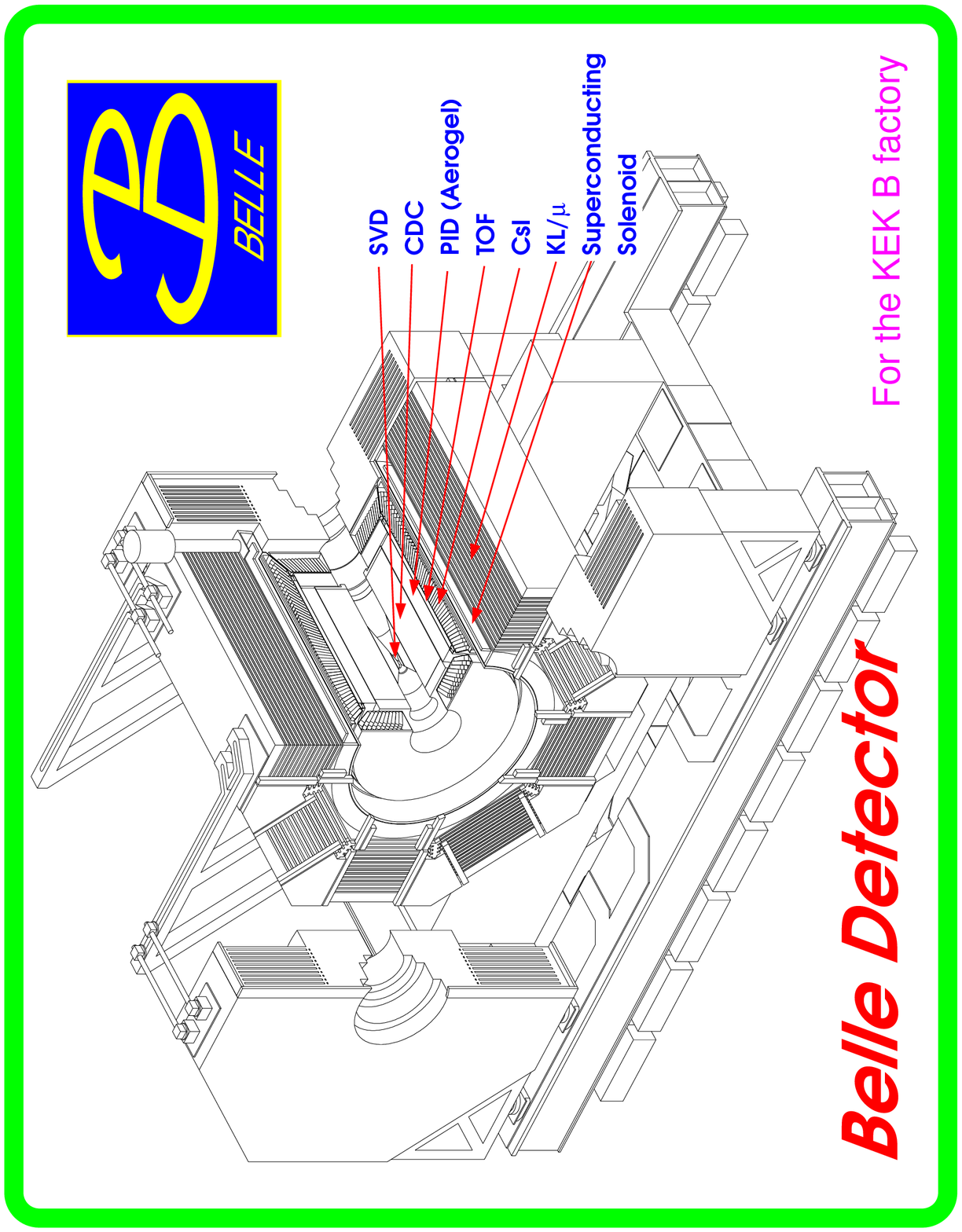,height=6.0in,bbllx=0bp,bblly=80bp,bburx=600bp,bbury=760bp,clip=}}
\fcaption{\label{belle_detector}Diagram of the Belle detector.}
\end{figure}

For an experiment to do frontier $B$ physics the following components appear to 
be necessary:
\begin{itemize}
\item Silicon vertex detector
\item Charged particle tracking with magnetic analysis
\item Cherenkov identification of charged hadrons
\item Electromagnetic shower detection
\item Muon detection with iron
\end{itemize}
A precision vertex detector is necessary to use the long $B$ lifetime to
reject background. Silicon is the current technology of choice; it can be
realized as strips or as pixels. Charged particle tracking with magnetic
analysis is important for momentum measurement as it is in most experiments.
In order to pick out specific $B$ decay modes, such as $K^+\pi^-$ from 
$\pi^+\pi^-$ or $\rho\gamma$ from $K^*\gamma$, it is crucial to have 
kaon and pion identification. Currently this is best provided using Cherenkov
radiation.\cite{rich} Electromagnetic shower detection and muon identification
are required to study semileptonic decays and provide flavor tags.
The BELLE experiment, shown in Fig.~\ref{belle_detector} is an example of a 
detector that has all of these elements.

There are important constraints on all of these detection elements. Radiation 
damage implies various limits and certain technologies. The number of 
interactions per second implies a rate limit on detector elements. It appears that the 
maximum rate on any detector element is about $10^7$/sec. The total detector 
readout rate is limited to about 10-100 MB/sec. (The smaller number is given by 
current technology and the larger number is based on expected improvement.) For 
an event size of 100 KB, this gives a maximum readout rate of 1000 events/sec.

Next, I will discuss the trigger. $e^+e^-$ experiments have a distinct
advantage  here, since they merely trigger on everything. Experiments at hadron
collider must  trigger very selectively, or the data transmission rate will be swamped by
background.  There are several trigger strategies which have been developed.
The one with the  highest background rejection is $B\to \psi X$,
$\psi\to\ell^+\ell^-$. Unfortunately the branching ratio  for the former is
only 1.1\% and the latter 12\%, giving a maximum triggerable $B$ event rate of
only $2.6\times 10^{-3}$. This must be reduced by efficiency of  the apparatus
and kinematic cuts.

Another strategy is to trigger on semileptonic decays, where the 10\% branching 
ratio to both muons or electrons is  attractive. Furthermore, for CP violation 
measurements through mixing, this trigger also provides a tag.  It has been 
traditionally easier to trigger on muons because electrons can easily be faked by 
photon conversions near the vertex or Dalitz decays of the $\pi^o$. 

\begin{figure}[htb]
\vspace{-1.1cm}
\centerline{\psfig{figure=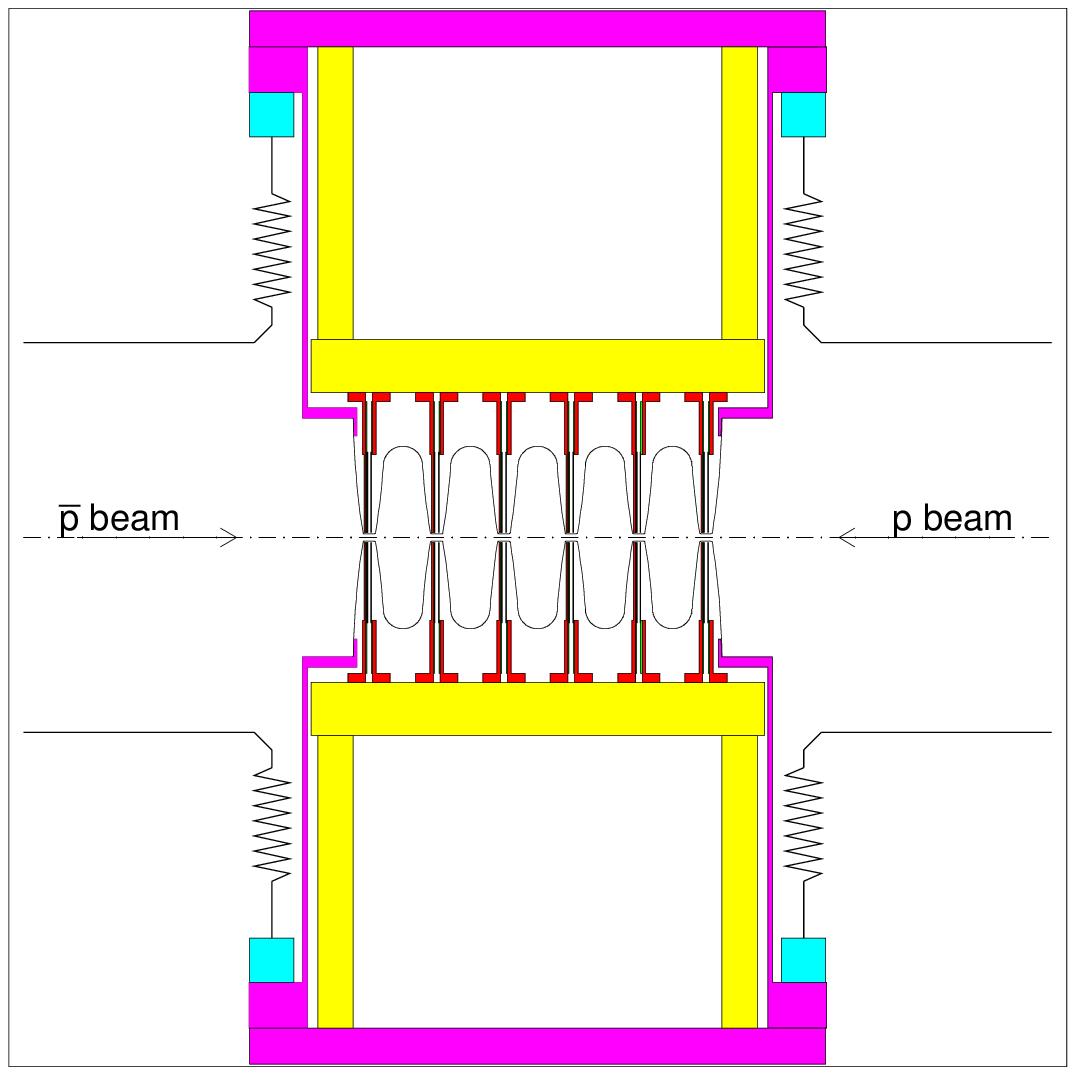,height=3.5in,bbllx=0bp,bblly=0bp,bburx=600bp,bbury=550bp,clip=}}
\vspace{-.05cm}
\fcaption{\label{p238}Side view of the P238 silicon detector
assembly and Roman pots. The 6 silicon planes are the vertical lines just above
and below the beam line. The bellows (zig-zag lines) allow movements in the
vertical direction of the pots, which are the thin vertical lines close to the
bellows (they have 2 mm wall thickness). The edges of the 200 $\mu$m-thick
aluminum RF shields closest to the beam (shown as the thin curved lines near
the silicon detectors) normally ran at a distance of 1.5 mm from the
circulating beams. The black horizontal pieces at top and bottom are the vacuum
bulkheads bolted to the Roman pots. }
\end{figure}

The most progressive strategy is to trigger on detached vertices. Recent
simulations  for BTEV have shown that it is possible to achieve a good
efficiency $>70$\% on $B$ decay events with a rejection on light quark 
background in excess of 100:1. To achieve this it is necessary to use a forward
geometry with the silicon vertex detector inside the beam pipe.\cite{butler} A
test of this concept was done at CERN by experiment P238.\cite{p238p} A sketch
of the silicon detector arrangement is shown in
Fig.~\ref{p238}.

It is also possible to consider triggering on specific low multiplicity final states such 
as $B^o\to \pi^+\pi^-$ by using hadrons with $p_t > $1 GeV/c. 

The crucial issue in all of the trigger strategies is what the background rates
are for  a high signal efficiency. Does this give enough signal events with
simultaneously  rejecting background at the 100:1 level?

\subsection{Hadron Geometries}

\begin{figure}[htb]
\vspace{-.2cm}
\centerline{\psfig{figure=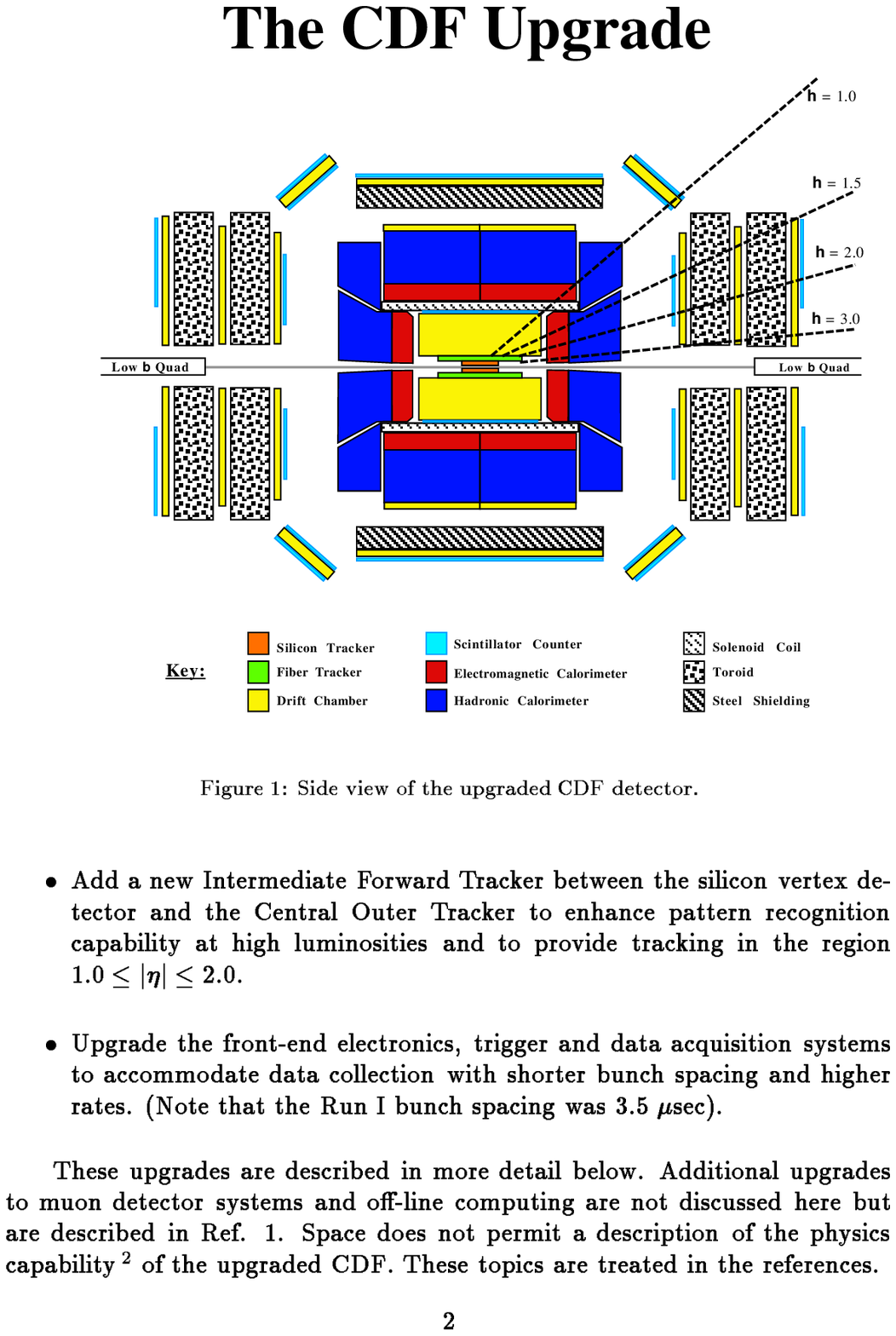,height=4.4in,bbllx=-0bp,bblly=340bp,bburx=600bp,bbury=603bp,clip=}}
\vspace{-.05cm}
\fcaption{\label{cdf}A schematic diagram of the CDF upgrade. The symbol
`h' refers to rapidity. Note that
the fiber tracker may change to a different technology.}
\end{figure}

\begin{figure}[htb]
\vspace{-1.1cm}
\centerline{\psfig{figure=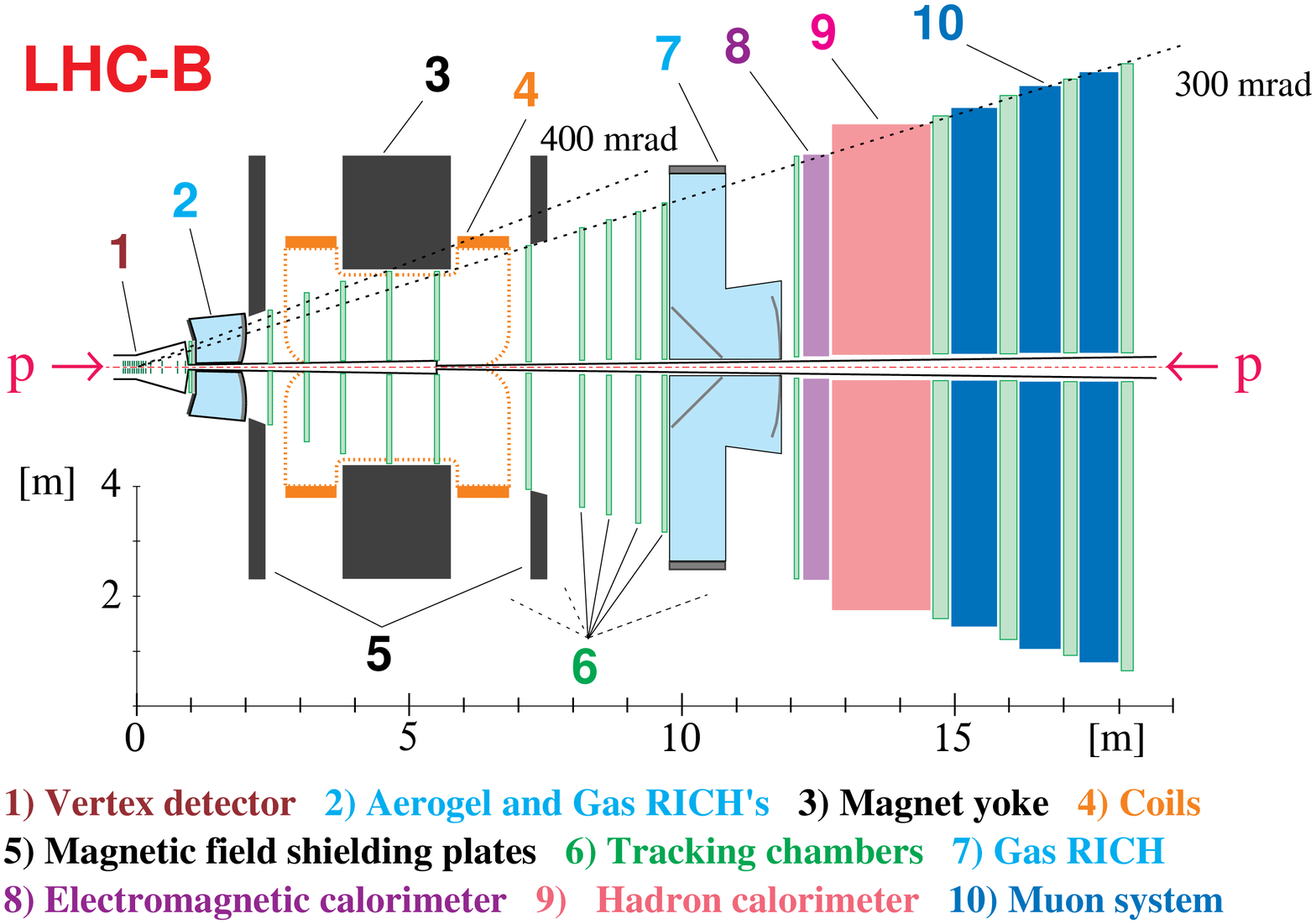,height=4.4in,bbllx=-80bp,bblly=100bp,bburx=700bp,bbury=700bp,clip=}}
\vspace{-.05cm}
\fcaption{\label{LHC_B}A schematic diagram of the proposed LHC-B detector.}
\end{figure}

There is a choice between two basic geometrical configurations that can be used
for  collider hadron $B$ experiments. One is a central detector. An example is
given by  the planned upgraded CDF detector, shown in Fig.~\ref{cdf}.
Here the detector  elements are arranged in an almost cylindrical manner about
the beam pipe, so that  the detector is very good near $\eta$ equals zero.
Notice that there are no detector  elements for particle identification, though
some information may be available from dE/dx measurements in the tracking
chamber.  An example of a forward detector is the proposed LHC-B experiment
shown in  Fig.~\ref{LHC_B}. Here the vertex detector is inside a flared beam
pipe. There are three  different radiators for the RICH detectors.

\begin{figure}[htb]
\vspace{-1.8cm}
\centerline{\psfig{figure=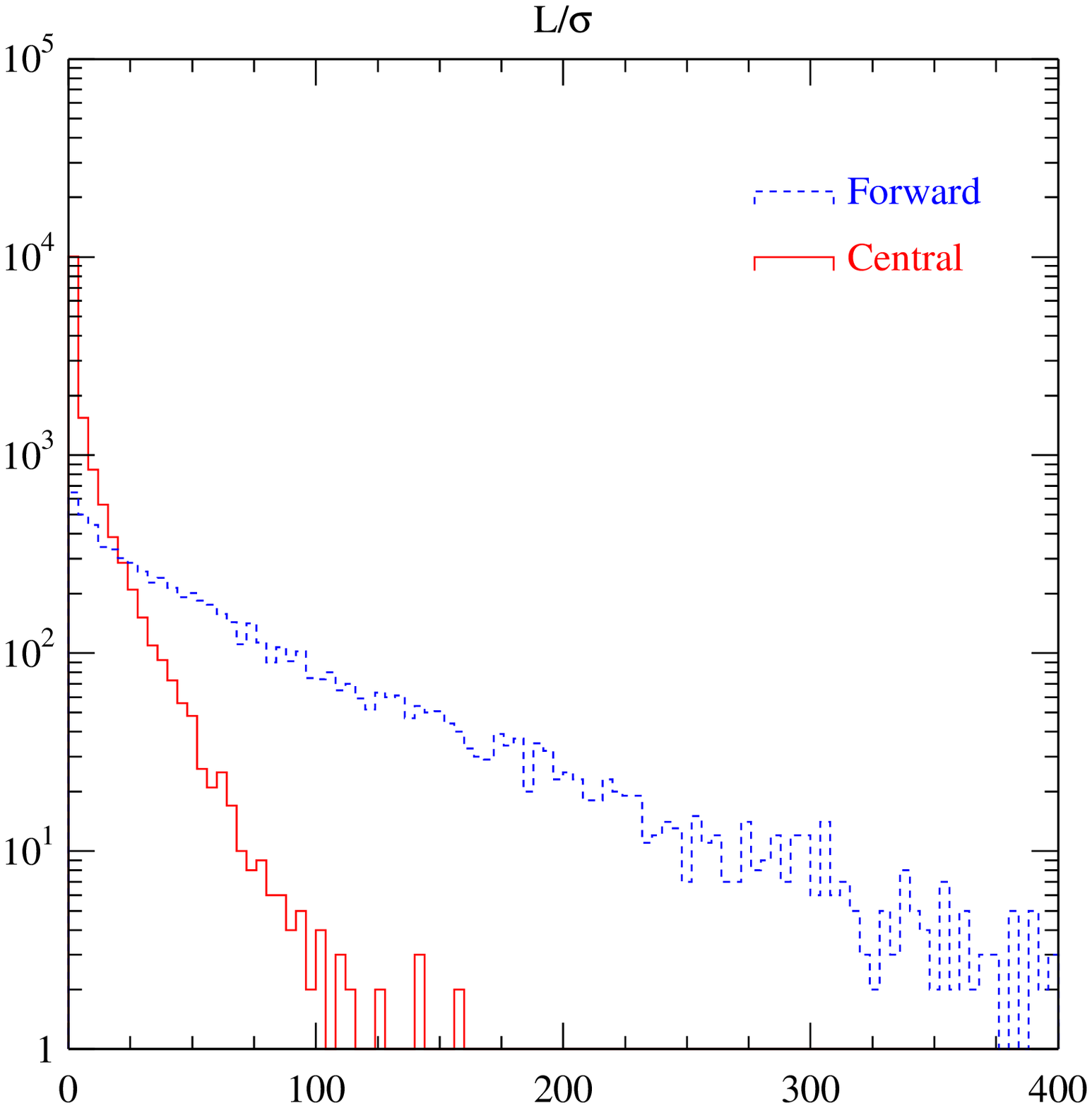,height=4.4in,bbllx=0bp,bblly=0bp,bburx=600bp,bbury=630bp,clip=}}
\vspace{-0.1cm}
\fcaption{\label{l_over_sig} Comparison of the $L/\sigma$ distributions
for the decay $B^o\to\pi^+\pi^-$ in central and forward detectors
produced at a hadron collider with a center of mass energy of 1.8~TeV.}
\vspace{1 cm}
\end{figure}

\begin{figure}[htb]
\vspace{-2.1cm}
\psfig{bbllx=0pt,bblly=70pt,bburx=600pt,bbury=600pt,file=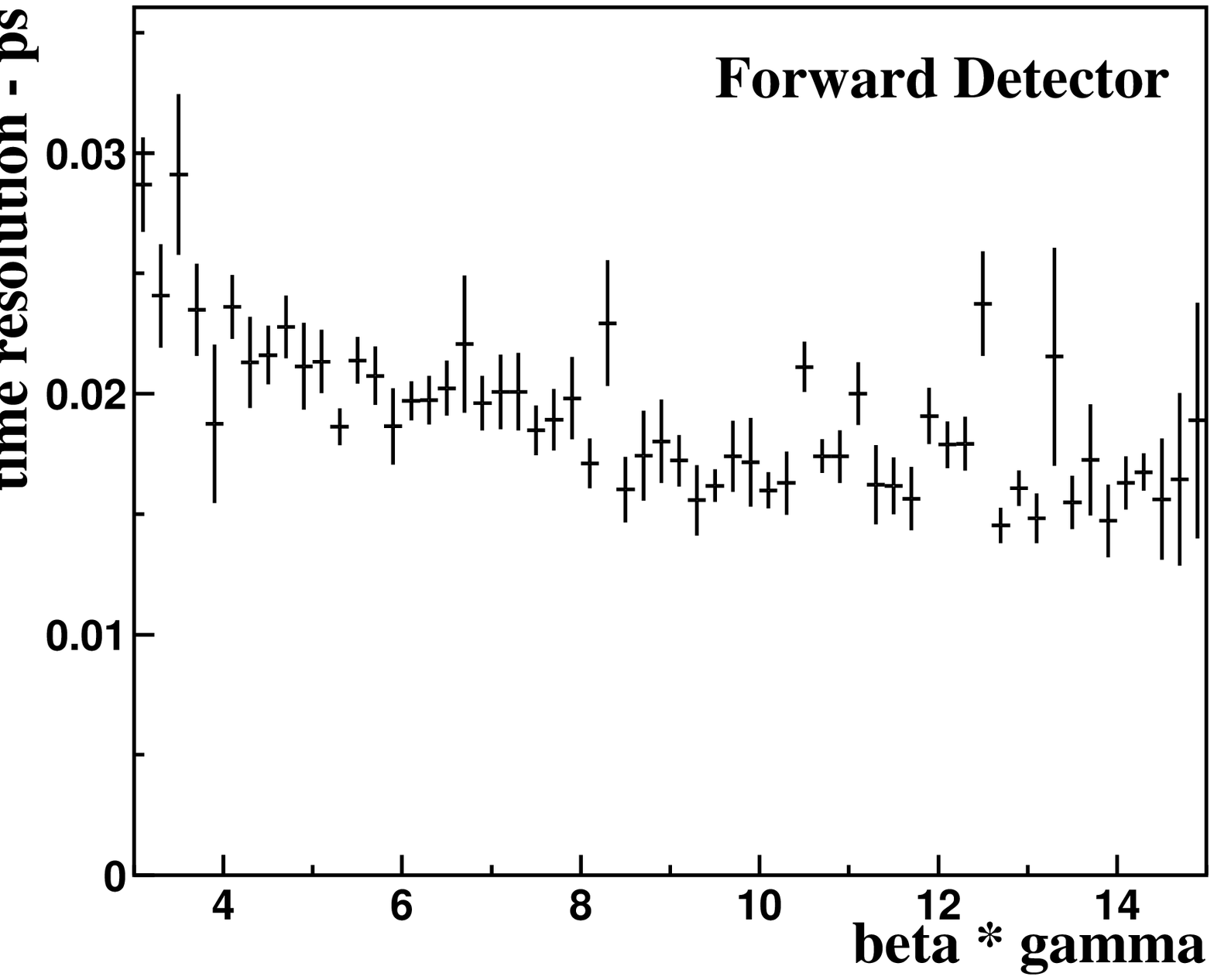,width=2.95in}
\vspace{-6.6cm}\hspace{3.in}
\psfig{bbllx=0pt,bblly=70pt,bburx=600pt,bbury=600pt,file=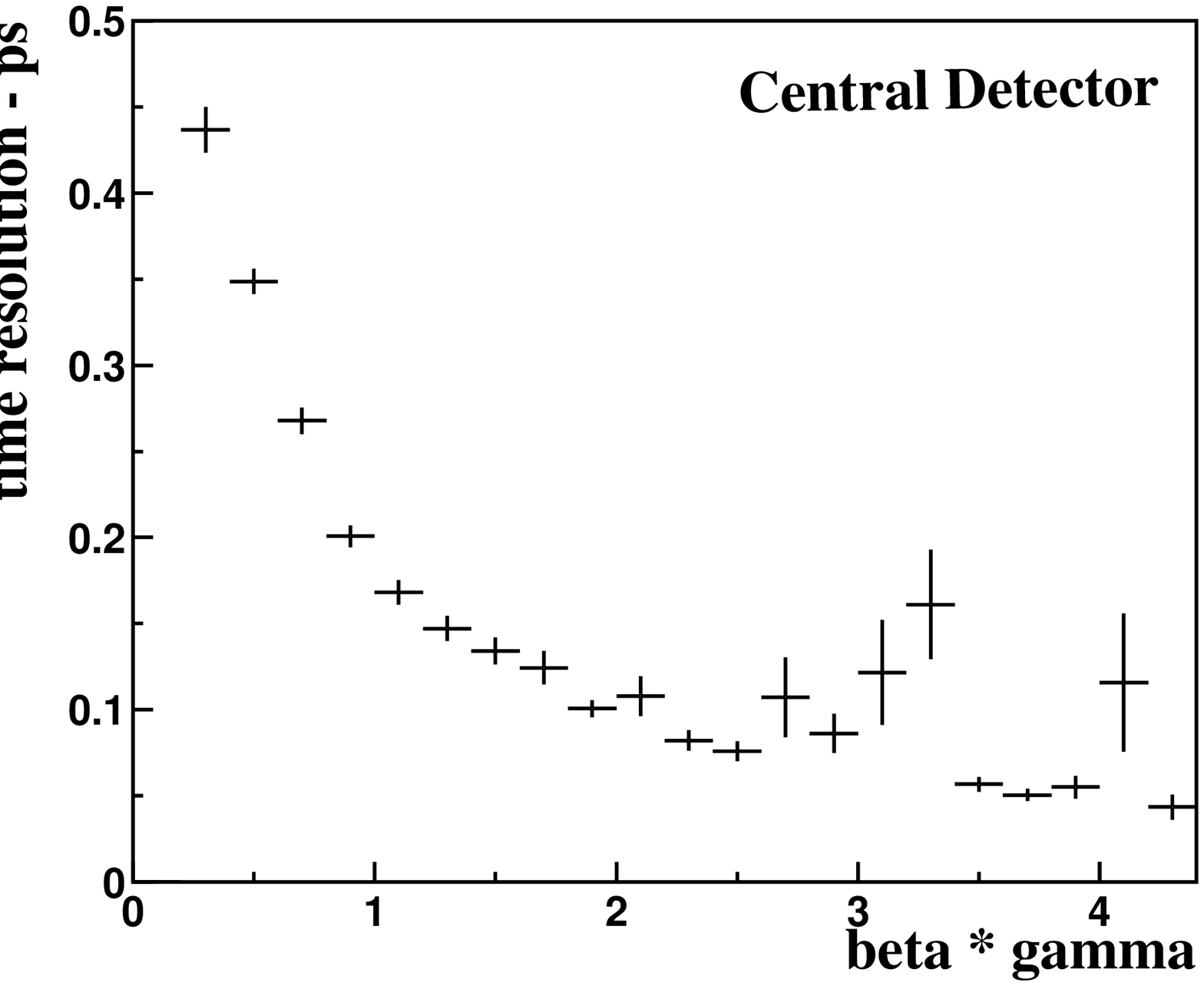,width=2.95in} 
\vspace{.8cm}
\fcaption{\label{cdf_dt_bg}The time resolution plotted as a function of
$\beta \gamma$ for a forward detector $( 2.0 < \eta < 4.5)$ and a
central detector $(|\eta | < 1.5)$ for the decay $B_s\to\psi \overline{K}^*$
produced at a hadron collider with a center of mass energy of 1.8~TeV.}
\end{figure}

In hadron colliders the most important rejection of non-$B$ background is 
accomplished by seeing a detached decay vertex. In Fig.~\ref{l_over_sig} I show the normalized 
decay length expressed in terms of $L/\sigma$ where $L$ is the decay length and 
$\sigma$ is the error on $L$ for the $B^o\to \pi^+\pi^-$ decay.\cite{procario}
 This study was done for the Fermilab Tevatron. The forward 
detector clearly has a much more favorable $L/\sigma$ distribution. In
Fig.~\ref{cdf_dt_bg} we 
show the time resolution in picoseconds for the forward and central detectors for 
the reaction $B_s\to\psi Ks$, which has been suggested as a possible way to
measure $B_s$ mixing.\cite{bstopsiks} Remarkably the time resolution is a factor of 10 smaller 
for the forward detector.

A comparison of  different $B$ experiments is shown in Table~\ref{table:det}.

\begin{table}[th]\centering\tcaption{Comparison of $B$ decay detectors}
\label{table:det}
\vspace*{2mm}
\begin{tabular}{cccccc}\hline\hline
Experiment & Particle & Vertex  & Photon  & $\sigma(b)$ & $\sigma(b)$ \\
                        & I. D.       & detection& detection&              & 
                       $\overline{\sigma(T)}$\\
\hline
Babar    & Excellent & Good &Excellent & 1 nb & 0.25\\
Belle    & Good & Good &Excellent & 1 nb & 0.25\\
CLEO    & Excellent & Mediocre$^{\dagger}$ &Excellent & 1 nb & 0.25\\
CDF    & Poor & Good &Poor & 50 $\mu$b & 10$^{-3}$\\
D0    & Poor & Good &Poor & 50 $\mu$b & 10$^{-3}$\\
HERA-B    & Excellent& Excellent &Poor & 6 nb  & 10$^{-6}$\\
LHC-B    & Excellent& Excellent &Poor & 300 $\mu$b  & $3\times 10^{-3}$\\\hline\hline
\multicolumn{6}{l}{$^{\dagger}$ detector is excellent but low $B$ velocity compromises
vertex detection}
\end{tabular}\end{table}

\section{Conclusions}

$B$ decay physics started in the 1980's and the first generation of experiments at 
CESR, DORIS, PEP, PETRA, LEP and CDF have made great contributions 
including the first fully reconstructed $B$'s and precise measurement of the $B$ 
meson masses,  measurement of the $B$ lifetimes, discovery of  $B^o -\bar{B^o}$ 
mixing, the measurement of the CKM
parameters  $V_{cb}$ and $V_{ub}$ and the sighting of the first rare decays.

Many mysteries, however, remain to be untangled. Measuring independently all
sides and angles of the CKM triangle may point us beyond the Standard Model
if the data are inconsistent. This will require measuring all three CP
violating angles, measuring $B_s$ mixing and precisely determining
$V_{ub}/V_{cb}$. Furthermore, observation of rare $B$ decays may also
point us beyond the Standard Model.\cite{beyond}

$e^+e^-$ threshold machines are great for future $B$ physics. They will surely 
produce precision measurements of $V_{ub}$ and $V_{cb}$ and the important 
measurement of $\sin(2\beta)$ using the $\psi K_S$ decay mode. Posssibly 
$\sin(\gamma)$ can be measured using charged $B$ decays and there are some who 
think these machines can measure $\sin(2\alpha)$, but I find that unlikely.
However, these experiments are limited by the total number of $B$ mesons. Even 
if these machines reach luminosities of $10^{34}$cm$^{-2}$s$^{-1}$, there are not 
enough $B$'s to probe most rare phenomena. The prospects for $B_s$ mixing,
$\Lambda_b$ and $B_c$ studies are dim.

There is a fantastic potential for studying CP violation phenomena and rare $B$ 
studies in hadronic machines but it's not easy. Let us consider the calculation of the 
error on an asymmetry measurement:
\begin{equation}
\sigma (a_{asy})={1 \over D\sqrt{N_{eff}\cdot \epsilon \cdot B}},
\end{equation}
where
\begin{equation}
 N_{eff}= N{Signal \over {Signal + Background}}, 
\end{equation}
$B$ is the branching ratio of the final state of interest, $\epsilon$ is the overall 
efficiency including the tagging efficiency. $D$ is the dilution factor caused by 
anything which causes a wrong-sign tag to be found, such as away side mixing, 
lepton misidentification etc.. A sample calculation is shown in
Table~\ref{table:acp_calc}.

\begin{table}[th]\centering
\tcaption{Sensitivity Calculation for Observing a CP asymmetry in $\psi K_S$}
\label{table:acp_calc}
\vspace*{2mm}
\begin{tabular}{lc}\hline\hline
CM energy & 2 TeV\\
Cross-section & 50 $\mu$b\\
Luminosity & $10^{32}$cm$^{-2}$s$^{-1}$\\
$N_{B^o}$/`Snowmass' year & $3.75\times 10^{10}$\\
${\cal B}(B^o\to\psi K_S)$ & $5.5\times 10^{-4}$\\
${\cal B}(B^o\to\psi(\mu^+\mu^-) K_S(\pi^+\pi^-))$ & $2.2\times 10^{-5}$\\
$N(B^o\to\mu^+\mu^-\pi^+\pi^-)$/year & $8.2\times 10^5$\\
Semi-leptonic decay of away side tag & 0.10\\
Tagged $N(B^o\to\mu^+\mu^-\pi^+\pi^-)$/year &  $8.2\times 10^4$\\
Triggering efficiency & 0.8\\
Reconstruction efficiency of $\mu\mu\pi\pi$ & 0.25\\
Reconstruction efficiency  $\mu$ tag& 0.25\\
Vertex finding efficiency & 0.9\\
Cleanup \& analysis cuts & 0.7\\
Dilution factors: & \\
~~~~~Shape dependence $D_{t-int}$ & 0.47\\
~~~~~mixing of muon tag & 0.75\\
~~~~~muon tag misidentification & 0.9\\
~~~~~Time resolution and cuts & 0.95\\
~~~~~Background & 0.95\\
Total sensitivity & 0.07\\
\hline\hline
\end{tabular}\end{table}

\begin{figure}[htb]
\vspace{-.4cm}
\centerline{\psfig{figure=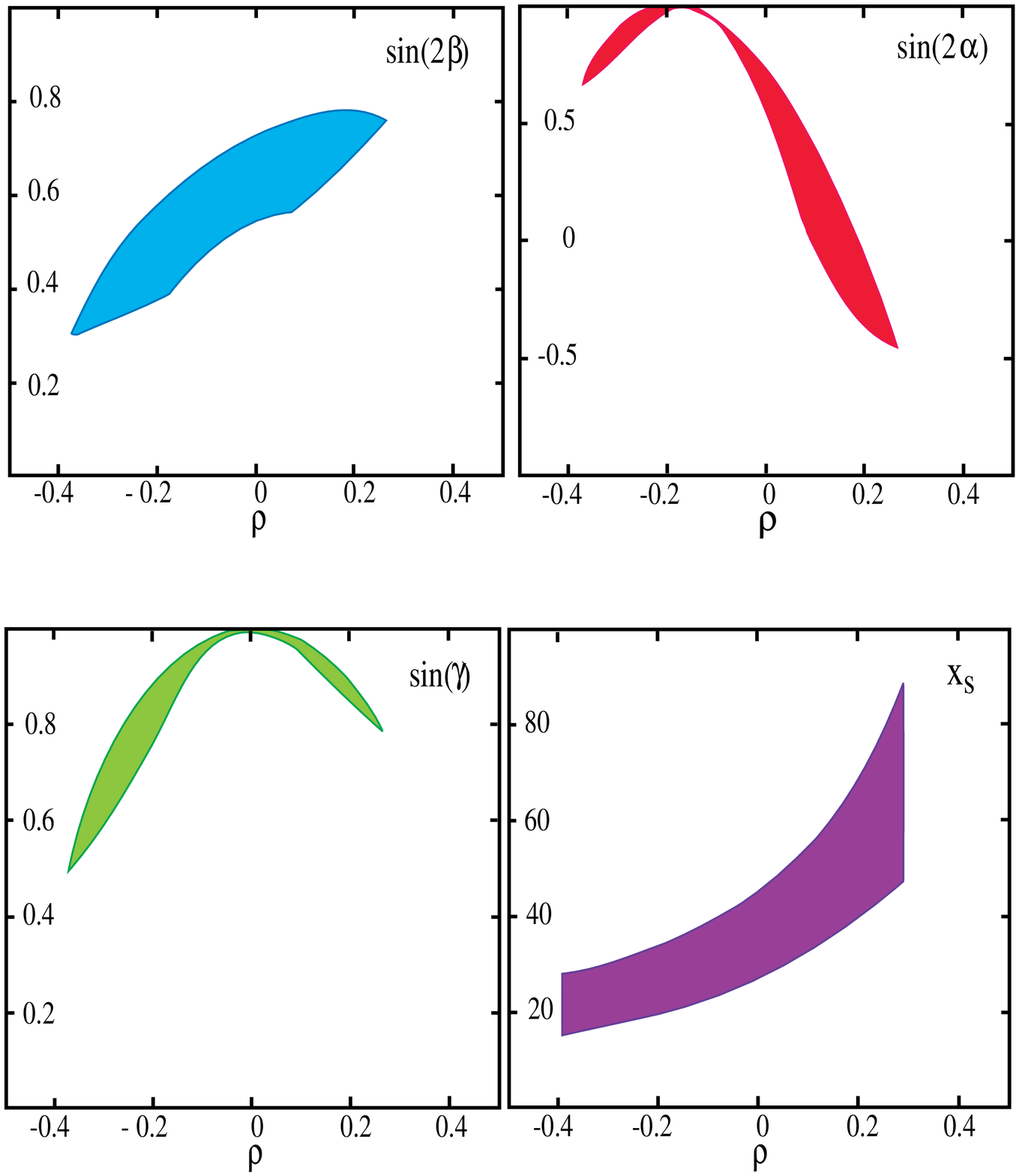,height=5.8in,bbllx=0bp,bblly=0bp,bburx=600bp,bbury=800bp,clip=}}
\vspace{-.8cm}
\fcaption{\label{allowed}The allowed values of three CP violating angles
and the $B_s$ mixing parameter $x_s$ as a function of $\rho$, taken from the
allowed region in Fig.~\ref{ckm_fig}.}
\end{figure}

This calculation shows an error in the asymmetry of 7\%. To see if that is in
the  range of interest, I show in Fig.~\ref{allowed} the expectations for the three CP
violating  angles and $x_s$. These plots merely reflect the ``allowed" region
shown in Fig.~\ref{ckm_fig}. It should be emphasized that this is not the
result of a sophisticated analysis,  which is difficult to do because of the
non-Gaussian nature of the theoretical  errors.

The decay modes which will probably be used to measure the CP violating angles 
are given in Table~\ref{table:modeCP}, with their branching ratios.
\begin{table}[th]\centering
\tcaption{Branching ratios for decay modes used in measuring CP violation}
\label{table:modeCP}
\vspace*{2mm}
\begin{tabular}{cccc}\hline\hline
CKM angle & Modes & ${\cal B}$ & Product ${\cal B}$\\
\hline
$\beta$ & $\psi K_S$ & $0.4\times 10^{-3}$ & $3.7\times 10^{-5}$\\\hline
$\alpha$ & $\pi^+\pi^-$ & $0.9\times 10^{-5}$ & $0.9\times 10^{-5}$\\\hline
$\gamma$\cite{GW} & $D^o K^-$ & $3.3\times 10^{-4}$ & $4.0\times 10^{-5}$\\
               & $\overline{D}^o K^-$ & $4.1\times 10^{-6}$ & $4.9\times 10^{-7}$\\
 & $D^o_{CP} K^-$ & $2.2\times 10^{-5}$ & $2.6\times 10^{-6}$\\\hline
$\gamma$\cite{GRLL} & $K^{\pm}\pi^o$, $\pi^{\pm}\pi^o$ & $\approx 10^{-5}$ & $\approx 
10^{-5}$\\
 & $K^o\pi^{\pm}$, $K^{\pm}\eta^{(')}$ & each & each\\
\hline\hline
\end{tabular}\end{table}

Finally, I list  in Table~\ref{table:expCP} the CP violation and $B_s$ mixing measurements of 
prime importance and my guess on which experiments, should they be built, are 
likely to perform these measurements and which could possibly perform them.

\begin{table}[th]
\centering
\tcaption{Prospects for CP violation and $B_s$ mixing measurements}
\label{table:expCP}
\vspace*{2mm}
\begin{tabular}{cccc}\hline\hline
Quantity& Modes & Possible & Likely\\\hline
$\sin(2\alpha)$ & $\pi^+\pi^-$ & Babar, Belle & LHC-B, BTEV\\
$\sin(2\beta)$ & $\psi K_S$ & HERA-B, CDF, CLEO & Babar, Belle, LHC-B, 
BTEV\\
$\sin(2\gamma)$ & $K \pi$ & Babar, Belle, CLEO & \\
$\sin(2\gamma)$ & $D^o K^-$ & Babar, Belle, CLEO & LHC-B, BTEV\\
$x_s$   & $\psi K^*$ & &LHC-B, BTEV\\
\hline\hline
\end{tabular}\end{table}

The $B$ system challenges us with the possibility of very diverse and important
measurements. Hopefully this physics will be done by the machines and
experiments in the next and future decades.
\newpage 
\section{Acknowledgements}
I have benefited greatly by physics discussion with many of my colleagues,
most recently with  M. Artuso, K. Berkelman, T. Skwarnicki, M. Witherell, J. Rosner and M. Gronau, M.
Neubert, C. Sachrajda, A. Ali and
A. Buras. I have also learned a lot from people associated with new $B$ efforts
including, J. Butler, C. Bebek, M. Procario, P. Mcbride, T. Ypsilantis
and P. Schlein. I also thank Thomas Ferbel and Barbara Ferbel for arranging such
an interesting  school. I learned a lot and so did Julia. My special thanks
to J. Rosner, K. Berkelman, M. Witherell and T. Skwarnicki for reading through
the manuscript and making many useful comments.


\begin{thebibliography}{99}
\bibitem{neut} 
Since the resolution on is about 2.5 worse than the
resolution on $|p_{miss}|$ ($\approx$110 MeV), the neutrino four-vector is
defined as $(E_{\nu},\overrightarrow{p}_{\nu})=
(|\overrightarrow{p}_{miss}|,\overrightarrow{p}_{miss}).$
\bibitem{pierre}
The use of asymmetric beam energies to study CP violation in $e^+e^-$ machines
operating the $\Upsilon (4S)$ was first suggested by Pierre Oddone.
\end{thebibliography}
\end{document}